\documentclass[preprint,linenumbers]{aastex631}
 \nolinenumbers
\usepackage{amsmath}
\def\f{\frac}

\def\l{\left}
\def\r{\right}
\def\d{{\mathrm{d}}}

\def\e1i{\epsilon_{1\mathrm{i}}}

\begin{document}

\begin{flushright}
  \footnotesize
  KCL-2023-61
\end{flushright}

\title{Importance of cosmic ray propagation on sub-GeV dark matter constraints}

\author{Pedro De la Torre Luque}
\affiliation{Instituto de Física Teórica, IFT UAM-CSIC, Departamento de Física Teórica,\\
Universidad Autónoma de Madrid, ES-28049 Madrid, Spain}
\affiliation{The Oskar Klein Centre, Department of Physics, \\
Stockholm University, Stockholm 106 91, Sweden}

\author{Shyam Balaji}
\affiliation{Physics Department, King’s College London, Strand, London, WC2R 2LS, United Kingdom}
\affiliation{Laboratoire de Physique Th\'{e}orique et Hautes Energies (LPTHE), \\ 
UMR 7589 CNRS \& Sorbonne Universit\'{e}, 4 Place Jussieu, F-75252, Paris, France}

\author{Jordan Koechler}
\affiliation{Laboratoire de Physique Th\'{e}orique et Hautes Energies (LPTHE), \\ 
UMR 7589 CNRS \& Sorbonne Universit\'{e}, 4 Place Jussieu, F-75252, Paris, France}



\begin{abstract}
We study sub-GeV dark matter (DM) particles that may annihilate or decay into SM particles producing an exotic injection component in the Milky Way that leaves an imprint in both photon and cosmic ray (CR) fluxes. Specifically, the DM particles may annihilate or decay into  $e^+e^-$, $\mu^+\mu^-$ or $\pi^+\pi^-$ and may radiate photons through their $e^\pm$ products. The resulting $e^\pm$ products can be directly observed in probes such as {\sc Voyager 1}. Alternatively, the $e^\pm$ products may produce bremsstrahlung radiation and upscatter the low-energy galactic photon fields via the inverse Compton process generating a broad emission from $X$-ray to $\gamma$-ray energies observable in experiments such as {\sc Xmm-Newton}. We find that we get a significant improvement in the DM annihilation and decay constraints from {\sc Xmm-Newton} (excluding thermally averaged cross sections of $10^{-31}$ cm$^3$\,s$^{-1} \lesssim \langle \sigma v\rangle \lesssim10^{-26}$ cm$^3$\,s$^{-1}$ and decay lifetimes of $10^{26}\,\textrm{s}\lesssim \tau \lesssim 10^{28}\,\textrm{s}$ respectively) by including best fit CR propagation and diffusion parameters. This yields the strongest astrophysical constraints for this mass range of DM of 1 MeV to a few GeV and even surpasses cosmological bounds across a wide range of masses as well.
\end{abstract}



    
\section{Introduction}

The Dark Matter (DM) problem is one of the main open issues in modern physics. DM composes around 85\% of the total matter in the Universe and yet its nature is currently unknown (see, e.g.,~\cite{Arbey_2021}). If DM is a particle, the most well studied candidate is the weakly interacting massive particle (WIMP). However WIMPs are yet to be detected and therefore other candidates for particle DM have recently attracted significant attention. In particular, particles that are lighter than WIMPs ($m_\chi \simeq 1$ MeV-few GeV), known in the literature as sub-GeV or \emph{light} DM, have been proposed and are theoretically well motivated \citep{Feng:2008ya, Hochberg:2014dra, Petraki:2013wwa}. We refer the reader to~\citet{Cirelli:2020bpc} which covers these aspects in more detail. We also note that indirect detection of light DM is challenging for two reasons: i) due to solar activity, the heliosphere is effectively a barrier to low-energy charged particles and therefore the flux of such particles is greatly reduced at Earth, ii) current $\gamma$-ray observatories have a lack of sensitivity in the $E_\gamma \simeq$ 100 keV-100 MeV energy range, commonly named the ``MeV gap.'' These two issues mitigate detection of the expected signal from light DM annihilation or decay and therefore need to be circumvented in order for stronger constraints to be derived.

Light DM particles inject low energy electrons (from now on, unless stated otherwise, we use ``electrons" to refer to both $e^\pm$) via their decay/annihilation which can leave imprints in the diffuse electron flux that we detect at Earth, as well as the $X$-ray to $\gamma$-ray Galactic diffuse emission because of bremsstrahlung and inverse Compton scattering (ICS) of the injected electrons. 
In this paper, building on studies of light DM such as~\citet{Boudaud:2016mos, Cirelli:2023tnx} and feebly interacting particles such as~\citet{DelaTorreLuque:2023nhh,DelaTorreLuque:2023huu}, we compute updated limits on the electron injection from light DM. For example, in recent years the {\sc Voyager 1} probe has measured the local flux of $e^\pm$ outside the heliosphere \citep{cummings2016galactic} at energies below tens of MeV. These measurements have the key advantage that they are not significantly affected by the solar modulation effect \citep{Potgieter:2013pdj}, which greatly reduces the flux of low-energy charged cosmic rays (CRs) at Earth. In this manner, using {\sc Voyager 1} electron flux data enables mitigation of the aforementioned solar modulation issue.

Additionally, we consider the production of secondary $X$-rays, in particular due to ICS of DM-produced electrons on galactic ambient radiation. This method was proven to be powerful in~\citet{Cirelli:2020bpc, Cirelli:2023tnx}. Therefore, we follow a similar approach, namely by using the most constraining dataset in~\citet{Cirelli:2023tnx} i.e. {\sc Xmm-Newton}. The main goal of this work is to perform a more realistic computation of the $X$-ray and $\gamma$-ray signals that would be produced by DM annihilation or decay and a reevaluation of the constraints derived in~\citet{Cirelli:2023tnx} from {\sc Xmm-Newton} data by considering the full-fledged CR transport in the Milky Way, particularly the impact of CR reacceleration on the constraints. In the same manner, we also update the limits obtained in~\citet{Boudaud:2016jvj} using {\sc Voyager 1} $e^\pm$ data. For this study we use a fully numerical approach that does not need approximations and utilise current state-of-the-art propagation setups.

The paper is arranged as follows, in Sec.~\ref{sec:Methodology} we outline the propagation methodology and secondary photon emission, in Sec.~\ref{sec:Results} we outline the main DM signals and constraints, in Sec.~\ref{sec:Discussion} we discuss the key findings and compare our results with other relevant light DM constraints and finally we conclude in Sec.~\ref{sec:Conclusion}.

\section{Electron-positron propagation and secondary radiations}
\label{sec:Methodology}
While the main novelty of this work revolves around a more detailed characterization of the propagation of electrons produced by light DM and their interactions with the Galactic environment, we refer the reader to~\citet{Cirelli:2020bpc} for a detailed description of the computation of their production spectra. In this section, we briefly review the main framework for generating the propagated spectra of electrons and $X$-ray signals from DM annihilation/decay. 

As a first step, in order to compute the distribution and energy spectrum of electrons produced by DM annihilation/decay in the Galaxy, we use a customized version of the {\tt DRAGON2} code~\citep{DRAGON2-1, DRAGON2-2}, which includes a few updates compared to the current public version, such as the possibility of production of line-like signals from annihilation and decay (like from the $\chi \overline{\chi} \rightarrow e^+e^-$ process) or the inclusion of antinuclei propagation and production. We have made this version publicly available~\citep{de_la_torre_luque_2023_10076728}.
{\tt DRAGON2}\footnote{\url{https://github.com/cosmicrays/DRAGON2-Beta\_version}} itself is a dedicated CR propagation code designed to self-consistently solve the diffusion-advection-loss equation~\citep{Ginz&Syr, DRAGON2-1}

\begin{equation}
\begin{split}
\nabla \cdot \left( D\vec{\nabla} f_i + \vec{v}_c f_i \right) &- \frac{\partial}{\partial p_i} \left[ \dot{p}_i f_i - p_i^2 D_{pp} \frac{\partial}{\partial p_i}\left( \frac{f_i}{p_i^2} \right) \right. \\
&\left. - \frac{p_i}{3} \left( \vec{\nabla} \cdot \vec{v}_c f_i \right) \right] = Q_i + \sum_j \left( c\beta_i n_{\textrm{ism}}\sigma_{j\rightarrow i} + \frac{1}{\gamma_j\tau_{j\rightarrow i}} \right) f_i \\
&- \left( c\beta_i n_{\textrm{ism}}\sigma_i + \frac{1}{\gamma_i\tau_i} \right) f_i
\end{split}
\label{eq:CRtransport}
\end{equation}

which describes CR transport for species $i$ with density per momentum unit $f_i \equiv dn_i/dp_i$ involved in the CR network, including particles of both astrophysical and exotic origin (e.g. via DM annihilation or black hole evaporation). The equation takes into account spatial diffusion through the coefficient $D$. $\dot{p}_i$ takes into account momentum losses due to interactions with the interstellar medium (ISM), $D_{pp}$ is the momentum diffusion coefficient (discussed in detail in Sec.~\ref{sec:Results}), $\vec{v}_c$ the Galactic wind velocity responsible for CR advection/convection. $Q_i$ describes the ``primary" source of species $i$, while the second right-hand side term takes into account the production of $i$ from the interaction with the ISM and decays of heavier particles. Here $c\beta_i$ is the velocity of individual particles $i$, $n_\textrm{ism}$ is the ISM number density, $\sigma_{j\rightarrow i}$ the cross-section of the destruction of heavier species $j$ from interactions with the ISM which produces $i$, $\gamma$ is the Lorentz boost factor and $\tau_{j\rightarrow i}$ is the half-life of sole decays of $j$ into $i$. The last right-hand side term describes the destruction of species $i$ from its interaction with the ISM and decays. $\sigma_i$ is the cross-section of the destruction of $i$ from interactions with the ISM and $\tau_i$ is the half-life of the decays of $i$. Since we are studying transport of electrons ($i =e$) produced by light DM, only the primary source term $Q_e$ is relevant in the right-hand side of Eq.~\eqref{eq:CRtransport}. In this case
\begin{equation}
    Q_e (\vec{x},E_i) = 
    \left\{
\begin{array}{ll}
	\displaystyle  
        \frac{\langle \sigma v\rangle}{2} \left(\frac{\rho_\chi(\vec{x})}{m_\chi}\right)^2\frac{dN_e^{\textrm{ann}}}{dE_e} &\text{(annihilation)}\\
	\\
        \displaystyle
        \ \; \Gamma \ \; \left(\frac{\rho_\chi(\vec{x})}{m_\chi}\right)\frac{dN_e^{\textrm{dec}}}{dE_e} &\text{(decay)}
    \end{array}
\right.
,
\end{equation}
where $\langle \sigma v\rangle$ and $\Gamma$ are the DM thermally averaged annihilation cross section and the DM decay rate, respectively. $\rho_\chi(\vec{x})$ is the DM energy density at the position $\vec{x}$ and $dN_e/dE_e$ the injection spectrum of DM-produced electrons. 
The three following DM annihilation/decay channels are relevant for the sub-GeV DM case that we study here
\begin{align}
\chi \overline{\chi} &\to e^+e^-,&
\chi \overline{\chi} &\to \mu^+\mu^-,&
\chi \overline{\chi}&\to \pi^+\pi^-, 
\label{eq:channels}
\end{align}
which are kinematically open whenever $m_\chi > m_i$ (annihilations) or $m_\chi > 2 m_i$ (decays) where $i=e,\mu,\pi$ and produce electrons as some point in the particle cascade. The injection spectra $dN_e/dE_e$ for each channel are calculated following the prescriptions given in~\citet{Cirelli:2020bpc}. Also, an equally important component of CR transport that is not shown in the equation is the ``halo height'' $H$, which describes the maximal height (boundary) of diffusion of species $i$, such that $\frac{dn_i}{dp_i}(R, z > H) = 0$, where $R$ and $z$ are the galactocentric cylindrical coordinates.

The transport equation features fully position- and energy-dependent transport coefficients in two-dimensional (assuming cylindrical symmetry) and three-dimensional configurations of the Galaxy structure (i.e. considering the spiral arm pattern of the Galaxy).
We solve the CR transport equation assuming cylindrical symmetry of the Galaxy, which is most commonly used in CR studies \citep{Taillet:2002ub,Strong:1998pw}. Cylindrical symmetry also matches the general shape of the Galaxy, which comprises a thin disk surrounded by a thick halo. Since the DM density in the Galaxy follows a radial distribution, it is reasonable to solve the CR transport equation with this assumption.
In these simulations, we compute the signals produced by light DM annihilations from $m_{\chi} = m_{i}$ for annihilation or decays from $m_{\chi} = 2 m_{i}$ up to $m_{\chi}=5$~GeV (where $i=e,\mu,\pi$ depending on the channel of interest), taking $3$ masses per decade. For the galactic DM energy distribution, we adopt a Navarro-Frenk-White (NFW) profile~\citep{Navarro:1995iw}. However, our results are sensitive to the choice of the DM profile since we are looking at regions of interest near the GC. In~\cite{Cirelli:2021uoc}, they show that this choice can impact the final limits up to $\mathcal{O}(60\%)$ and we expect that it would be the same in our study. The bounds from {\sc Voyager 1} are expected to be a bit less sensitive to the DM profile, since they come from local data, where the dark matter density is known relatively well. We use a spatial grid with a resolution of $\simeq 150$~pc and an energy resolution of $5\%$, from $50$~keV to $10$~GeV. We have tested that our results do not change appreciably for lower energy or spatial resolution.
At these energies, the relevant energy losses are mainly due to Coulomb and ionisation interactions with the interstellar gas, as well as bremsstrahlung losses, whose time-scales become dominant below energies of a few tens of MeV (see Fig.~4 in the Supp. Material of~\cite{Boudaud:2016mos} for a comparison of the relevant energy-loss time scales for $e^{\pm}$). In the {\tt DRAGON2} code, all relevant sources of energy losses (ionization, Coulomb, bremsstrahlung, adiabatic, inverse Compton and synchrotron losses) are included and solved numerically (without using approximations). We use the Galactic gas distribution model implemented by the {\sc Galprop} group~\citep{Moskalenko:2001ya, Ackermann:2012pya}\footnote{A recent update on the {\sc Galprop} gas model was published in~\cite{Porter_2022}, which should not differ very significantly from the one used in previous versions.}, with a gas composition assumed to be a mixture of hydrogen and helium nuclei with uniform density ratio $f_{\rm He} = 0.1$.

We parameterise the energy dependence of the diffusion coefficient as a broken power-law with a break at rigidity $R_b\simeq 312$~GV~\citep{genolini2017indications}
\begin{equation}
 D (R) = D_0 \beta^{\eta}\frac{\left(R/R_0 \right)^{\delta}}{\left[1 + \left(R/R_b\right)^{\Delta \delta / s}\right]^s}\;,
\label{eq:diff_eq}
\end{equation}
and assume for simplicity that this value is uniform everywhere in the Galaxy. 
The propagation parameters used are given in Tab.~\ref{tab:params}, which have been obtained from a combined fit to AMS-02 data~\citep{aguilar2018observation, Aguilar:2018njt, AMS_LiBeB, AMS_gen} for B, Be and Li ratios to C and O, using the Markov-chain Monte Carlo analysis reported in~\citep{DeLaTorreLuque:2021ddh, DeLaTorreLuque:2021nxb}. For ease of reproducibility, the full \texttt{DRAGON} input and outputs files are publicly available\dataset[here]{https://doi.org/10.5281/zenodo.10076728}
, where the comparisons to data are also shown. We provide details of the analysis and comparisons to CR data in Appendix~\ref{sec:appendix0}. While the parameters $R_b$, $\Delta\delta$ and $s$ are only important for the regime above $\sim312$~GeV, we fit the parameters $H$, $D_0$, $\eta$, $\delta$, $H$ and $v_A$ in this analysis, leaving $R_0$ fixed to $4$~GV, since it is just the rigidity where the diffusion coefficient is normalized. All of the parameters are reported in Tab.~\ref{tab:params}. We discuss the impact of the uncertainties involved in the diffusion model employed in Sec.~\ref{sec:Discussion} and dedicate an appendix to show how variations of different parameters affect our results in App.~\ref{sec:appendixB}. 
\begin{table}[t!]
    \centering
    \begin{tabular}{|c|c|c|}
    \hline
Halo height&      $H$&$8.00^{+2.35}_{-1.96}~\rm{kpc}$\\
Norm. of Diffusion coeff.&    $D_{0}$  & $1.02^{+0.12}_{-0.10}\times10^{29}~\rm{cm}^{2}\rm{s}^{-1}$ \\
Norm. rigidity    &    $R_{0}$ & $4~\rm{GV}$ \\
Diffusion spectral index &    $\delta$&$0.49\pm 0.01$  \\
$\beta$ exponent &  $\eta$&      $-0.75^{+0.06}_{-0.07}$\\
Alfvén velocity&  $v_{A}$&$13.40^{+0.96}_{-1.02}~\rm{km/s}$\\
Break rigidity &       $R_{b}$&$312 \pm 31~\rm{GV}$ \\
Index break &        $\Delta\delta$&$0.20 \pm 0.03$ \\
Smooth. param. &     $s$&$0.04 \pm 0.0015$\\
\hline
\end{tabular}
\caption{Main propagation parameters used in our analysis with the uncertainties related to their determination. Uncertainties in the parameters $R_b$, $\Delta\delta$ and $s$ come from \citep{genolini2017indications}. }
\label{tab:params}
\end{table}

Once the diffuse (steady-state) distribution of electrons in the Galaxy is obtained, we make use of the {\tt HERMES} code~\citep{Dundovic:2021ryb} to integrate along the line of sight the CR spatial and energy distributions obtained with {\tt DRAGON2}, using detailed interstellar radiation field (ISRF) models~\citep{Vernetto:2016alq} to get high-resolution sky maps of the diffuse $\gamma$-ray emission at the relevant energies. 
With this setup we compute the ICS emission from both electrons and positrons interacting with the different ISRFs. In general, the impact of the ISRFs in the $e^+e^-$ spectra is negligible for sub-GeV particles since ICS scales as $E^2$. They are, in turn, fundamental for the X-ray secondary emission, since IC emission is directly proportional to the number density of ISRF photons. In the energy range of interest for the {\sc Xmm-Newton} data ($\simeq 1$-$10$ keV), the main components are IC on CMB and IC on the radiation emitted by Galactic interstellar dust heated by stellar light. Therefore, following \cite{Vernetto2016prd}, we estimate that the uncertainties coming from the ISRF model used must be lower than $30\%$.
We have also checked that the bremsstrahlung has no impact on our results and have accounted separately for final state radiation (FSR) signals, which are computed in~\citet{Cirelli:2023tnx}, as discussed below.

\medskip

\section{Light dark matter signals and constraints}
\label{sec:Results}

In this work, we employ a state of the art propagation setup and an updated parametrisation of the energy dependence of the diffusion coefficient. However, we remind the reader that the current CR analyses used to characterize the propagation processes are mostly restricted to the availability of data on secondary CRs (mainly B, Be and Li), for which the AMS-02 collaboration~\citep{aguilar2018observation} has measured data only from a few hundred MeV. This implies that our knowledge on the transport process of charged particles in the Galaxy below hundreds of MeV is still very limited and there is no robust estimation of the diffusion coefficient below $\simeq 100$ MeV since different assumptions of the diffusion setup are able to reproduce the current local data with relative accuracy (see \cite{Weinrich_2020} and \cite{silver2024testing}). In fact, the recent data from AMS-02 on B, Be and Li has revealed a statistically significant change on the energy dependence of the diffusion coefficient at sub-GeV energies~\citep{Weinrich_2020, DeLaTorreLuque:2021nxb} which could be explained by damping of Alfvénic waves~\citep{Ptuskin_2006, Reichherzer:2019dmb, Fornieri:2020wrr}, and different regimes of turbulence can also appear at lower energies. 

Here, we include the change on the trend of diffusion at sub-GeV energies, which can result in sizeable differences on the fluxes of electrons produced by low-mass DM particles.
In particular, we employ a diffusion coefficient that includes a factor $\beta^{-0.75}$ (Eq.~\eqref{eq:diff_eq}, being $\beta$ the dimensionless speed of the particles as a ratio of the speed of light) that implies a rise in diffusion at sub-GeV energies. 
Another important novelty in our calculations is the study of the effect of reacceleration by magnetohydrodynamic (Alfvénic) turbulence in the Galaxy, which is regulated in our evaluations by the Alfvén speed and is directly linked to the momentum diffusion coefficient $D_{pp}$ in Eq.~\eqref{eq:CRtransport}~\citep{Strong_1998, osborne1987cosmic, seo1994stochastic}.
\begin{equation}
    D_{pp} = \frac{4}{3}\frac{1}{\delta(4-\delta^2)(4-\delta)}\frac{v_A^2p^2}{D}\;.
\end{equation}
We emphasise that the diffusive motion of CRs grants an unavoidable level of reacceleration on the particles (that can be thought as the exchange of energy of charged particles with plasma waves). Both~\citet{Boudaud:2016mos, Cirelli:2023tnx} were the first sub-GeV DM constraints from {\sc Voyager 1} and {\sc Xmm-Newton} data respectively, however they did not study how these signals are affected by reacceleration in detail. However, as we show below, this may cause dramatic changes in the sub-GeV electron signals that we study here.

We analyse each annihilation/decay channel separately, even though, in principle, DM could annihilate or decay to a mix of modes in specific models. In the case of the $X$-ray signals generated by the electrons injected by DM, we consider that the total photon flux has two parts: (i) the immediate emission from the charged particles in the final state (commonly known as FSR emission) and (ii) the delayed emission of photons from ICS by the energetic $e^\pm$ injected by DM processes. The immediate emission has two components: FSR from the charged leptons or pions in the final state, and radiation from muons or pions that decay with an extra photon ($\mu \to e\nu_e\nu_\mu\gamma$, $\pi \to \ell \nu_l\gamma$, with $\ell=e,\mu$ covering both particles and antiparticles and can be modified accordingly).


\subsection{Galactic electron-positron flux induced by dark matter}
\label{sec:El_flux}

\begin{figure}[ht!]
\includegraphics[width=0.54\linewidth]{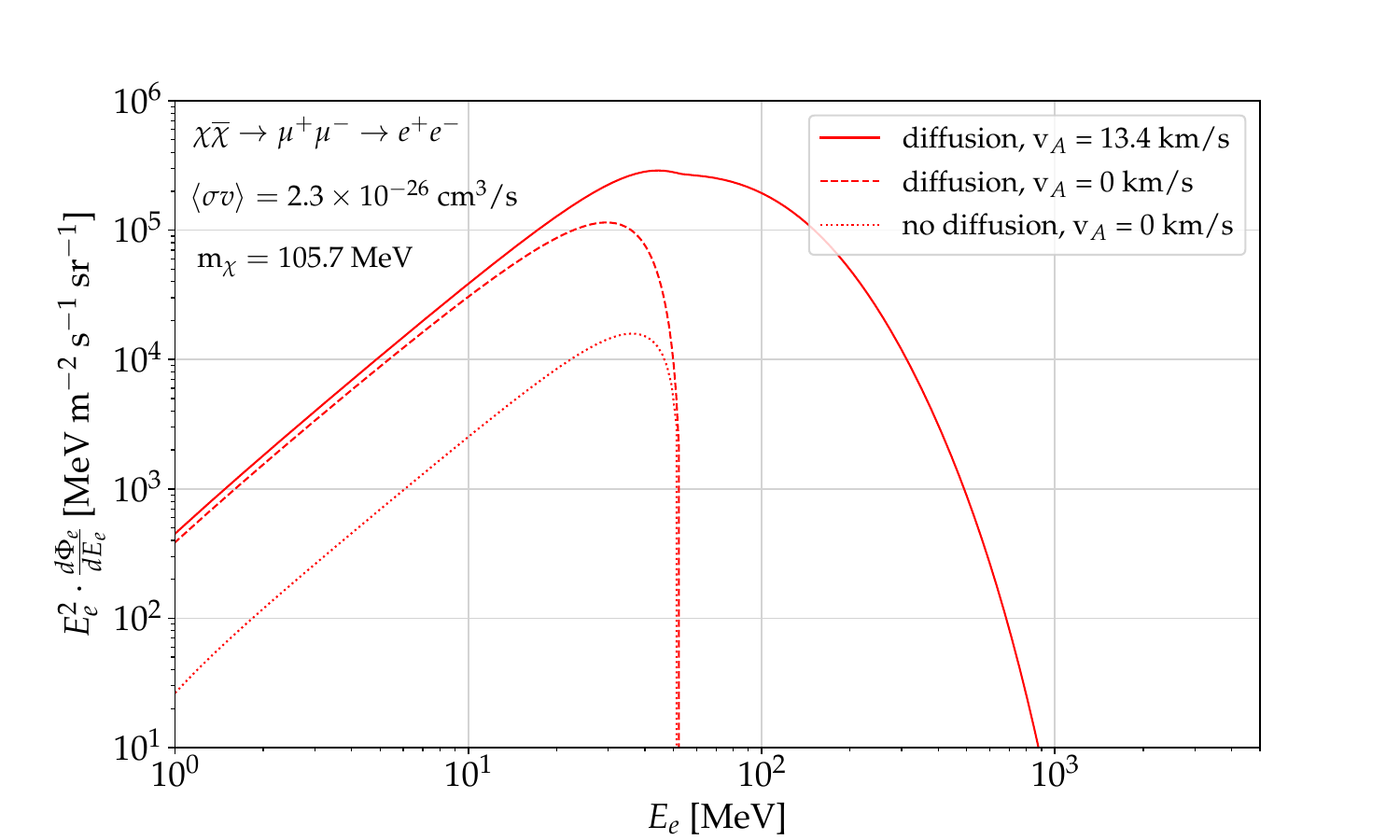} \hspace{-0.7cm}
\includegraphics[width=0.54\linewidth]{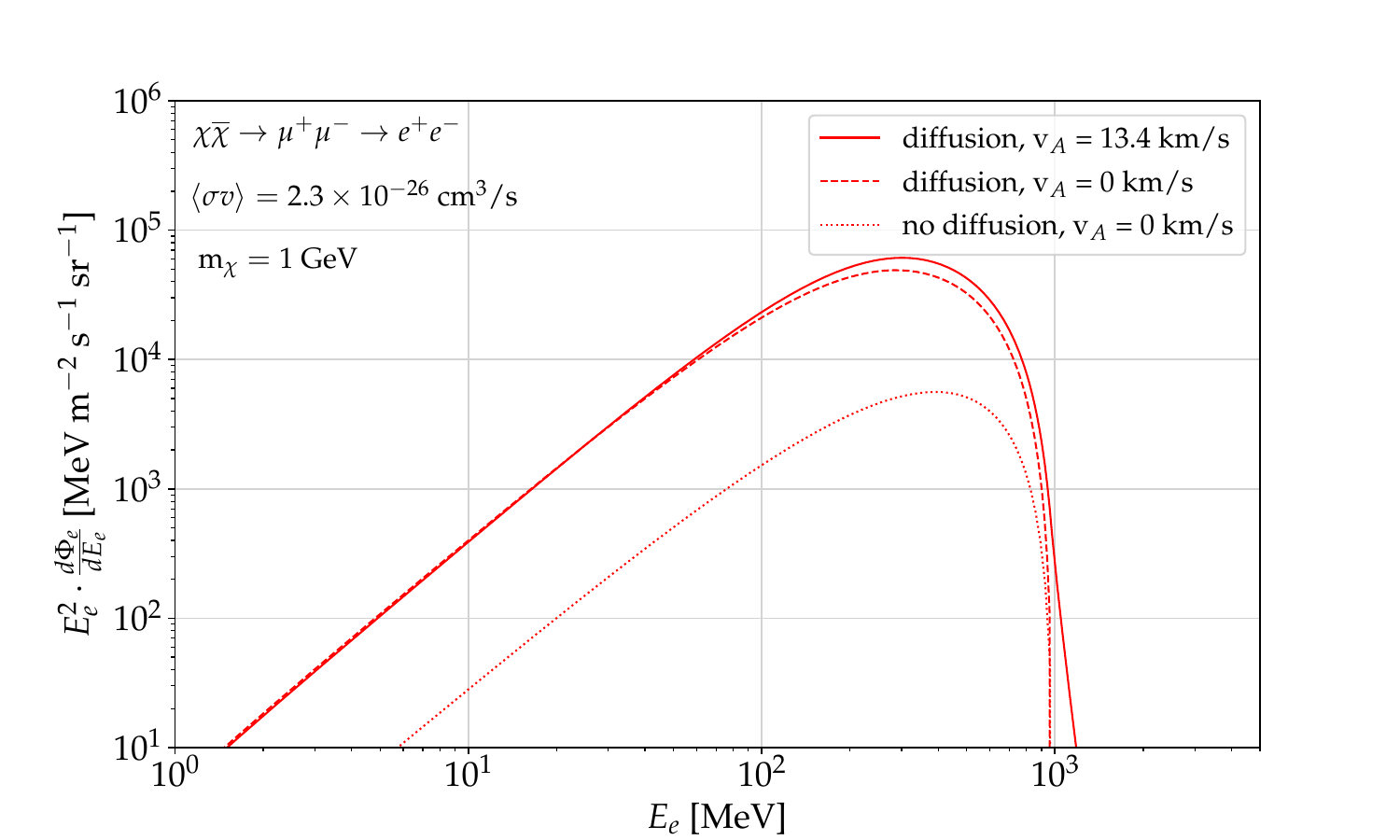}
\caption{Comparison of the predicted $e^\pm$ fluxes at Earth from DM annihilation to muons for $m_{\chi} = m_\mu = 105.7$ MeV (left panels) and $m_{\chi} = 1$ GeV (right panels) showing the impact of spatial diffusion and reacceleration in the electron flux. The solid lines describe the scenario including $e^\pm$ diffusion and reacceleration, the dashed lines the predicted signal with no reacceleration included and the dotted line represents the case where no propagation of $e^\pm$ is considered (only energy losses).}
\label{fig:diff+vA}
\end{figure}

To emphasise the importance of including all the relevant effects associated with the propagation of electrons~\cite{1974Ap&SS..29..305B} in our light DM constraints we show, in Fig.~\ref{fig:diff+vA}, the DM-induced local interstellar electron spectrum from DM annihilation  for a fixed thermally averaged cross section of $\langle \sigma v\rangle= 2.3\times 10^{-26}$ cm$^3$\,s$^{-1}$ for different propagation setups. In the left panel, we fix the DM mass to the muon mass of $105.7$~MeV while in the right panel we fix it to $1$~GeV. Here, we are comparing a setup (labelled ``no diffusion") where we switch off any particle diffusion effects, but include energy losses of the particles injected by a NFW DM distribution. Then, we consider a more realistic situation where we enable diffusion of particles, with the parameters in Tab.~\ref{tab:params}, but no reacceleration. Finally, we allow all these propagation effects to be active. As one can see, the different setups yield quite different predictions that would lead to constraints that differ significantly. However, it is important to mention that the difference in the results from different setups (with and without diffusion) depend on the exact parameters in our diffusion coefficient and, given the uncertainties present in current CR studies, they can be as high as an order of magnitude. We illustrate the effect of changes in the diffusion parameters, and the halo height (keeping the ratio $D_0/H$ to be the best-fit value found from the analysis of secondary CRs -- see Appendix.~\ref{sec:appendix0}) in our predictions in Figs.~\ref{fig:ScanParams} and~\ref{fig:halosize} (Appendix.~\ref{sec:appendixB}), respectively. From these variations, we observe that the predicted $e^{\pm}$ DM signals are significantly more affected by changes in the halo height, producing variations of up to a factor of a few. This parameter regulates the volume of propagation of charged particles in the Galaxy and constitute one of the main uncertainties in current CR studies~\citep{Ginzburg_H, Evoli_H, DeLaTorreLuque:2021yfq, Maurin_H}.
Additionally, from Fig.~\ref{fig:diff+vA} we see that reacceleration promotes low energy electrons to energies well above the mass of the DM particle producing them (left panel). This, in fact, can cause huge changes in the signals predicted, even for moderate levels of reacceleration (i.e. values of $v_A\gtrsim5$~km/s). In the higher mass case (right panel) there is no significant broadening of the spectrum due to reacceleration since much more energy is needed to promote particles above the GeV scale. 
On top of this, we highlight that the effect of possible spatial dependence of diffusion in the Galaxy is not easily quantifiable, since constraints are still very weak and the theoretical models of plasma-CR interactions are far from offering clear predictions. Therefore, further uncertainties may be present and change our predictions in zones of the Galaxy far from us (where the diffusion coefficient may be different to the local one).

\begin{figure*}[t!]
\includegraphics[width=0.5\linewidth]{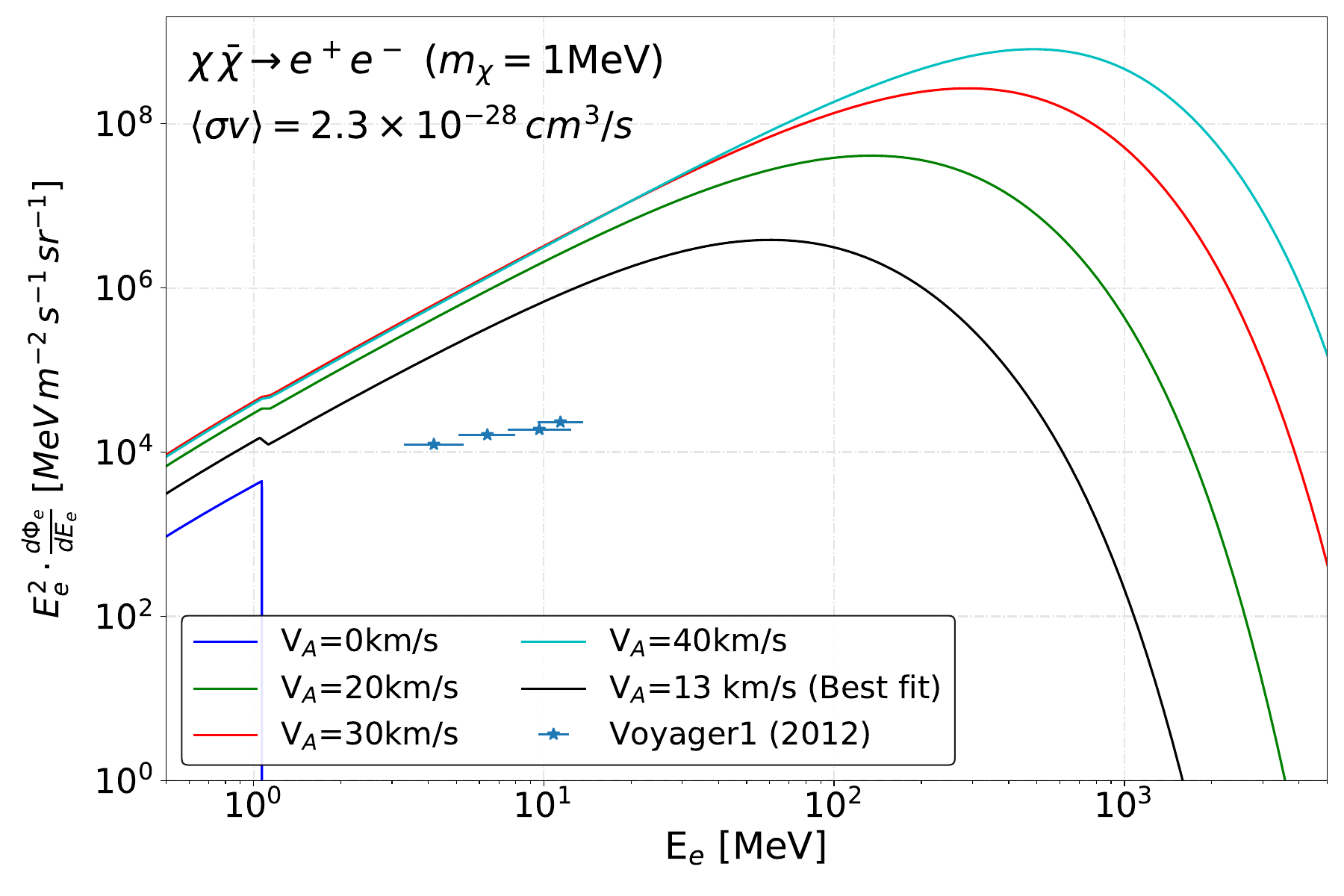}
\includegraphics[width=0.5\linewidth]{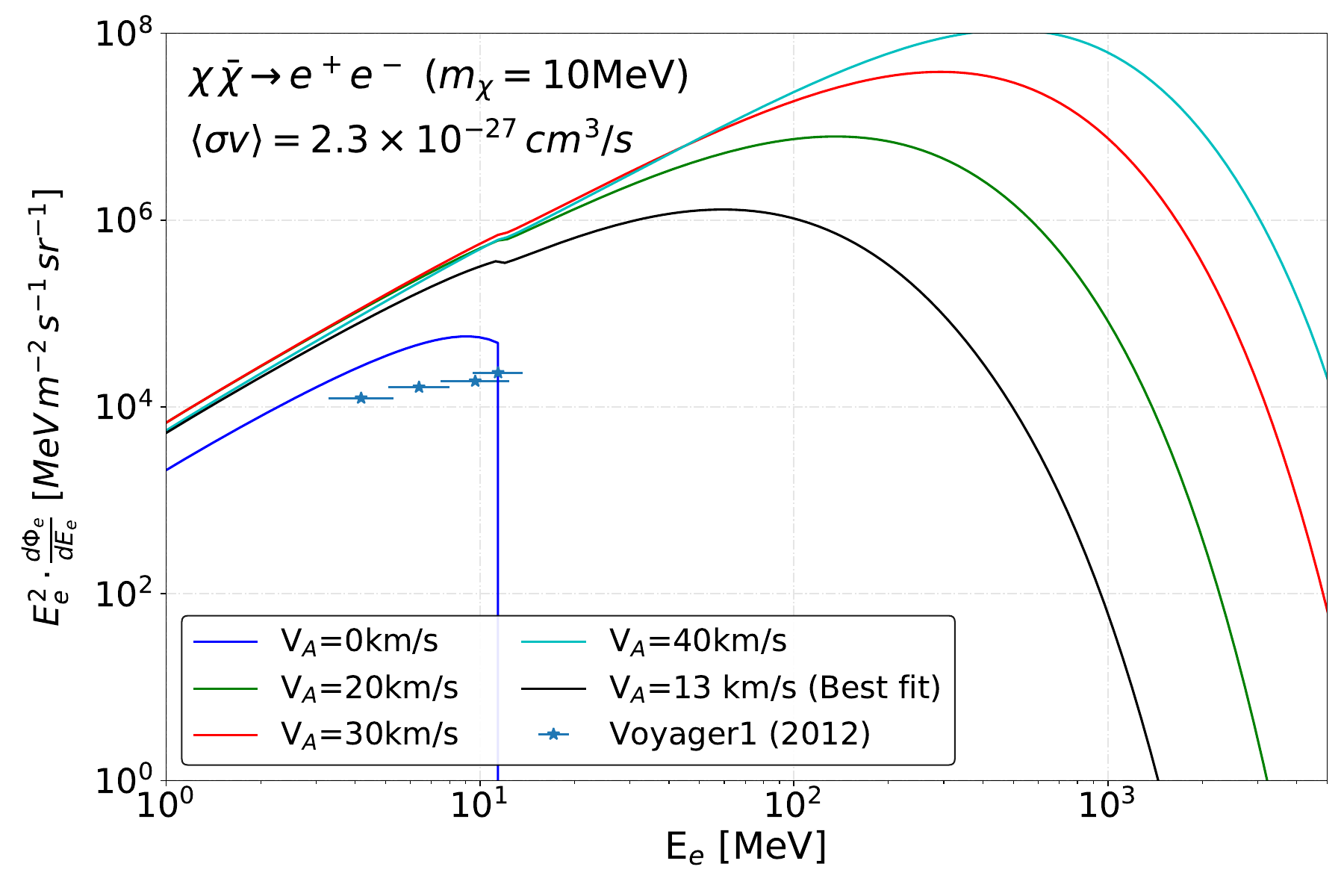}

\includegraphics[width=0.5\linewidth]{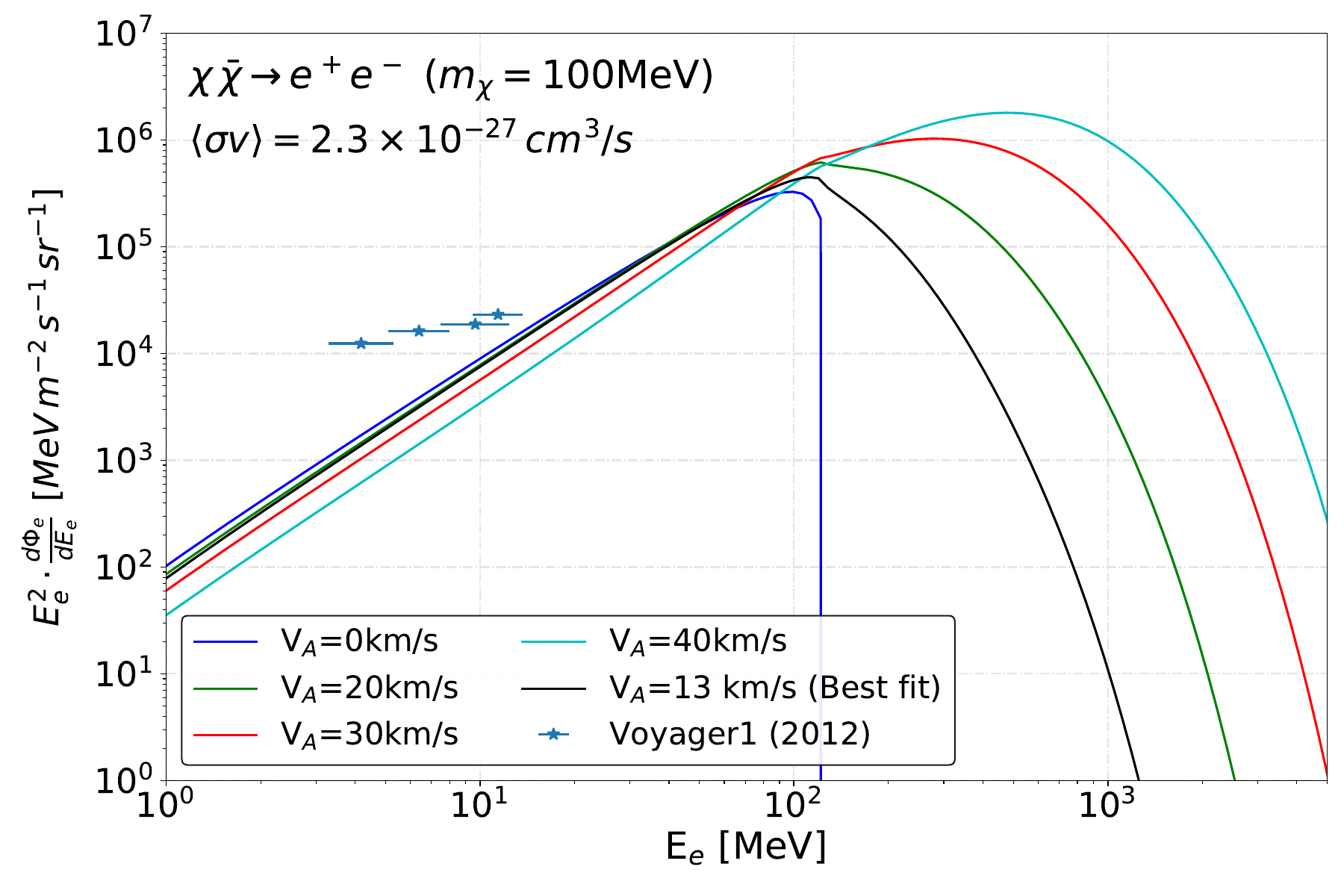}
\includegraphics[width=0.5\linewidth]{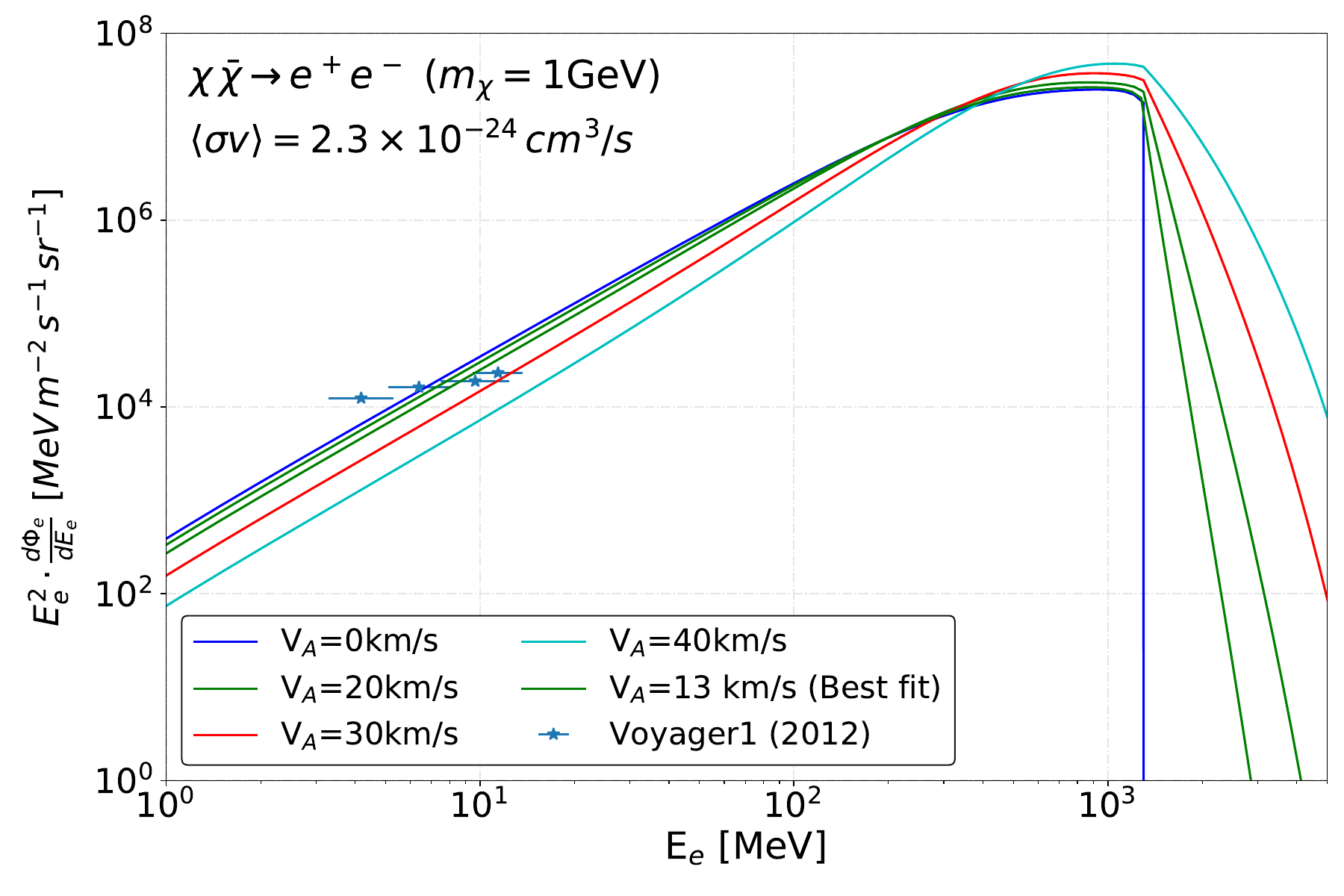}
\caption{Comparison of the predicted $e^\pm$ flux measured by {\sc Voyager 1} as a function of electron energy from dark matter annihilation directly in $e^+e^-$ states. We consider dark matter masses, $m_\chi$, of $\simeq 1$~MeV (top left panel), $\simeq 10$~MeV (top right panel) and $\simeq 100$~MeV (bottom left) and $\simeq 1$ GeV (bottom right panel). We show the {\sc Voyager 1} data as blue points and the various re-acceleration scenarios with curves derived for Alfvén velocities of $0$ (blue), 13 (black), 20 (green), 30 (red) and 40 (cyan) km/s. }
\label{fig:Va_ee_Scan_Voy}
\end{figure*}

The mass dependence of the DM electron signals is shown in Appendix.~\ref{sec:appendixA}, where we depict, in Fig.~\ref{fig:Voy_Comp}, the predicted local interstellar electron spectra ($e^\pm$) produced from annihilation (left panels) and decay (right panels) of DM of different masses from our best-fit propagation parameters, compared to {\sc Voyager 1} data. 
As we expect, similar masses yield very similar signals, with the difference being a bit more pronounced in the direct $e^\pm$ channel, which would simply be a line in the absence of energy losses. 
We remind the reader that the {\sc Voyager 1} measurements consist of the sum of $e^\pm$ outside the heliosphere, therefore, the predicted spectra that we are showing is the predicted $e^\pm$ local interstellar flux (i.e. no effect of solar modulation in the electrons is expected). In addition, since our simulation includes in-flight annihilation losses of positrons and triplet pair production~\citep{Gaggero:2013eik}, the amount of $e^+$ is not exactly the same as that of $e^-$, although they do not differ by more than a few percent.

Given the importance of reacceleration in the constraints of low-mass DM particles, we show, in Fig.~\ref{fig:Va_ee_Scan_Voy}, the local electron spectrum produced by annihilation of DM particles, in the direct $e^+ e^-$ channel, for masses from $\simeq1$~MeV (top left panel) to $\simeq 1$~GeV (bottom right panel) at different values of the Alfvén velocity, ranging from $v_A=0$~km/s to $v_A=40$~km/s.
In each case, the predicted best-fit $v_A$ ($v_A\simeq 13$~km/s, see Tab.~\ref{tab:params}) is indicated as a black line, and constitute our main prediction. 
We do not consider values greater than $v_A=40$~km/s because recent detailed CR analyses do not predict values larger than this~\citep{DeLaTorreLuque:2021nxb, Weinrich_2020} and because a larger $v_A$ would imply that the CR acquire more energy via interactions with interstellar turbulence than from their injection in supernova remnants~\citep{Ptuskin_2006}.
This illustrates how dramatic the effect of including reacceleration is: first, for DM masses below the energy range of {\sc Voyager 1} data, even a low level of reacceleration ($v_A\gtrsim 5$~km/s) is enough to boost the electrons beyond the {\sc Voyager 1} data range of energies, thus strengthening DM constraints. Second, we can see that different levels of reacceleration (i.e. different values of $v_A$) very significantly impact the flux of electrons and the shape of the spectrum at high energy as well. 

\subsection{Secondary emissions from DM annihilation/decay products}
\label{sec:secondaryemission}

Electrons from DM annihilation/decay interact with ISRFs and gas as they traverse the Galaxy. ISRFs are mainly composed by three components (light from stars, dust and the CMB) which produce light at different wavelengths such as optical, UV, IR and microwave. These interactions produce a continuous emission from $X$-rays to $\gamma$-rays, depending on the energy of the $e^\pm$ and the ISRF photons. The emission comes from two processes: ICS, which increases the energy of the ISRF photons, and bremsstrahlung, which occurs when $e^\pm$ collide with the interstellar gas. By observing the emission at MeV in $\gamma$-rays, we can deduce important information about the population or particles injected by DM annihilation or decay.

We can calculate the differential flux of the prompt emissions by integrating the emissions along the line of sight (l.o.s.) in a given direction $\theta$, which is the angle from the direction to the Galactic Center (GC). We use $s$ as a parameter for the l.o.s. The formula is

\begin{equation}
\label{eq:promptflux}
   \frac{d\Phi_{{\rm prompt} \, \gamma}}{dE_\gamma \, d\Omega}=\frac{1}{4\pi}\frac{dN_{{\rm prompt} \, \gamma}}{dE_\gamma}
  \times \left\{
\begin{array}{ll}
	\displaystyle  
        \frac{\langle \sigma v\rangle}{2} \int_\text{l.o.s.}ds\,\left(\frac{\rho_\chi(r(s,\theta))}{m_\chi}\right)^2 &\text{(annihilation)}\\
	\\
        \displaystyle
        \ \; \Gamma \ \; \int_\text{l.o.s.}ds\,\left(\frac{\rho_\chi(r(s,\theta))}{m_\chi}\right) &\text{(decay)}
    \end{array}
\right.    
,
\end{equation}
where the expressions for the photon spectra $dN_{{\rm prompt}\,\gamma}/dE_\gamma$ (where prompt = FSR or Rad) are given in~\citet{Cirelli:2020bpc}. 

\begin{figure*}[t!]
\includegraphics[width=0.5\linewidth, height=0.22\textheight]{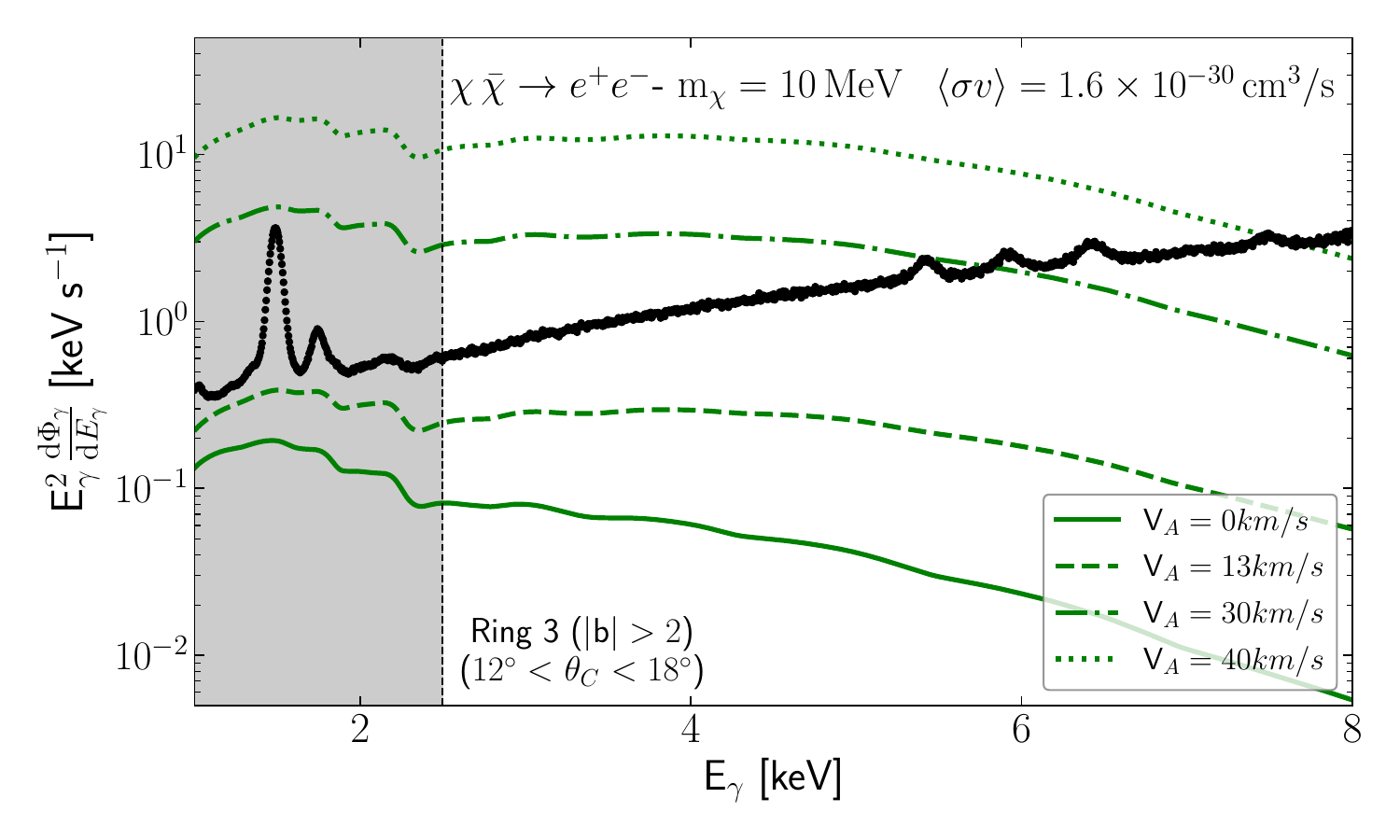}
\includegraphics[width=0.5\linewidth, height=0.22\textheight]{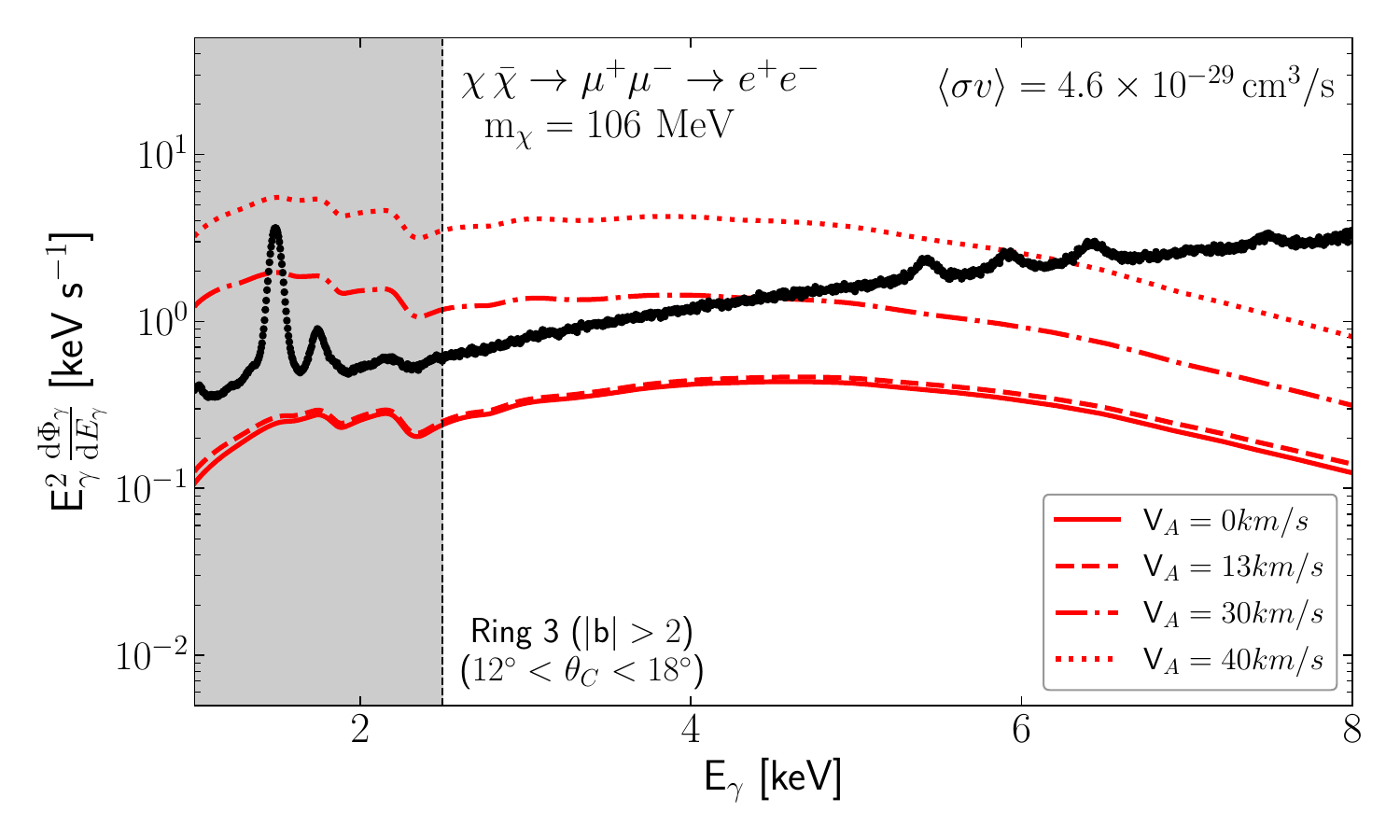}
\caption{Comparison of MOS data at Ring 3~\citep{Foster:2021ngm} with the predicted DM-induced $X$-ray signal in that galactic region, for DM with $m_\chi=10$~MeV annihilating in $e^+e^-$ (left panel) and with $m_\chi \simeq100$~MeV annihilating via the $\mu^+ \mu^-$ channel (right panel) for different levels of reacceleration ($v_A$ values of $0$~km/s, $13$~km/s (best-fit value), $30$~km/s and $40$~km/s).}
\label{fig:XMM_VA}
\end{figure*}

Given the high impact of reacceleration and propagation in the electron signals produced by light DM, the associated secondary radiations (mainly ICS and bremsstrahlung at the energies of interest in this paper) will be similarly affected. Here we focus on the effect that the diffusion setup has on the constraints obtained from {\sc Xmm-Newton} data, since the authors of~\citet{Cirelli:2023tnx} demonstrated that this dataset leads to the most constraining DM limits among all the current astrophysical experiments sensitive to the keV-MeV energy range. In particular, we use data from the MOS detector onboard {\sc Xmm-Newton}, which was provided in~\citet{Foster:2021ngm} in the energy range from $1$~keV to above $10$~keV, and are divided into 30 galactocentric rings, $6^{\circ}$ wide, around the GC. 
We illustrate the predicted $X$-ray signals at the Ring 3 (the most constraining one~\citep{Cirelli:2023tnx, DelaTorreLuque:2023nhh}) for different values of $v_A$ in Fig.~\ref{fig:XMM_VA}, in the case of a DM particle with $m_{\chi} = 10$~MeV annihilating into $e^+e^-$ (left panel) and for a DM particle with $m_{\chi} \simeq 100$~MeV annihilating via the $\mu^+ \mu^-$ channel.
Since we consider only the range of 2.5-8 keV to obtain our constraints, due to the background noise in the detector at lower and higher energies respectively, we shade the energy region below $2.5$~keV in these 
plots.

As one can see, for the low mass case, representative of DM mass below $m_\chi\simeq$ 100 MeV, the $X$-ray signals change very significantly for different $v_A$ values, while, for the high mass case, the difference between the predicted signals using no reacceleration become similar to the ones predicted from our best-fit setup. More extreme $v_A$ values still lead to sizeable differences up to GeV DM masses. This means that our limits including the effect of reacceleration will dramatically strengthen our limits for low DM masses but the limits for the $\pi^+ \pi^-$ and $\mu^+ \mu^-$ channels are not expected to change considerably when including reacceleration. We have tested that bremsstrahlung has a negligible contribution (across the whole mass range) to these signals, given that the MOS detector data that we use masks the galactic plane ($|b|<2^{\circ}$), where this emission can be more important. More details about the secondary emissions from these signals can be found in~\citet{Cirelli:2023tnx}.

For completeness, we show in Fig.~\ref{fig:XMM_Comp} the predicted $X$-ray signals at Ring 3 produced from annihilation (left panels) and decay (right panels) of DM for different masses for the $\pi^+ \pi^-$ (top row), $\mu^+ \mu^-$ (middle row) and direct $e^+ e^-$ (bottom row) channels, using our best-fit propagation parameters. The contribution of FSR becomes irrelevant in all the cases when including reacceleration and is important only below $m_\chi \simeq 10$~MeV for the cases without reacceleration. We show the FSR and secondary spectrum (predicted in the case of no reacceleration) for DM with $m_\chi = 4$~GeV annihilating into $e^+ e^-$ in the left panel of Fig.~\ref{fig:FSR}. We also illustrate, in the right panel of this figure, the effect of including FSR and reacceleration to the annihilation bounds in the direct $e^+ e^-$ channel. As one can see, reacceleration completely dominates the bounds below $m_\chi\simeq 30$~MeV. The details on how these limits are derived are given in the section below.

\subsection{Impact of the diffusion setup in sub-GeV DM constraints}
\label{sec:Comp_Limits}

In order to derive the $95\%$ confident limits, we determine the best-fit $\langle\sigma v\rangle$ value and the $2\sigma$ statistical error associated through a $\chi^2$ test (in particular, we use the \verb|curve_fit| function in the \verb|scipy.optimize| Python package) for every mass and channel. In the case of MOS data, the fits are performed in the region from $2.5$ to $8$~keV, as in~\citet{Foster:2021ngm,Cirelli:2023tnx}, due to the background noise in the detector. Following the strategy from~\citet{Cirelli:2023tnx}, besides analysing every ring dataset, we obtain the bounds from a combined analysis of all the rings, using the following expression: $\chi^2 = \sum_i \left( \frac{\textrm{Max}\left[\phi_{\textrm{DM}\gamma, i}(p, m_{\chi}) - \phi_{i , 0} \right]}{\sigma_i}\right)^2 $, where $i$ denotes the ring data set and $p$ can be either $\langle\sigma v \rangle$ or $\Gamma$ in case of annihilation or decay, respectively. $\phi_i$ is the observed flux in each ring and $\sigma_i$ the associated standard deviation of the measurements. We then impose a $2\sigma$ bound on the parameter $p$ (for each value of $m_{\chi}$) whenever we obtain $\chi^2 = 4$.
We emphasise that these bounds are very conservative, both for the case of {\sc Xmm-Newton} and {\sc Voyager 1}, since the several backgrounds that are expected, such as diffuse emission or extragalactic background light, are not included. We do not perform a dedicated analysis of these backgrounds in this work because of the uncertainties involved, but leave such a development for a future work.

\begin{figure*}[t!]
\includegraphics[width=0.5\linewidth]{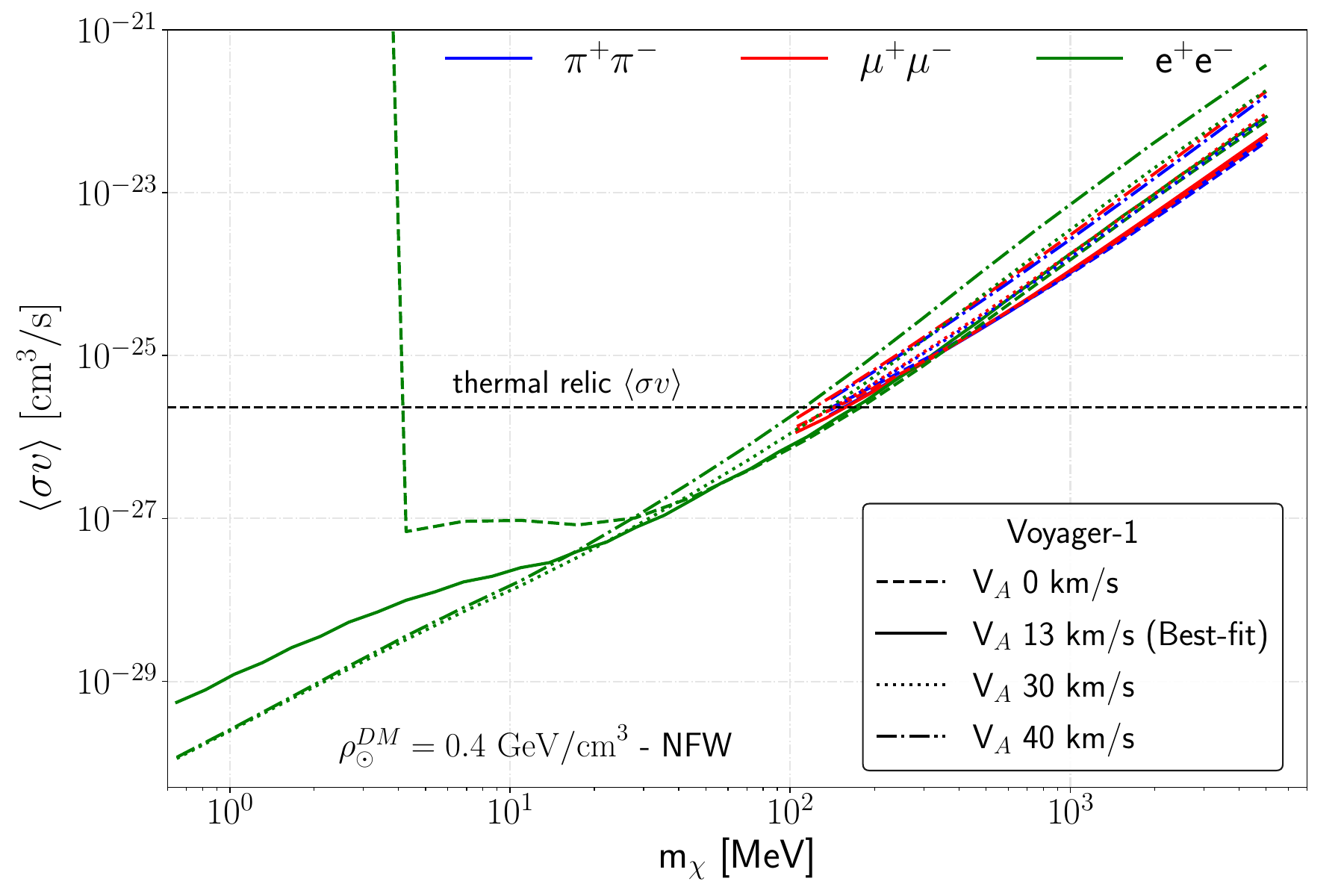}
\includegraphics[width=0.5\linewidth]{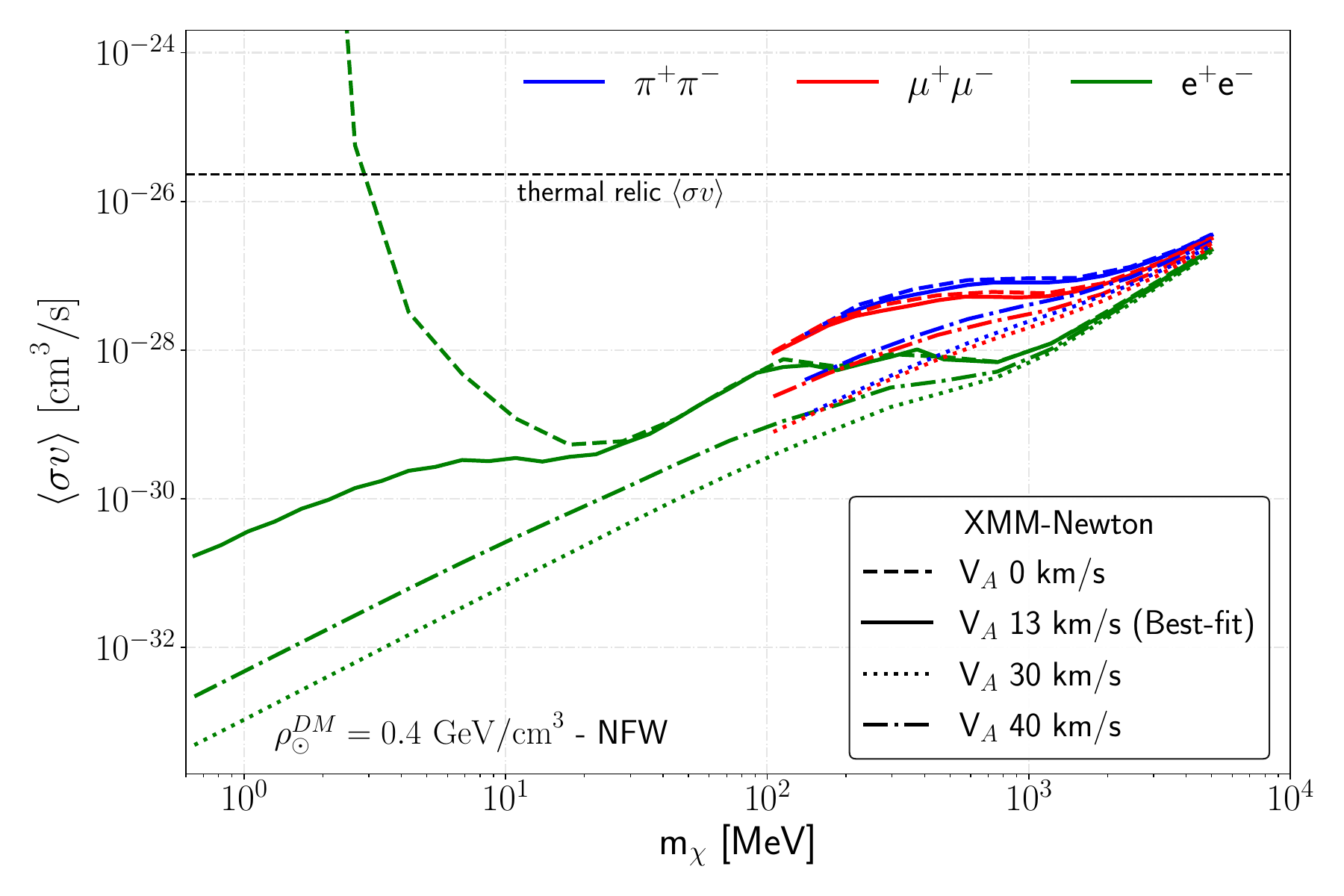}
\caption{Limits on the DM annihilation cross section $\langle \sigma v \rangle$ at the $95\%$ CL derived from {\sc Voyager 1} (left panel) and from the combined analysis of all MOS ring datasets (top left) for different levels of reacceleration and for the $\pi^+ \pi^-$ (blue), $\mu^+ \mu^-$ (red) and direct $e^+ e^-$ (green) channels. We consider Alfvén velocities $v_A$ of $0$ (dashed), 13 (i.e. our best-fit value; solid), $30$ (dot-dashed) and $40$ (dotted) km/s, respectively. To facilitate the comparison of the results for different levels of reacceleration, we do not include prompt emission in the right panel.}
\label{fig:lims_VA}
\end{figure*}

To emphasise the importance of using a more realistic model of electron diffusion, including their reacceleration by interstellar turbulence, we show in Fig.~\ref{fig:lims_VA} a comparison of the bounds derived from {\sc Voyager 1} data (left panel) and {\sc Xmm-Newton} (right panel) for different values of $v_A$, from $0$~km/s (no reacceleration) to $40$~km/s, leaving the rest of propagation parameters involved untouched. Bounds on the signals from the $\pi^+ \pi^-$, $\mu^+ \mu^-$ and direct $e^+ e^-$ channel are indicated as blue, red and green lines, respectively. 
It is remarkable that in the case of no reacceleration we can not constrain the annihilation rate for DM masses below a few GeV from {\sc Voyager 1}, given that the electron spectra lie below the {\sc Voyager 1} data (see the upper right panel of Fig.~\ref{fig:Va_ee_Scan_Voy}). The same occurs for {\sc Xmm-Newton}, since the lower energy of the particles (corresponding to their low masses) is not able to boost the ambient photons (via IC scattering) to the energies where {\sc Xmm-Newton} has measurements.
To clearly show what the effect of reacceleration in these limits is, we do not include the prompt emission of $\gamma$-rays (FSR) in the right panel ({\sc Xmm-Newton} bounds), which can only lead to any difference in the limits for the case of no reacceleration and for masses below $m_\chi\simeq10$~MeV (see also the right panel of Fig.~\ref{fig:FSR}). In both cases, the effect of reacceleration is huge towards low masses and, interestingly, as can be seen from Fig.~\ref{fig:Va_ee_Scan_Voy}, including some level of reacceleration allows us to constrain DM masses below $m_\chi \simeq20$~MeV.

\begin{figure*}[t!]
\includegraphics[width=0.54\linewidth]{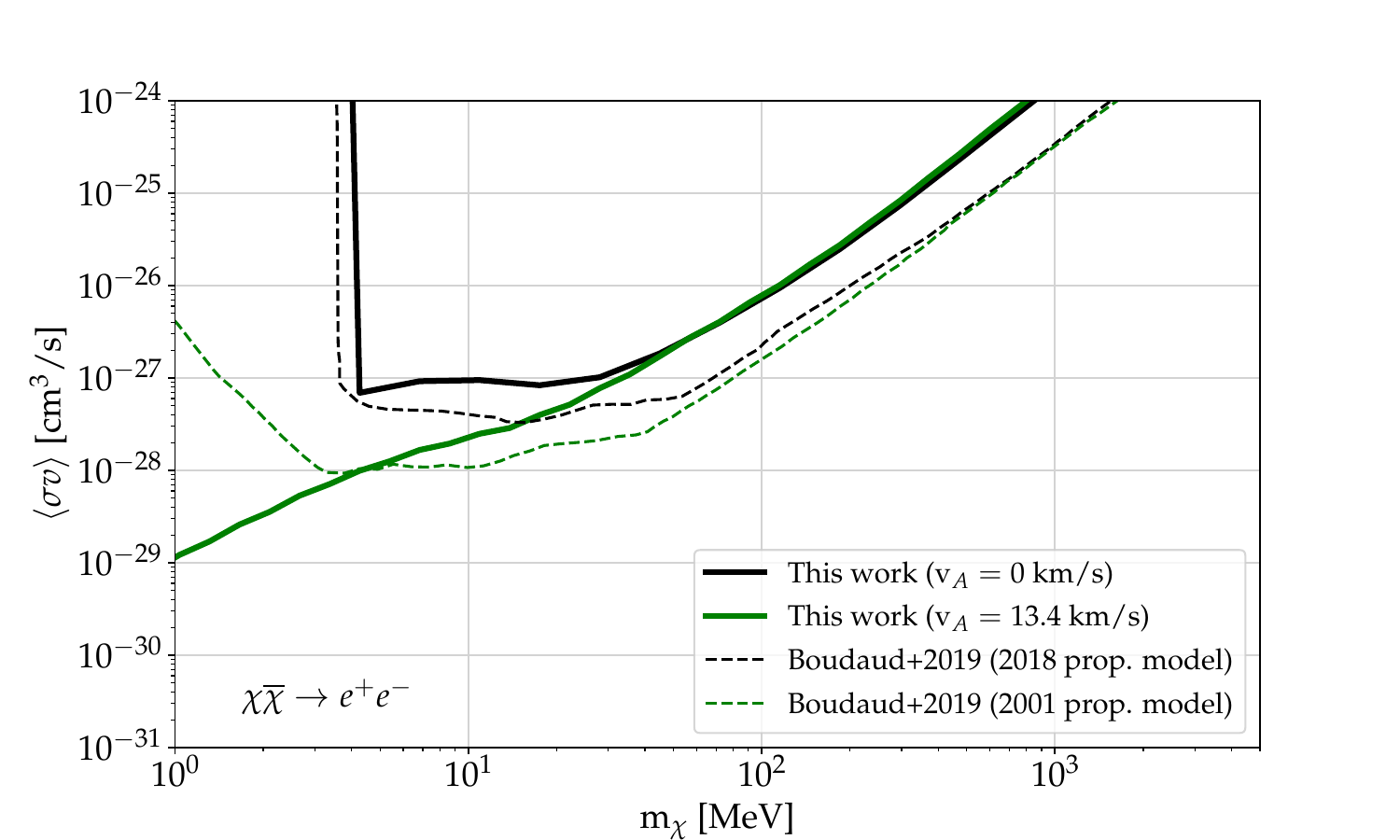} \hspace{-0.7cm}
\includegraphics[width=0.54\linewidth]{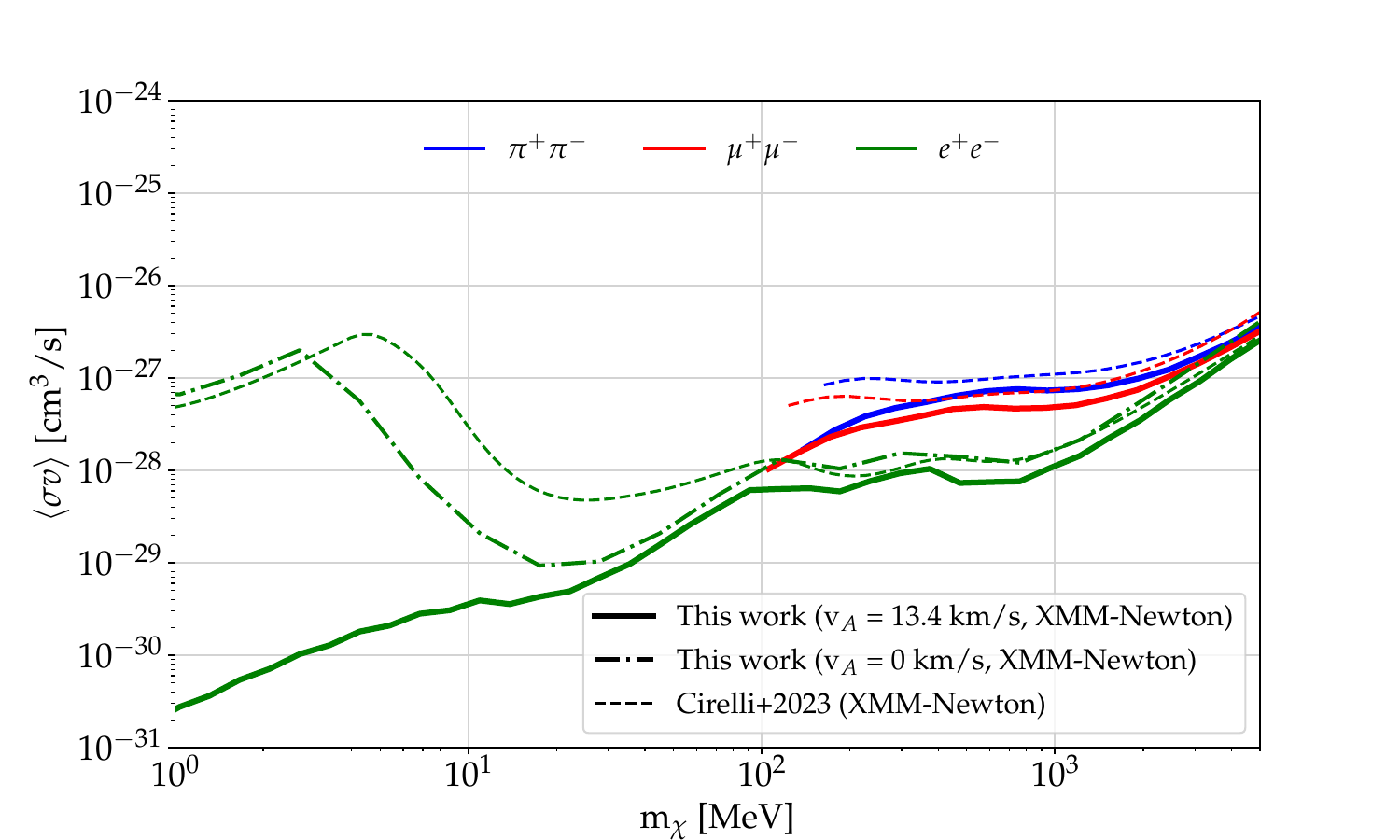}

\includegraphics[width=0.54\linewidth]{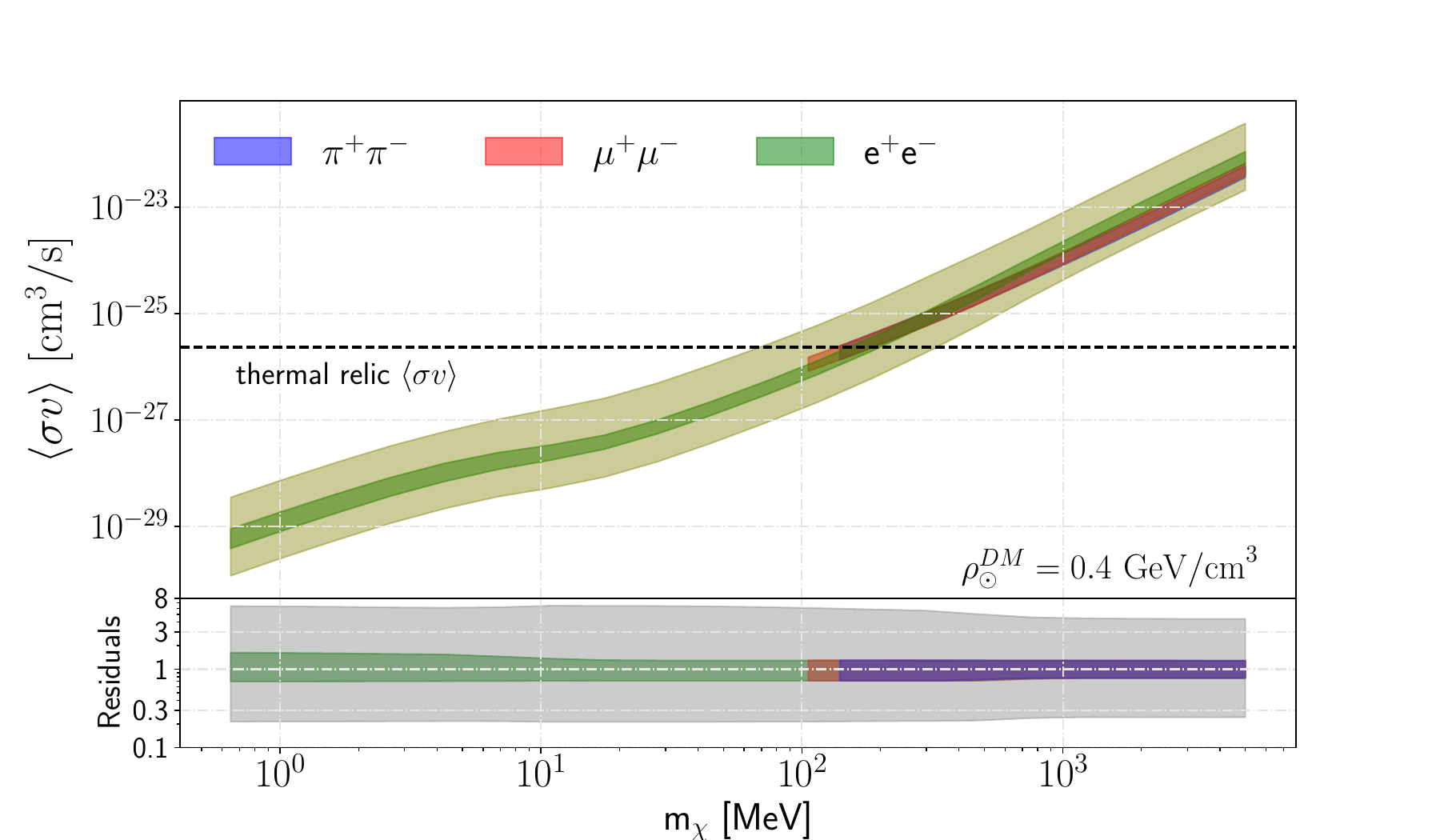} \hspace{-0.7cm}
\includegraphics[width=0.54\linewidth]{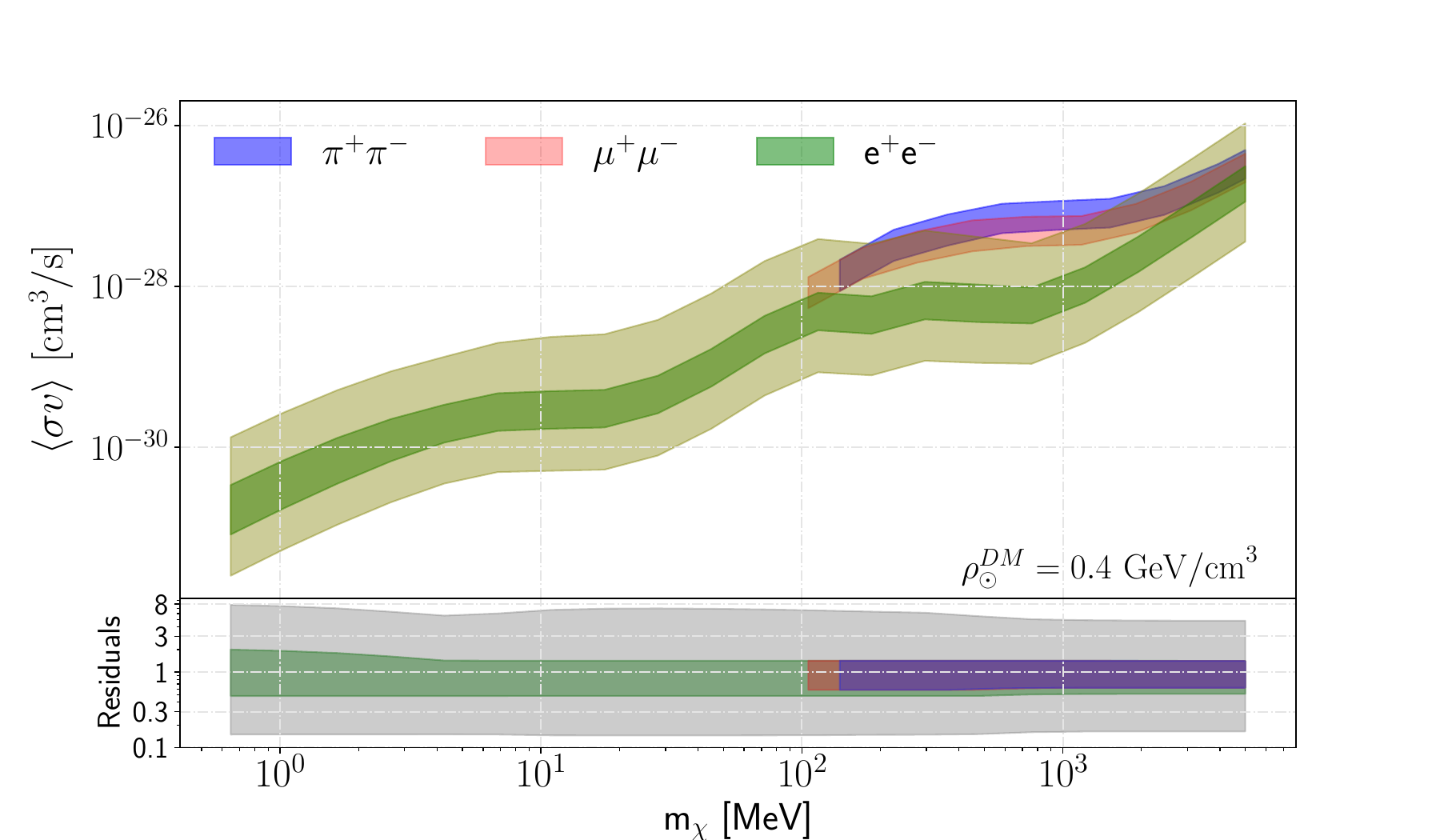}
\caption{The top panels show the limits on annihilating and decaying DM from {\sc Voyager 1} (top left panel) and from {\sc Xmm-Newton} (top right panel), compared with previous constraints using the same datasets. In the left panel we show our {\sc Voyager 1} bounds for two different $v_A$ values: 0 (solid black line) and the best-fit value 13.4 km/s (solid green line). The bounds are plotted alongside with previous {\sc Voyager 1} bounds (see main text for more details) where two propagation models were used: one without reacceleration from~\citet{Reinert:2017aga} (dashed black line) and one with reacceleration from~\citet{Maurin:2001sj, Donato:2003xg} (dashed green line). In the top right panel we show our {\sc Xmm-Newton} bounds for $v_A = 13.4$ km/s (solid lines) and 0 km/s (dot-dashed line) with the ones in~\citet{Cirelli:2023tnx} (dashed lines) for the $\pi^+\pi^-$ (blue line), $\mu^+ \mu^-$ (red line) and $e^+ e^-$ (green line) channels respectively. Bottom panels show the estimated uncertainties in our predictions, as discussed in the main text.}
\label{fig:new+oldbounds}
\end{figure*}

Then, we discuss the difference between our updated bounds from the {\sc Voyager 1} and {\sc Xmm-Newton} datasets with the previous constraints using the same datasets. The comparison is shown in Fig.~\ref{fig:new+oldbounds}. The authors of~\citet{Boudaud:2016mos} published the first constraints on low-mass DM using {\sc Voyager 1} limits. However, their evaluation of electron propagation differs significantly from the one we use here, which considers the most recent CR data. They use the semi-analytical code {\sc Usine}~\citep{Maurin:2018rmm} which employs an effective treatment to account for the energy losses, called the ``pinching method''~\citep{Boudaud:2016jvj}, (from Coulomb and ionisation processes that become the dominant channels below a few tens of MeV, a region that is especially relevant in this study). In addition, we also emphasise the effect of reacceleration in our constraints, which was also treated differently in {\sc Usine} (relying on an effective approximation). More importantly, the authors of~\citet{Boudaud:2016mos} used a parameterisation obtained from a study in 2001~\citep{Maurin:2001sj}, when the precision of data was more limited and the systematic uncertainties (e.g. with spallation cross sections) was under less control, meaning that their diffusion coefficient differs significantly from our updated one. Specifically, they use a parametrisation with fixed $\eta = 1$ (i.e. $D\propto \beta$, see Eq.~\eqref{eq:diff_eq}), while the one obtained in analyses of recent AMS-02 data when including this parameter in the fit is negative~\citep{DeLaTorreLuque:2021nxb, Weinrich_2020} and, in our case, $\eta = -0.75$, implying much more confinement at the relevant sub-GeV energies considered here. This means that, in the sub-GeV regime, our diffusion coefficient is markedly different, which can lead to important differences in our predicted signals and constraints. In particular, the limits from~\citet{Boudaud:2016jvj} are obtained with a halo height value of $H\sim15$~kpc, which explains why they are a factor of a few stronger than ours above a few tens of MeV. In addition, here we clearly observe how reacceleration affects our constraint differently than~\citet{Boudaud:2016jvj}. 

In the top-left panel of Fig.~\ref{fig:new+oldbounds}, we are also including the limit derived by the same authors of~\citet{Boudaud:2016jvj} using a more updated diffusion coefficient (from 2018, coming from~\citet{Reinert:2017aga}), and no reacceleration. We have taken these bounds from a \href{https://indico.in2p3.fr/event/18701/contributions/71758/attachments/53484/69702/IRN_annecy_boudaud.pdf}{2019 talk by M. Boudaud} which we refer to as ``Boudaud+2019" with the corresponding propagation models in Fig.~\ref{fig:new+oldbounds} shown in dashed black and dashed green respectively.
These comparisons evidence the fact that, given the uncertainties in the evaluation of the diffusion coefficient at low energies and the halo height, different propagation setups can differ significantly, and, therefore, it would be crucial moving forward to update the constraints we derive here with updated propagation setups as more CR data becomes available at low energies and uncertainties are further reduced.

To compare our results with those obtained from~\citet{Cirelli:2023tnx}, which we refer to as ``Cirelli+2023", shown as a dashed green line (top-right panel of Fig.~\ref{fig:new+oldbounds}), one can also expect differences due to the different diffusion setups. In particular, the authors of~\citet{Cirelli:2023tnx} did not include the propagation of the DM-induced signals nor reacceleration, only including injection of the particles and energy losses in their setups. In this case, the resultant bounds only differ slightly for DM masses above the some tens of MeV due to the use of older ISRF maps in~\citet{Cirelli:2023tnx}, which tend to have a less significant optical component. Remarkably, we are able to extend the constraining power to DM masses below $100$~MeV thanks to the incorporation of reacceleration in our scheme, which was absent in theirs. Reacceleration dramatically increases the energy of electrons produced by annihilation or decay of low-mass DM. The energy of the associated IC-scattered photons thus increase and therefore reach the energy range of the {\sc Xmm-Newton} data.

In the bottom-left panel of Fig.~\ref{fig:new+oldbounds} we show the uncertainties related to our derived limits for {\sc Voyager 1}, which contain the $1\sigma$ uncertainties in the determination of the propagation parameters from our analyses (which mainly comes from reacceleration and halo height), $1\sigma$ uncertainties in the normalization of the dark matter density at Earth (taken to be $\rho_{\odot}=0.420^{+0.011}_{-0.009} \pm0.025$~GeV/cm$^3$ from \cite{Pato_2015}) and a very conservative $10\%$ factor to account for uncertainties in the gas distribution and energy losses. In the case of limit from {\sc Xmm-Newton} (bottom-right panel), we include also a quite conservative $30\%$ factor that accounts for uncertainties in the ISRFs, as reported in \cite{Vernetto:2016alq}. In addition, we have estimated the uncertainties by considering two extremal DM density distributions: a cored profile~\citep{Burkert:1995yz}, which leads to significantly weaker constraints, and a contracted NFW profile (with slope $\gamma=1.26$), that leads to stronger constraints. We display in Fig.~\ref{fig:new+oldbounds} an other uncertainty band, in the bounds for the direct $e^{\pm}$ channel, representing the difference between these two extremal profiles on top of the other astrophysical uncertainties (see olive bands). Given that these uncertainties are expected to be roughly the same for also for the other channels, we indicate it as a grey band in the residual panels.

\section{Current constraints and discussion}
\label{sec:Discussion}

\begin{figure*}[t!]
\includegraphics[width=0.552\linewidth]{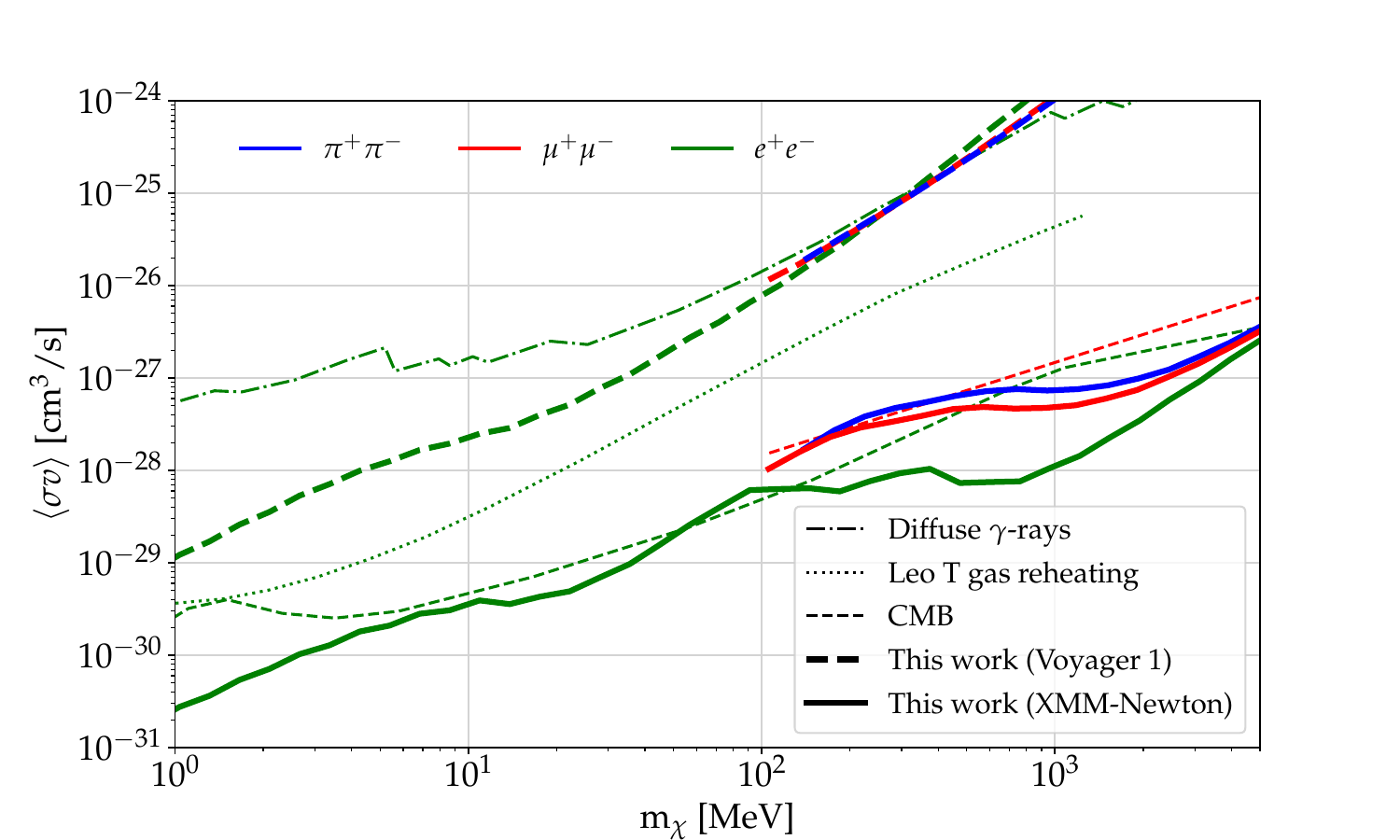}\hspace{-0.9cm}
\includegraphics[width=0.552\linewidth]{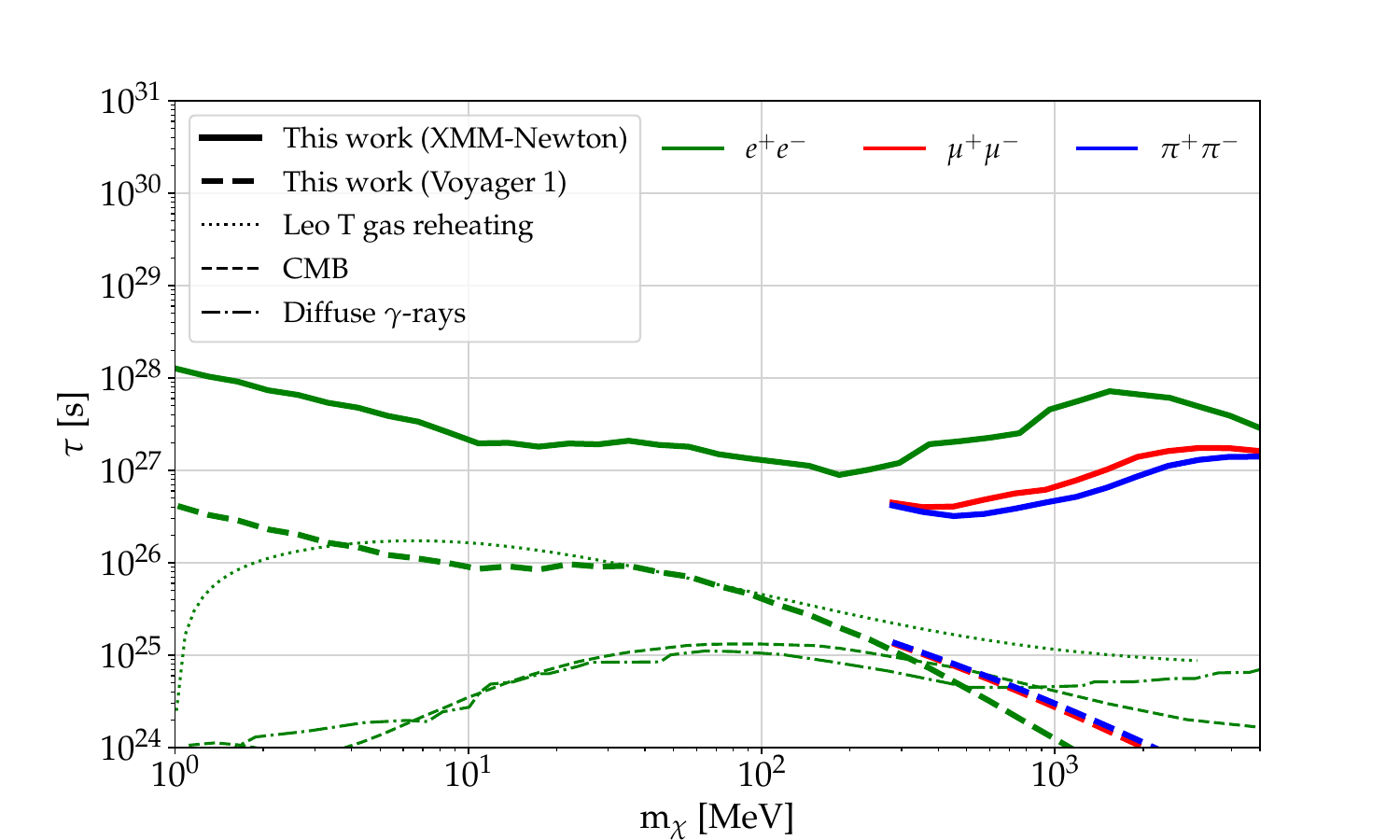}
\caption{Comparison of the $95\%$ confidence bounds on annihilating (left panel) and decaying (right panel) DM derived in this work with the best-fit propagation parameters (thick dashed for {\sc Voyager 1} and solid lines for {\sc Xmm-Newton} respectively) with other existing constraints. We show the diffuse $\gamma$-ray bound from~\citet{Essig:2013goa} (dot dashed line), the CMB bounds from~\citet{Slatyer:2015jla} and~\citet{Lopez-Honorez:2013cua} (dashed lines) and the bounds from gas reheating in the Leo T dwarf galaxy from~\citet{Wadekar:2021qae} (dotted line). Once again we show the channels $\pi^+\pi^-$ (blue), $\mu^+\mu^-$ (red) and $e^+e^-$ (green) channels, respectively.}
\label{fig:bounds+litterature}
\end{figure*}

In this section, we compare our results to other existing independent DM constraints in the literature, summarised in Fig.~\ref{fig:bounds+litterature}. We report the $e^+ e^-$ channel bounds from Essig et al. in~\citet{Essig:2013goa} as the dot-dashed line (that we label ``diffuse $\gamma$-rays"), where a compilation of diffuse $X(\gamma)$-ray data from {\sc Heao-1}, {\sc Integral} and {\sc Comptel} was used in their analysis. In this study only FSR is considered as the source of $X(\gamma)$-ray emission from light DM annihilating/decaying into $e^\pm$. In our setup FSR is a subdominant $X$-ray production process compared to ICS from reaccelerated $e^\pm$ on ambient photons and to ICS from unaccelerated $e^\pm$ for higher DM masses, thus the bound we derived from the {\sc Xmm-Newton} dataset is more stringent over the whole light DM mass range considered.

High gas content dwarf galaxies present systems that are relatively unaltered and have extremely low rates of gas cooling. This makes them very susceptible to heating by DM  that either annihilates or decays into Standard Model (SM) particles.  Such dwarfs are particularly sensitive to DM producing $e^\pm$ with energies in the range 1-100 MeV or photons with energies 13.6 eV-1 keV, because these products can be efficiently thermalised in the abundant neutral hydrogen gas of the dwarfs. Bounds can be derived by requiring the rate of heat injection by DM to not exceed the ultra-low radiative cooling rate of gas in the Leo T dwarf galaxy. This was done in~\citet{Wadekar:2021qae} which we show in Fig.~\ref{fig:bounds+litterature} as a dotted line labelled ``Leo T gas reheating." For both annihilating and decaying DM, the bound we derive using {\sc Xmm-Newton} is more stringent than the Leo T over the whole DM mass range of interest.

Finally, we also report the bounds derived by computing the impact of the $e^\pm$ injection from DM annihilation or decays on the CMB anisotropies, from~\citet{Slatyer:2015jla} and~\citet{Lopez-Honorez:2013cua} which are shown as short dashed lines in Fig.~\ref{fig:bounds+litterature} and are labelled ``CMB." These bounds are derived from the fact that injection of ionising particles from DM processes during the cosmic Dark Ages will increase the residual ionisation fraction, broadening the last scattering surface and modifying the anisotropies of the CMB (measured to exquisite sensitivity by the {\sc Planck} experiment). These DM bounds are known to be among the strongest and the most robust to date for annihilating DM over a wide mass range. 

Using our model of reacceleration, gas and ISRF maps and the usual NFW DM density profile, the bound we derive on DM annihilating into $e^+ e^-$ and $\mu^+ \mu^-$, using {\sc Xmm-Newton} (shown in the left panel of Fig.~\ref{fig:bounds+litterature}) is more stringent than the CMB one in across almost the entire DM mass range considered. However we emphasise that the CMB bounds are more robust (since they depend on fewer parameters and have smaller uncertainties). The bounds we derive depend on the CR propagation model, gas and ISRF maps normalisation and DM density profile, and are affected by the astrophysical uncertainties within these physical descriptions. This is highlighted by~\citet{Cirelli:2020bpc, Cirelli:2023tnx}, wherein they found that different parametrisations of the aforementioned ingredients (except the propagation model) can improve or worsen the bounds up to one order of magnitude when compared to the fiducial scenario. Additionally, there are other ingredients that would significantly affect our evaluations and whose modelling are far beyond the scope of this work, such as inhomogeneous diffusion~\citep{DeLaTorreLuque:2022buq} of particles in the Galaxy or advection by strong winds~\citep{Recchia_2020} towards the centre of the Galaxy.

For decaying light DM (shown in the right panel of Fig.~\ref{fig:bounds+litterature}), the CMB bounds remain strong but not among the strongest in the literature (unlike the annihilating DM case). This is due to the fact DM clusters at redshifts $z \lesssim 100$, which greatly enhances the DM annihilation rate and therefore the $e^\pm$ injection, whereas the decay rate remains constant. Hence our bound from {\sc Xmm-Newton} is orders of magnitude more stringent that the CMB counterpart in the decaying DM case.

One other important factor to highlight in the case of annihilating DM is that we only show the bounds derived under the assumption that DM is $s$-wave. If DM is assumed to be $p$-wave (i.e. $\langle \sigma v \rangle \propto v^2$), the CMB bounds weaken by around $ 4$ orders of magnitude~\citep{Diamanti:2013bia} since at the time of recombination ($z \simeq 1100$) all DM is cold, since DM halos start to form at $z \simeq 100$. Therefore energy injection from DM annihilation that ultimately impacts CMB anisotropies is very low. In the case of Leo T, the velocity dispersion of DM particles is expected to be low (around 7 km/s~\citep{Wadekar:2021qae} compared to 220 km/s in the Milky Way). This also lowers the $p$-wave DM annihilation rate and limits the reheating of the gas in Leo T, thus leading to around $2$ orders of magnitude weaker bounds.

In our study we limited ourselves to $s$-wave DM annihilations. In~\citet{Boudaud:2018oya} the authors set bounds on the $p$-wave annihilation cross section using the same {\sc Voyager 1} dataset and CR propagation scheme as in~\citet{Boudaud:2016mos}. By combining the results of the two aforementioned scenarios, one can show that their bounds can vary by less than an order of magnitude if $p$-wave DM is assumed. If the same analysis as~\citet{Boudaud:2018oya} was done in our case, we would expect the same variation in our bounds in the case of $p$-wave DM since we also studied DM annihilations in the Milky Way at the present time, so the DM velocity dispersion would also be the same. We therefore conclude that if DM was truly $p$-wave, our constraints would be orders of magnitude stronger than the CMB and Leo T ones, which are currently the most competitive with this work (assuming that DM is $s$-wave).

\section{Conclusion}
\label{sec:Conclusion}
We have explored the possibility of detecting signals of sub-GeV dark matter (DM) particles coupling to the Standard Model (SM) via their production of diffuse emissions when they either decay or annihilate in the Milky Way Galaxy. Given the low DM mass that we consider here, these SM particles can be electrons, muons or pions, which can eventually radiate photons through their $e^\pm$ products. The resulting diffuse $e^\pm$ flux originated can be directly observed in probes such as {\sc Voyager 1}. Alternatively, the $e^\pm$ products lead to secondary photon radiations from bremsstrahlung and may upscatter the low-energy galactic photon fields, namely via inverse Compton scattering (ICS), generating a broad diffuse emission from $X$-ray to $\gamma$-ray energies observable in experiments such as {\sc Xmm-Newton}. In particular, we have used the {\sc Xmm-Newton} $X$-ray telescope to search for this emission and derive strong constraints on light DM annihilation and decay rates.

We have tested a diffusion-reacceleration scheme for propagation of the DM-produced particles, and analysed how these ingredients affect these signals. We show that a detailed treatment of $e^\pm$ propagation is necessary to realistically describe the DM-induced signals, although the uncertainties in diffusion at sub-GeV energies are still high. We have highlighted the importance of including reacceleration of $e^\pm$ by turbulence in the interstellar medium. In addition, we have also used realistic maps of gas density and interstellar radiation fields (ISRFs) to calculate the bremsstrahlung and ICS emission from DM-induced $e^\pm$ particles. We have assumed a Navarro-Frenk-White (NFW) profile for the DM density distribution in the Galaxy. We have compared our model predictions with {\sc Voyager 1} and {\sc Xmm-Newton} flux data and performed a likelihood analysis to obtain upper limits on the DM annihilation cross section and decay lifetime as a function of DM mass.

We have found that we get a significant improvement in the DM annihilation and decay constraints from {\sc Xmm-Newton} by including best fit cosmic ray (CR) propagation parameters obtained from detailed analyses of the most recent CR data. We find that we get a significant improvement in the DM annihilation and decay constraints excluding thermally averaged cross sections of $10^{-31}$ cm$^3$\,s$^{-1}\lesssim \langle \sigma v\rangle \lesssim10^{-26}$ cm$^3$\,s$^{-1}$ and decay lifetimes of $10^{26}\,\textrm{s}\lesssim \tau \lesssim 10^{28}\,\textrm{s}$ respectively, in the DM mass range between $m_\chi\simeq1$~MeV and $5$~GeV, using our setup. This yields the strongest astrophysical constraints for this mass range of DM and surpasses cosmological bounds across a wide range of masses as well. We have also compared our results with other existing bounds from different sources, such as diffuse $\gamma$-rays, gas reheating in dwarf galaxies, and cosmic microwave background (CMB) anisotropies. We have found that our {\sc Xmm-Newton} bounds are more stringent than most of these bounds for DM masses between $1$~MeV and a few GeV.

Our work revisits the power of {\sc Voyager 1} and particularly {\sc Xmm-Newton} to probe light DM scenarios and complement other astrophysical and cosmological probes. Our results show that sub-GeV DM particles that can produce SM particles are more highly constrained by $X(\gamma)$-ray observations than previously considered and are unlikely to explain any anomalous signals in photon or CR fluxes. Our work also motivates further studies of light DM models with different annihilation or decay channels, such as neutrinos or photons, that may evade $X(\gamma)$-ray constraints but still leave observable signatures in other wavelengths or experiments. 
We finally emphasise that this work should be updated when better parametrisations of the diffusion coefficient at low energy are available, to reduce the uncertainties in these evaluations and improve the robustness of the bounds presented here.

\begin{acknowledgments}
\textbf{Acknowledgments:} We are grateful to Marco Cirelli for helpful feedback on the draft and at the early stages of this work.
SB and JK acknowledge the hospitality of the Institut d’Astrophysique de Paris ({\sc Iap}) where part of this work was done. PDL was supported by the European Research Council under grant 742104 and the Swedish National Space Agency under contract 117/19 during the preparation of this project and is currently supported by the Juan de la Cierva JDC2022-048916-I grant, funded by MCIU/AEI/10.13039/501100011033 European Union "NextGenerationEU"/PRTR. This project used computing resources from the Swedish National Infrastructure for Computing (SNIC) under project Nos. 2021/3-42, 2021/6-326, 2021-1-24 and 2022/3-27 partially funded by the Swedish Research Council through grant no. 2018-05973. JK is supported by funding from CNRS under the grant ``DaCo: Dark Connections'', in collaboration with Alma Mater Studiorum - Università di Bologna. 
\end{acknowledgments}

\appendix

\section{Fits to the local CR data}
\label{sec:appendix0}
\begin{figure*}[h]
\includegraphics[width=0.57\linewidth]{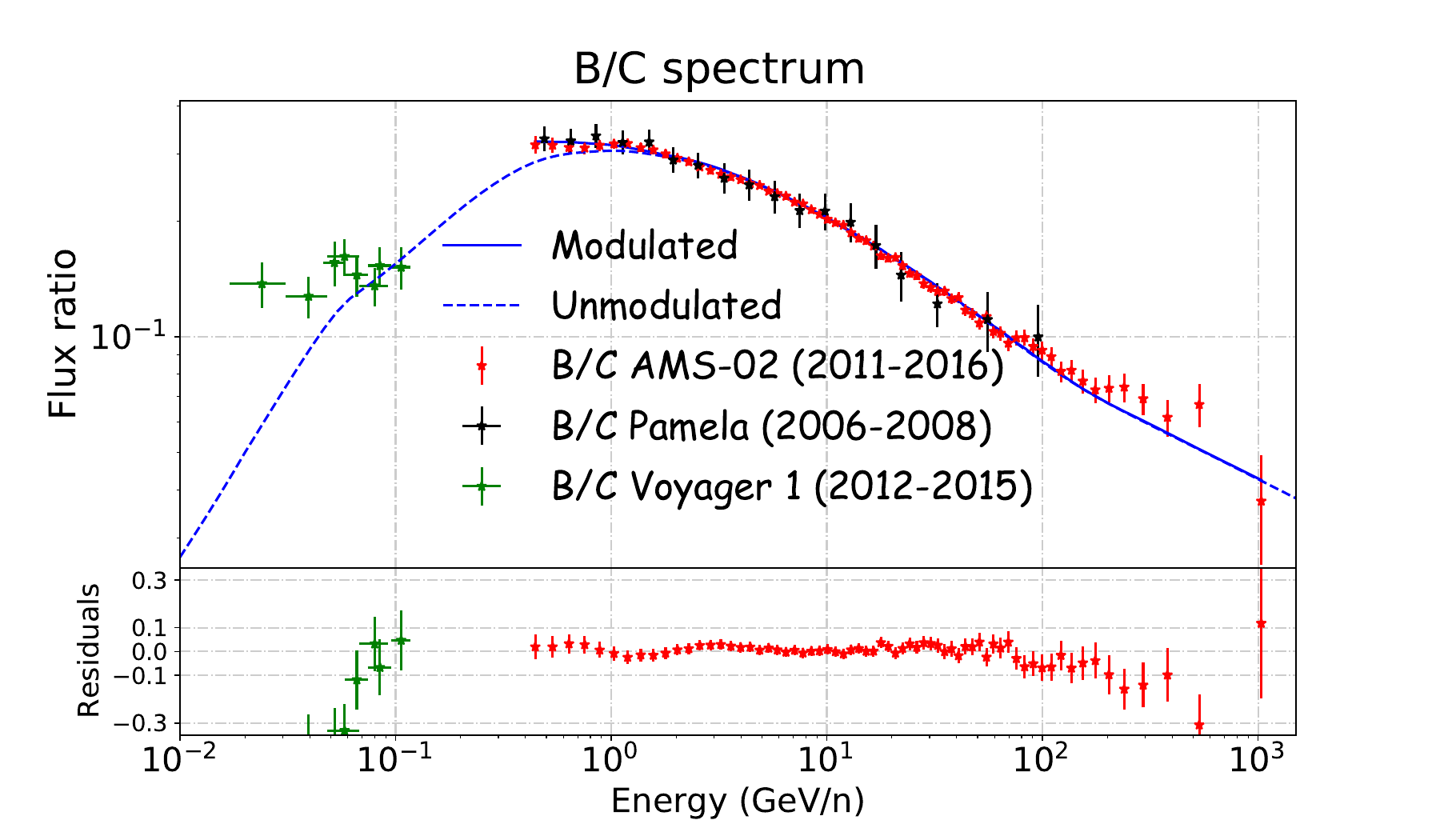} \hspace{-1cm}
\includegraphics[width=0.57\linewidth]{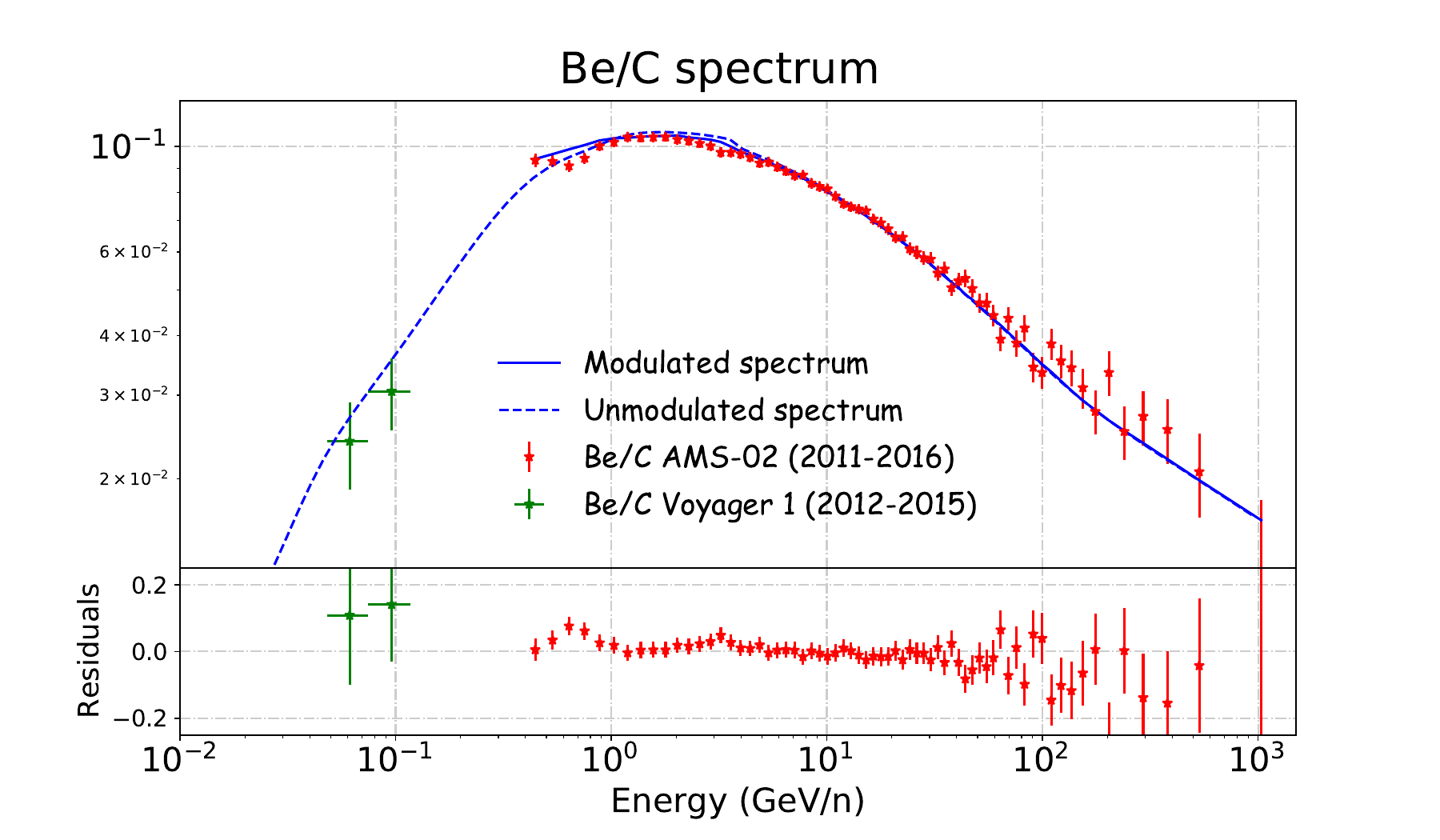}
\includegraphics[width=0.57\linewidth]{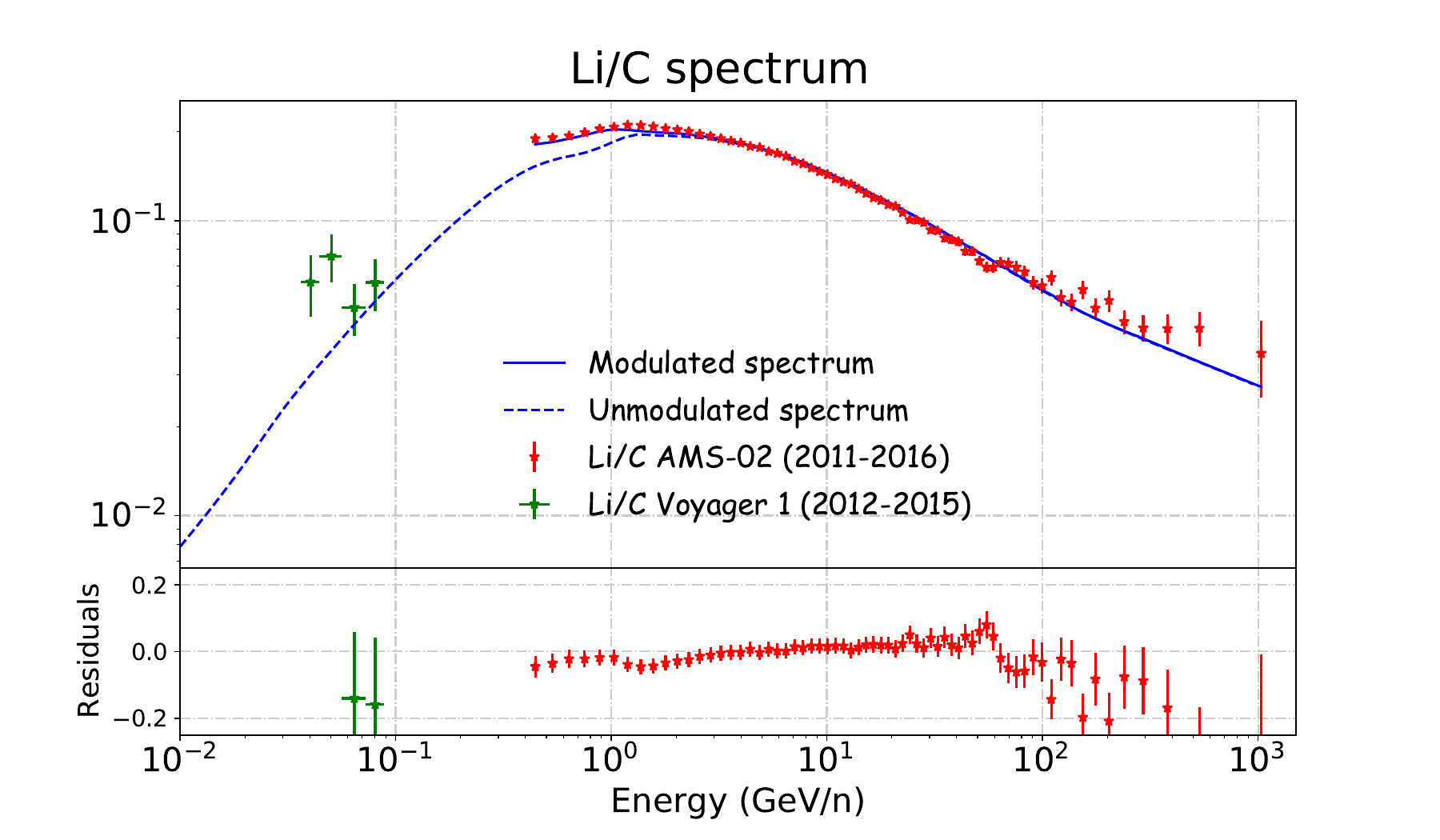}
\hspace{-1cm}
\includegraphics[width=0.57\linewidth]{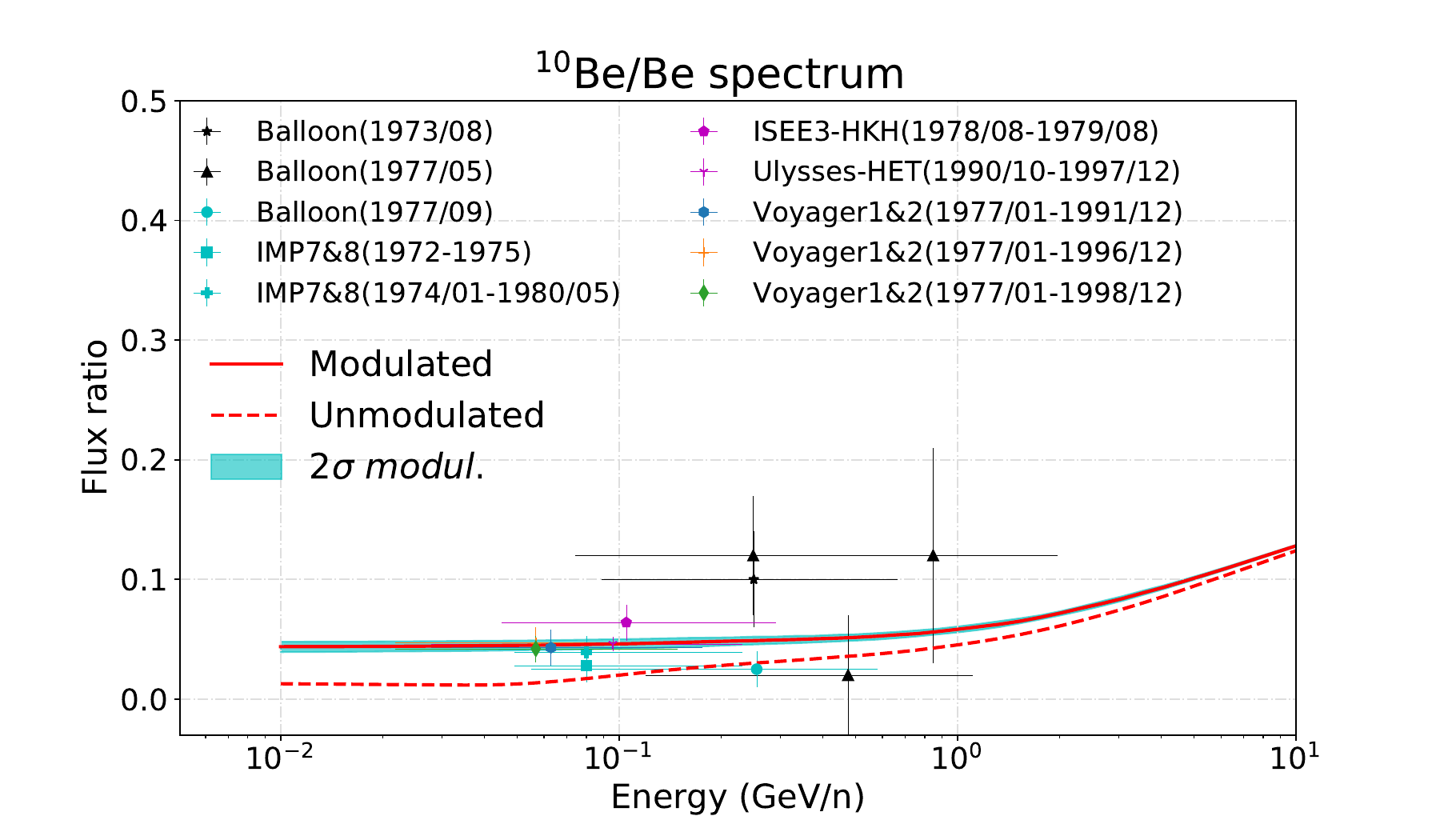}
\caption{Main CR observables used to determine the propagation parameters employed in this work. AMS-02 data are shown as red markers and Voyager-1 data as green markers.}
\label{fig:CRFit}
\end{figure*}

In this appendix, we show the main CR observables used to determine the propagation parameters employed in this work, as displayed in Fig.~\ref{fig:CRFit}. 
They are obtained from a Markov-Chain Monte Carlo (MCMC) analysis of a set of fluxes and flux-ratios of different CR species as measured by AMS-02. The parameters D$_0$, $\eta$, $\delta$, $H$ and $v_A$ are the main parameters obtained in this analysis. This analysis is similar to the one presented in~\citet{DeLaTorreLuque:2021nxb} (where we refer the reader for the technical details) but for a Steiman-Cameron 3D distribution (four-arm model~\citep{Steiman-Cameron:2010iuq}) of the gas density (implemented according to the radial and vertical dependence found in~\citet{ferriere2007spatial}) and source distribution (following~\citet{Lorimer_2006} and whose injection is parametrized with a broken power-law fitted to data). More details are given in~\citet{DRAGON2-1}.
The scaling factors corresponding to B, Be and Li are $\mathcal{S}_\textrm{B} = 0.99$, $\mathcal{S}_\textrm{Be} = 0.92$ and $\mathcal{S}_\textrm{Li} = 0.88$, respectively. 
In this analysis, we use inelastic cross-sections from the CROSEC parameterisation~\citep{Barashenkov:1994cp} and the DRAGON2 spallation cross sections for the production of secondary CRs~\citep{DelaTorreLuque:2021joz, Evoli:2019wwu}. Other common parameterizations are those from {\sc Galprop}~\citep{GALPROPXS, GALPROPXS1} or FLUKA~\citep{delaTorreLuque:2022vhm}. 
However, the DRAGON2 cross sections have recently shown to provide a better simultaneous fit to the B, Be and Li flux ratios, compared to other cross sections data-sets~\citep{DeLaTorreLuque:2021nxb}.

In this analysis, we consider uncorrelated AMS-02 errors, since the collaboration did not report the full covariance matrix~\cite{AGUILAR20211}. However, including correlations in AMS-02 error data only results in an increase in the uncertainty in the determination of the propagation parameters. This is not specially relevant here, since we find that the systematic uncertainties associated with resolving the Alfvén velocity completely dominate over the other related uncertainties.

\section{Mass dependence of dark matter induced signals}
\label{sec:appendixA}

\begin{figure*}[ht!]
\includegraphics[width=0.5\linewidth]{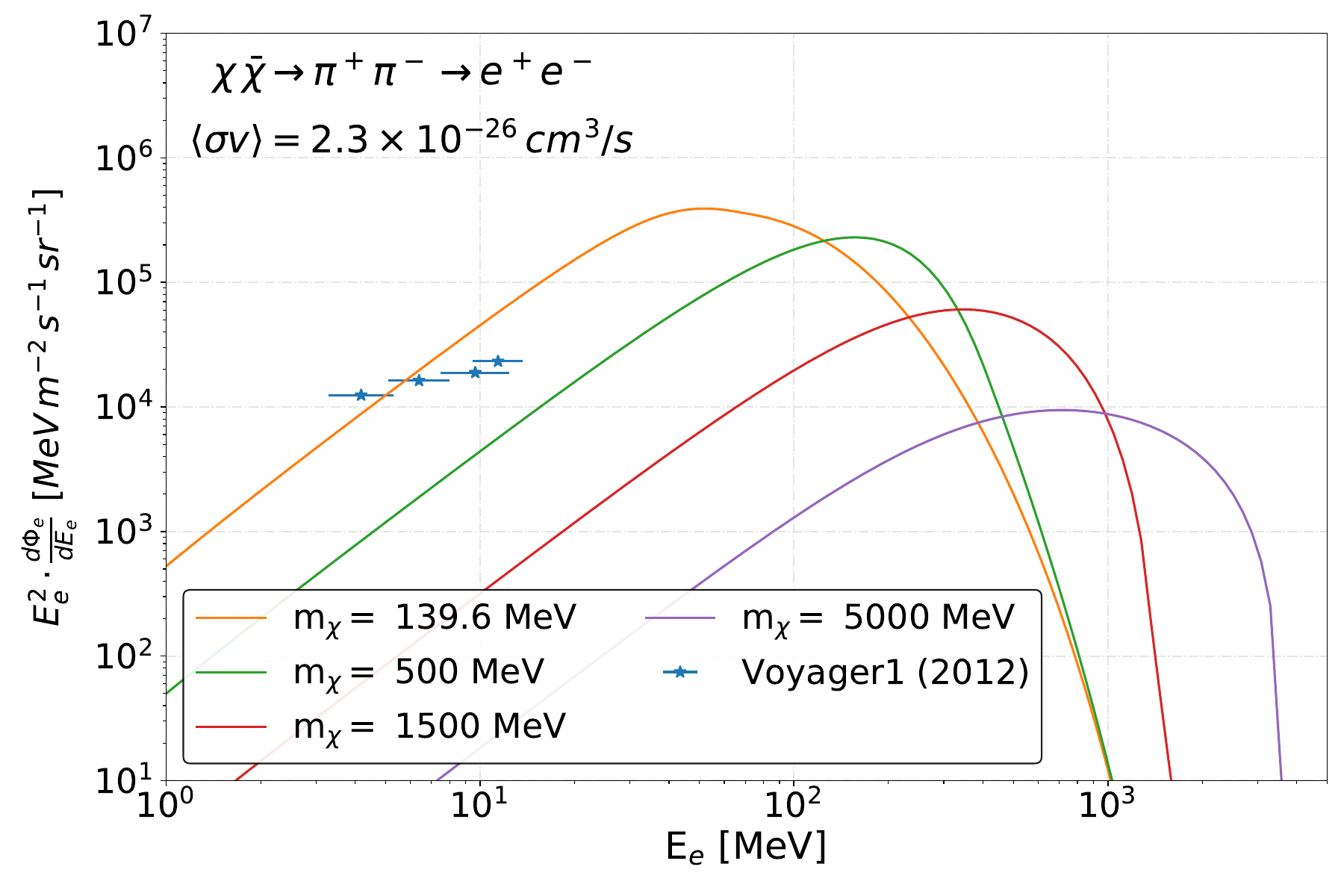}
\includegraphics[width=0.5\linewidth]{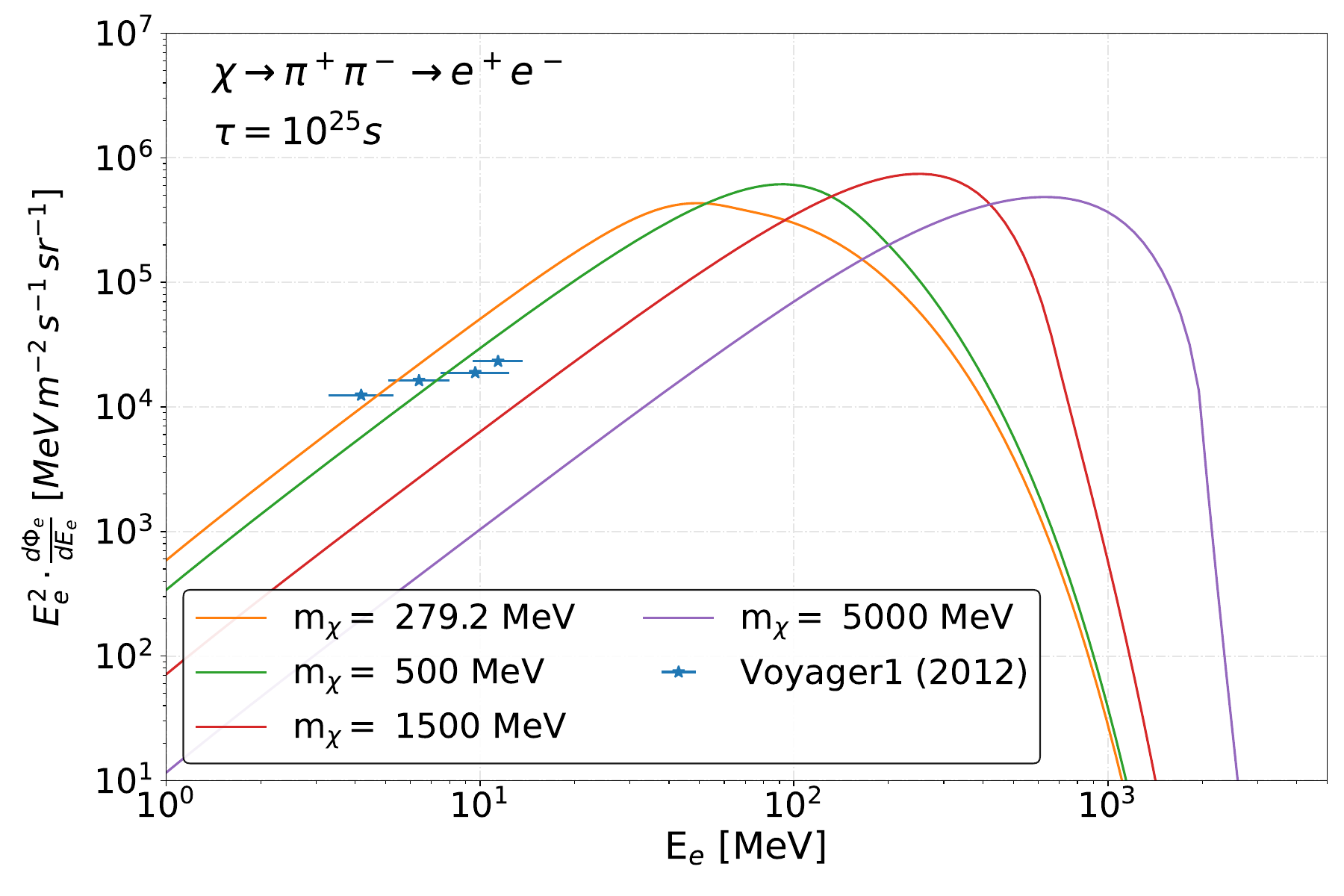}

\includegraphics[width=0.5\linewidth]{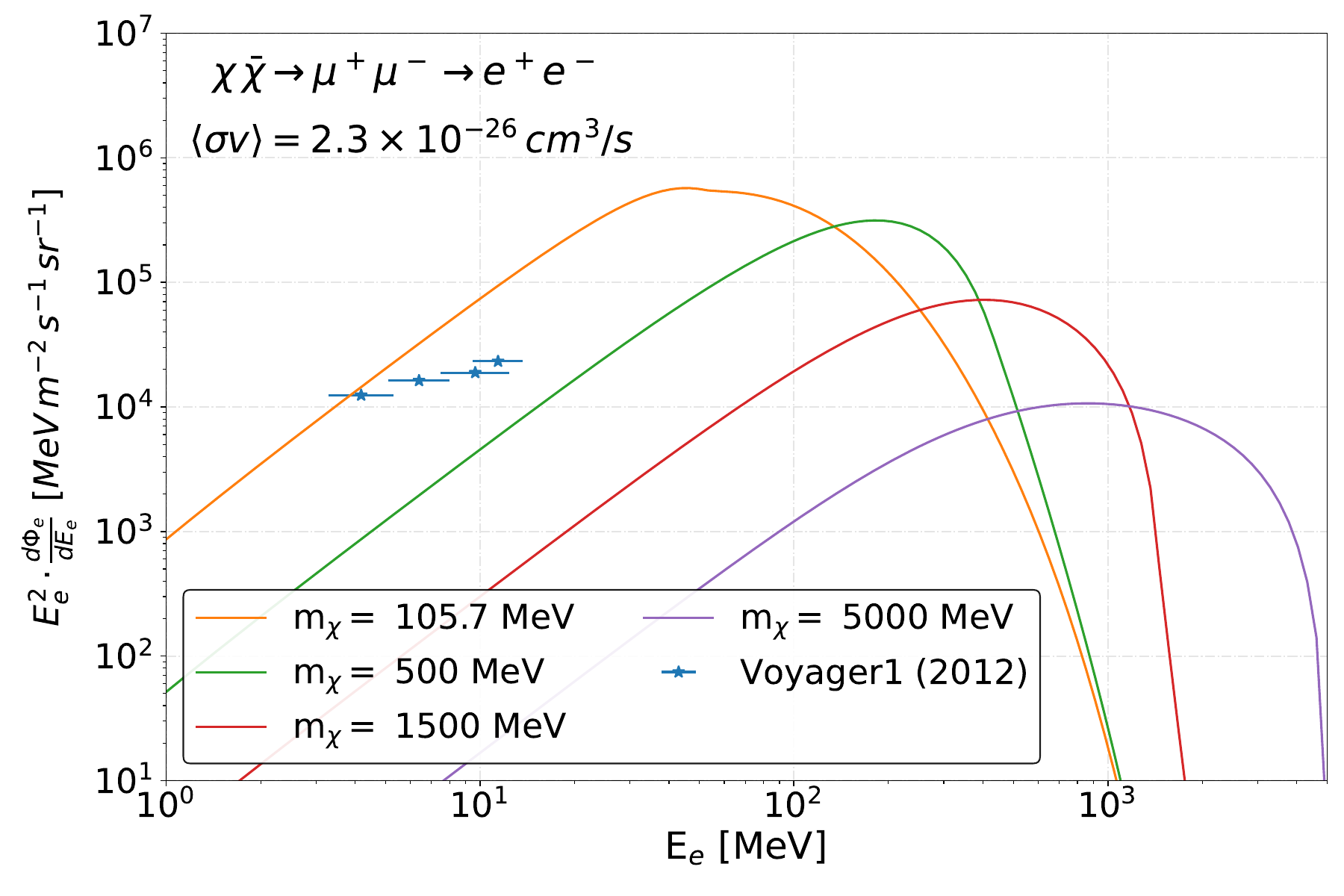}
\includegraphics[width=0.5\linewidth]{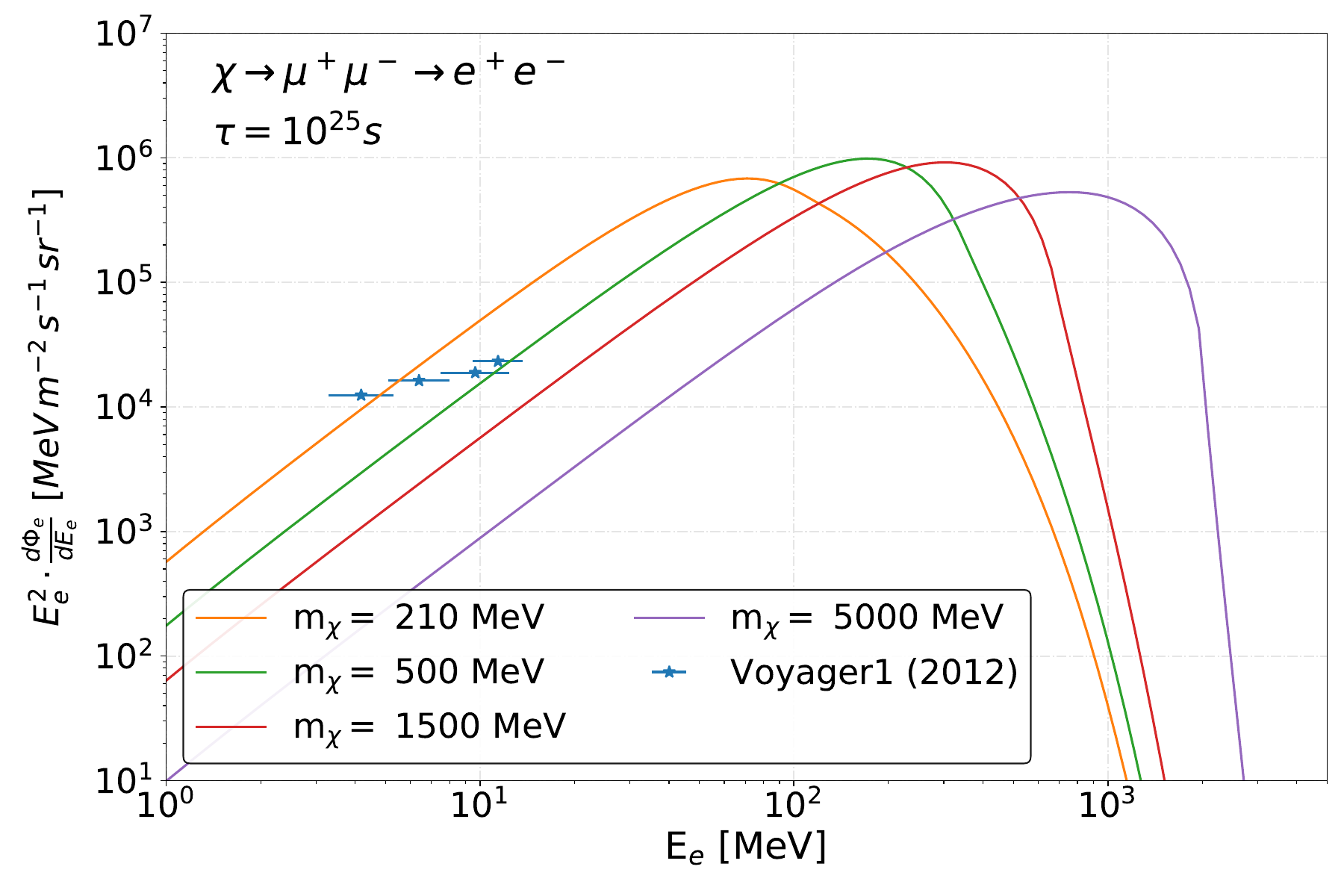}

\includegraphics[width=0.5\linewidth]{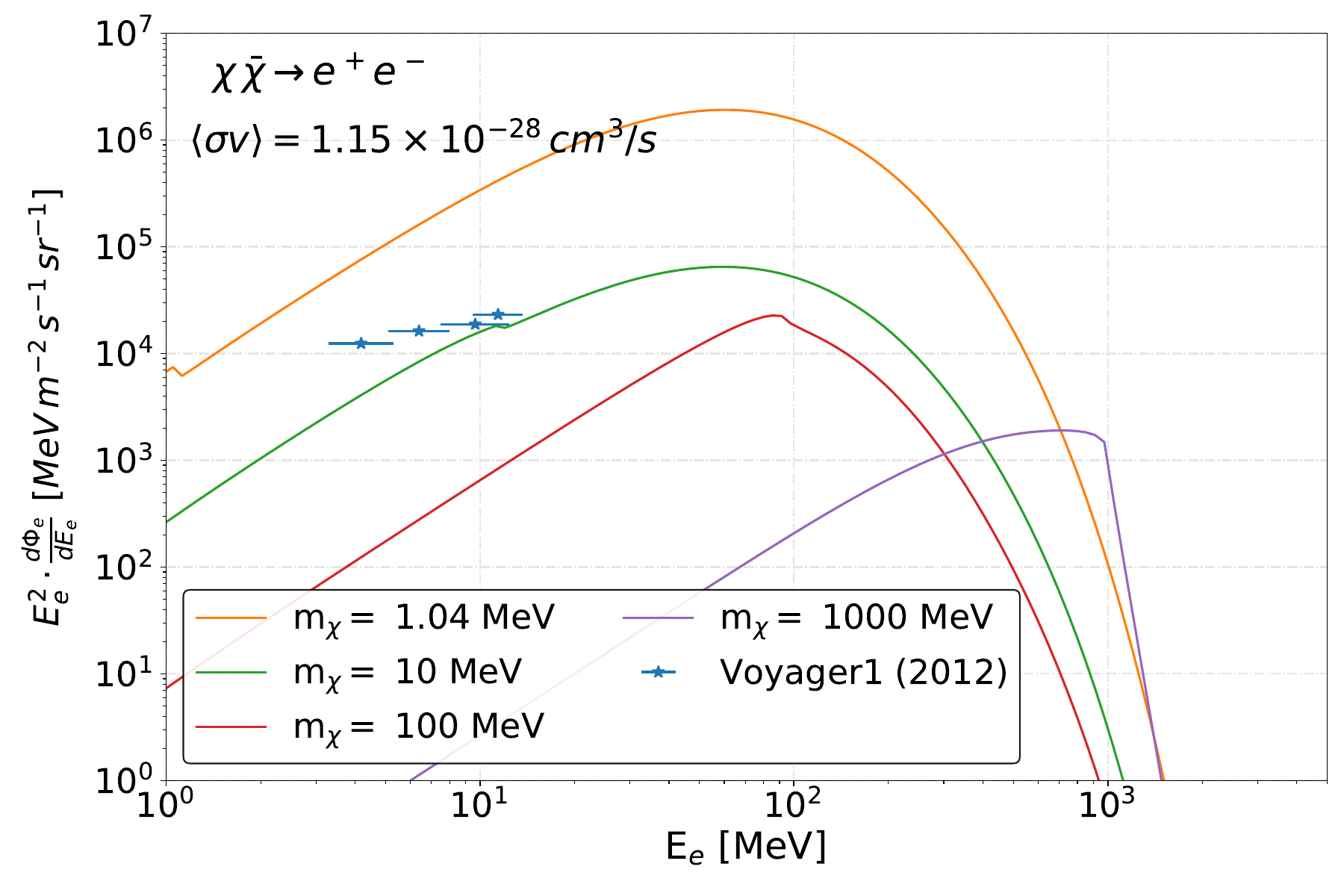}
\includegraphics[width=0.5\linewidth]{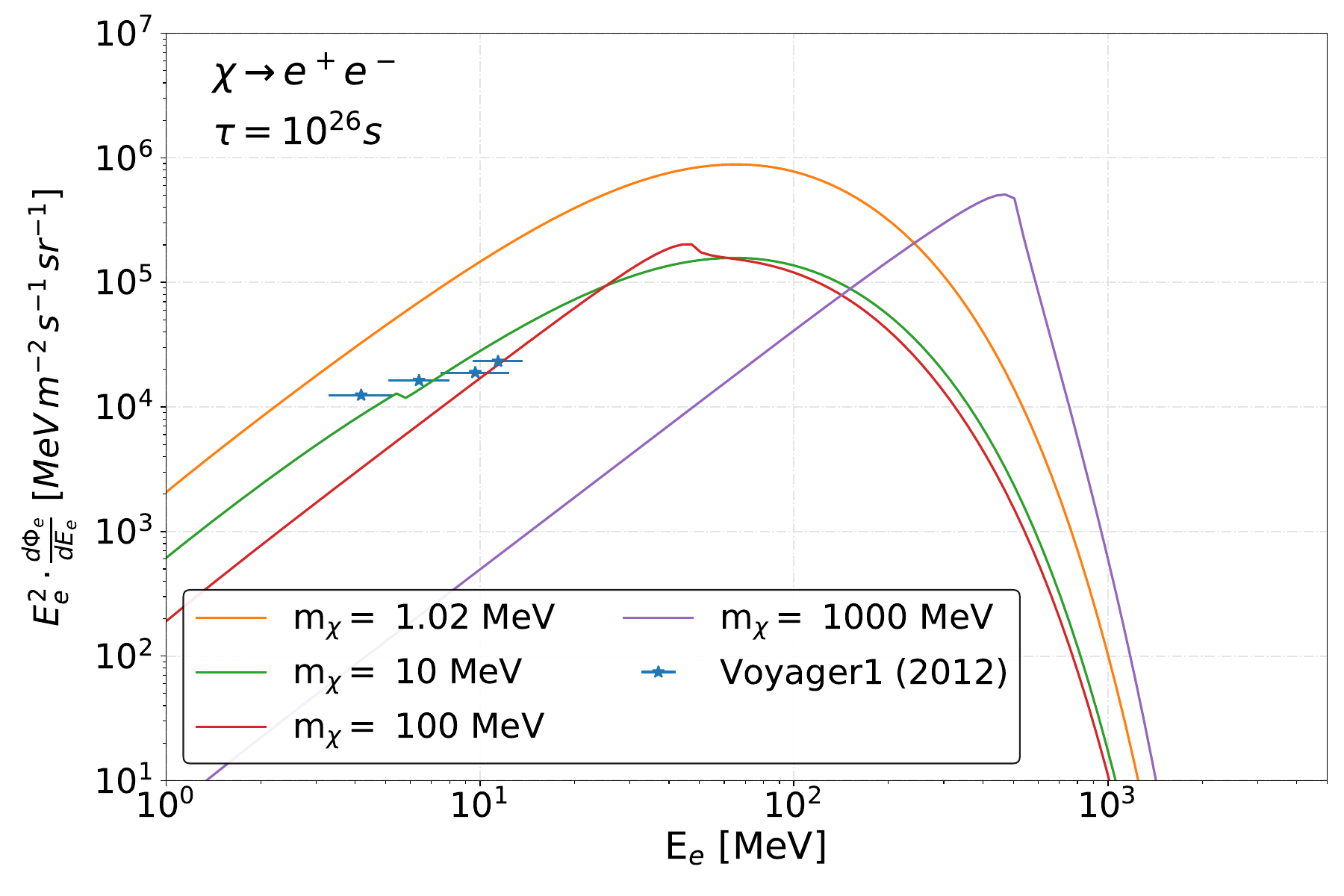}
\caption{Comparison of the predicted $e^\pm$ flux measured by {\sc Voyager 1} as a function of energy from dark matter annihilating (left column) and decaying (right column) into $\pi^+ \pi^-$ (top row), $\mu^+ \mu^-$ (middle row) and $e^+e^-$ (bottom row) for the specified annihilation cross section $\langle\sigma v\rangle$ and lifetime $\tau$ shown in the plots, and considering a Navarro-Frenk-White profile with dark matter density at Solar System position of $0.4$~GeV/cm$^3$. We show the {\sc Voyager 1} data as blue points, for reference. All the panels take our best-fit propagation parameters (Tab.~\ref{tab:params}). }
\label{fig:Voy_Comp}
\end{figure*}

In Fig.~\ref{fig:Voy_Comp} we show the DM-induced $e^\pm$ flux injected from DM annihilation (left plots) and decay (right plots) for various DM masses, $m_\chi$, after propagation, for the $\pi^+ \pi^-$ (top panels), $\mu^+ \mu^-$ (middle panels) and direct $e^+ e^-$ (bottom panels) channels. We consider the best-fit propagation parameters shown in Tab.~\ref{tab:params} and the values of $\langle \sigma v\rangle$ and $\tau$ are shown inside the figures in each case. We note that different masses considered lead to markedly different shaped spectra and their normalisation can be easily understood from their injection spectrum (see Eq.~\eqref{eq:promptflux}).  In addition, since we are comparing predictions directly to {\sc Voyager 1} data, we show the total $e^\pm$ flux.

As in Fig.~\ref{fig:Va_ee_Scan_Voy}, we observe that the propagated flux peaks at an energy slightly below the peak of the injected flux. The peak of the spectrum is once again shifted to higher energies as expected due to reacceleration. 
We also remark on the importance of energy losses (dominated here by Coulomb and ionisation losses, which mainly depend on the interstellar gas density distribution), in particular for the direct $e^+ e^-$ channel, all the $e^\pm$ flux lies below the DM mass threshold in energy due to this phenomenon. 

\begin{figure*}[t!]
\includegraphics[width=0.5\linewidth]{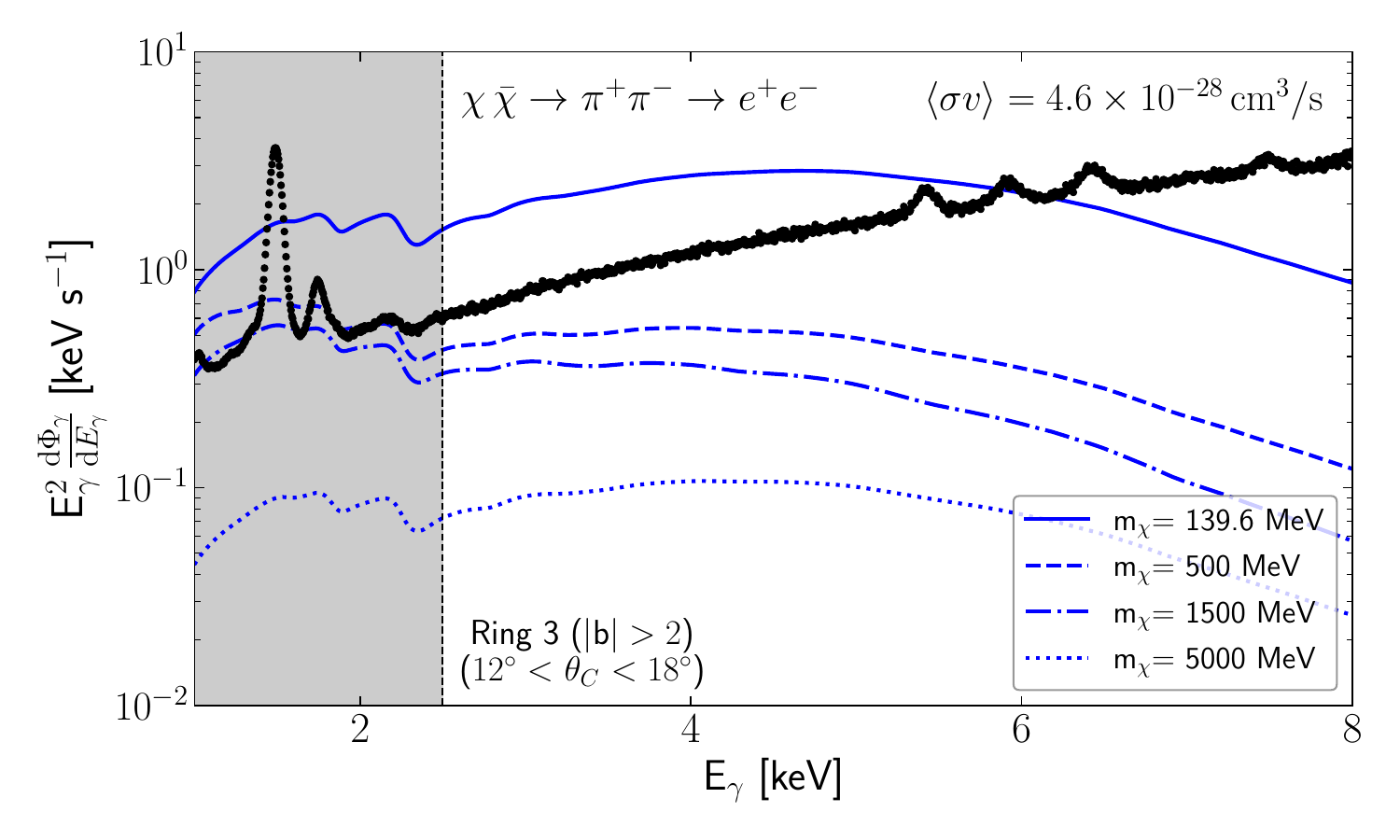}
\includegraphics[width=0.5\linewidth]{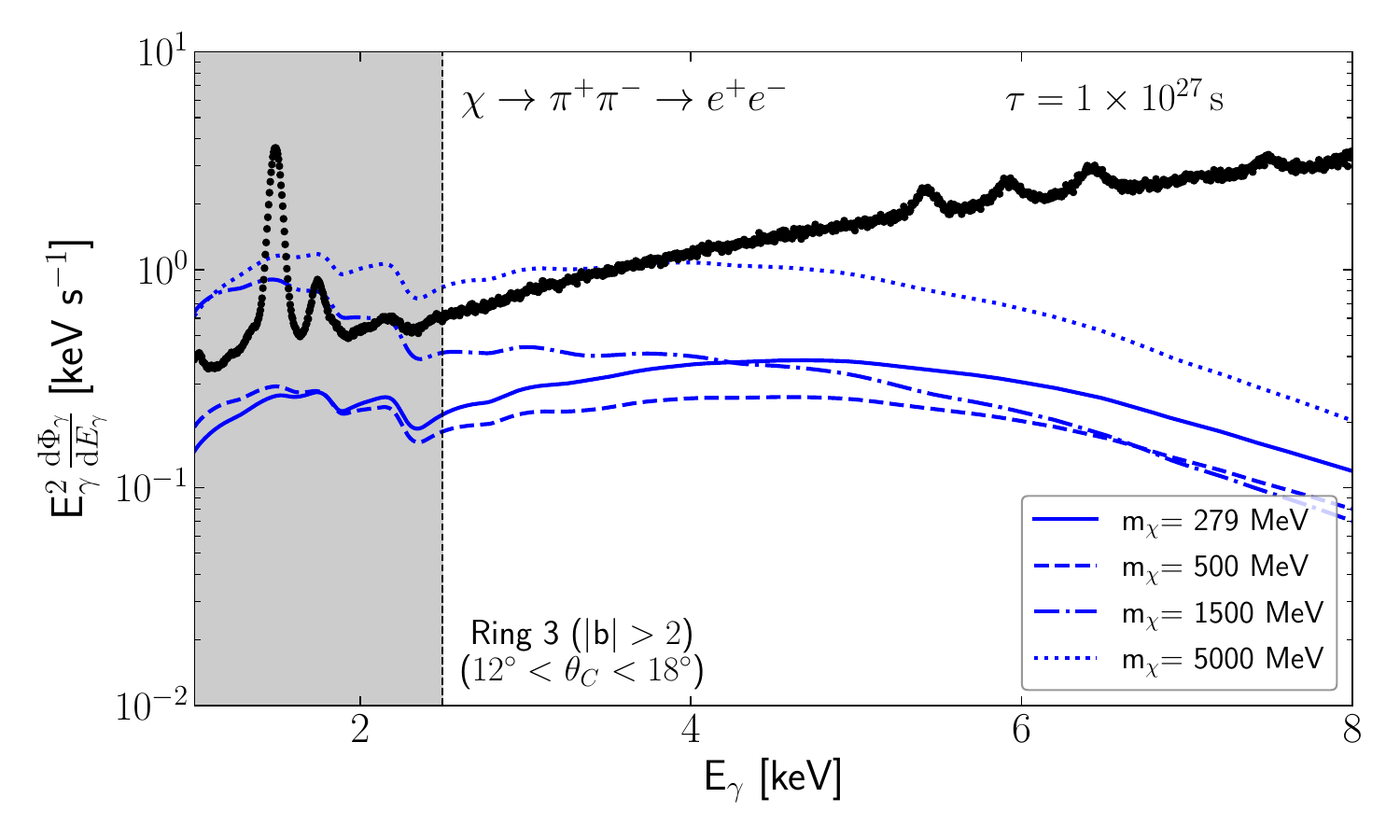}

\includegraphics[width=0.5\linewidth]{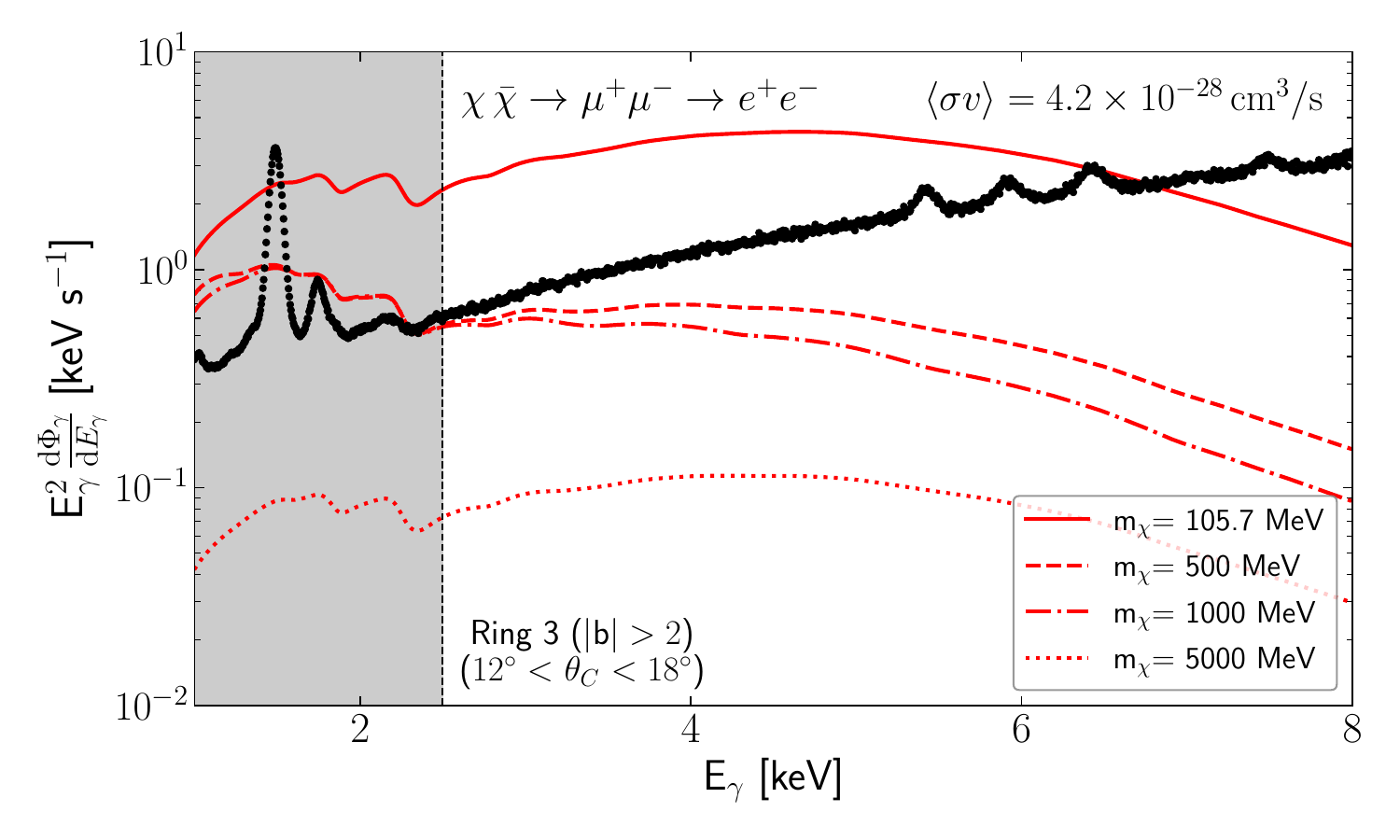}
\includegraphics[width=0.5\linewidth]{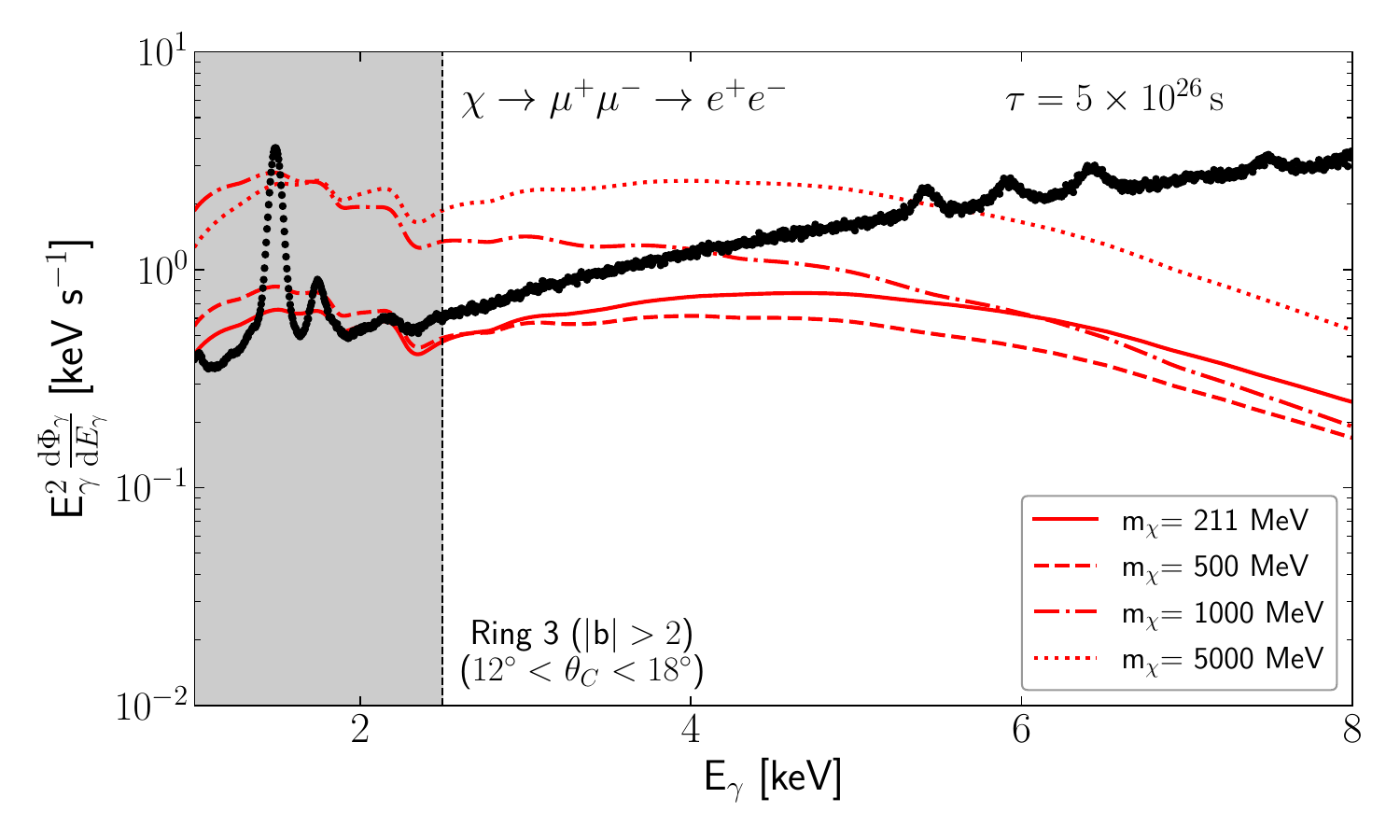}

\includegraphics[width=0.5\linewidth]{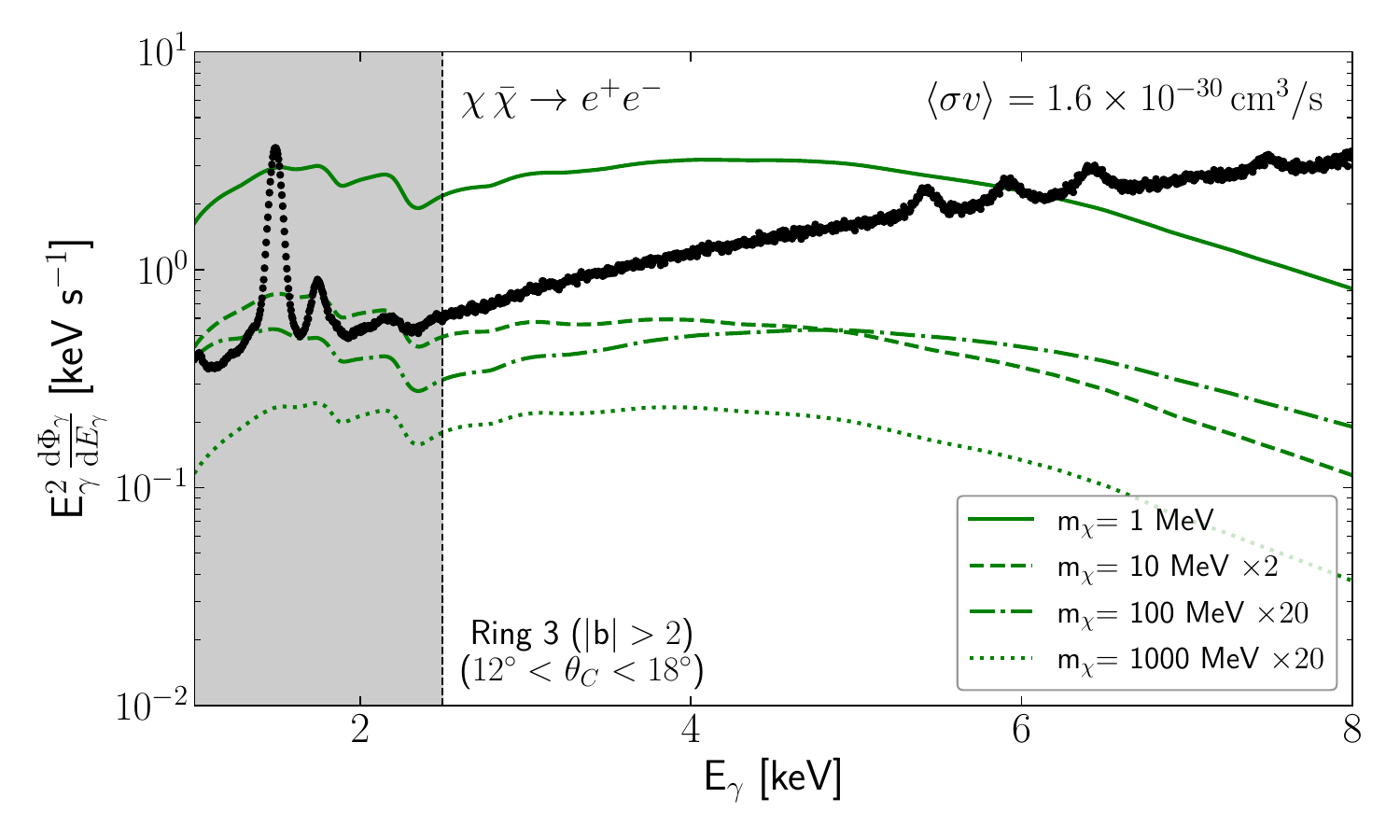}
\includegraphics[width=0.5\linewidth]{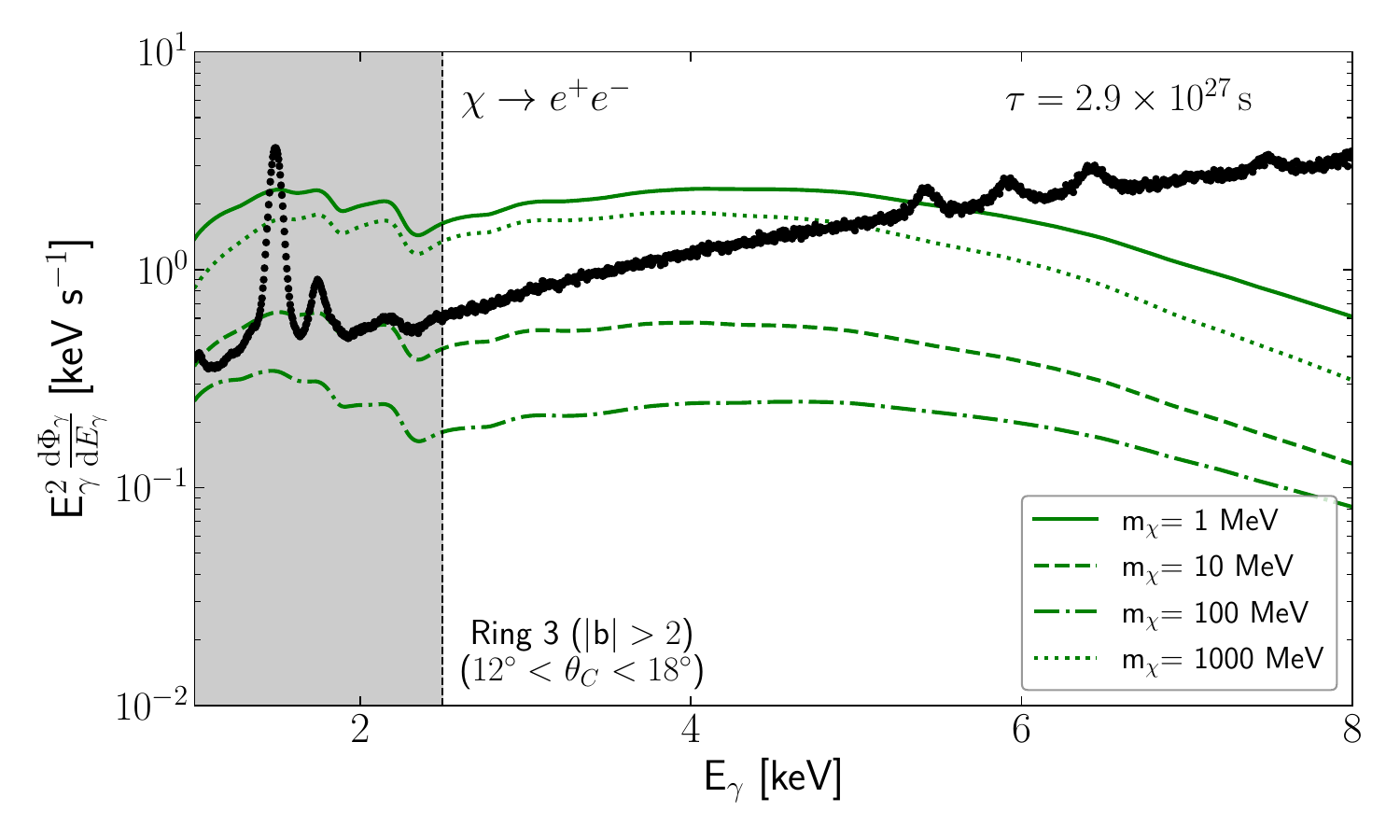}
\caption{Comparison of MOS data at Ring 3~\citep{Foster:2021ngm} with the predicted secondary signal for dark matter annihilating (left column) and decaying (right column) into $\pi^+\pi^-$ (top row), $\mu^+\mu^-$ (middle row) and $e^+ e^-$ (bottom row), for different masses. We show the total $1\sigma$ error bars for the data and shade the region not used in the fits to this data since we consider only the range $2.5$-$8$~keV.}
\label{fig:XMM_Comp}
\end{figure*}

Similarly, we show in Fig.~\ref{fig:XMM_Comp} the DM-induced $X$-ray signals generated by annihilation (left plots) and decay (right plots) for various $m_{\chi}$, in the Galactic region corresponding to the Ring 3 dataset (MOS detector). Again here, the $\pi^+ \pi^-$, $\mu^+ \mu^-$ and direct $e^+ e^-$ channels are shown in the top, middle and bottom panels respectively, and we consider the secondary radiations generated from the $e^\pm$ created by DM following a NFW profile.

\section{Uncertainties related to the propagation parameters}
\label{sec:appendixB}

\begin{figure*}[h!]
\includegraphics[width=0.5\linewidth]{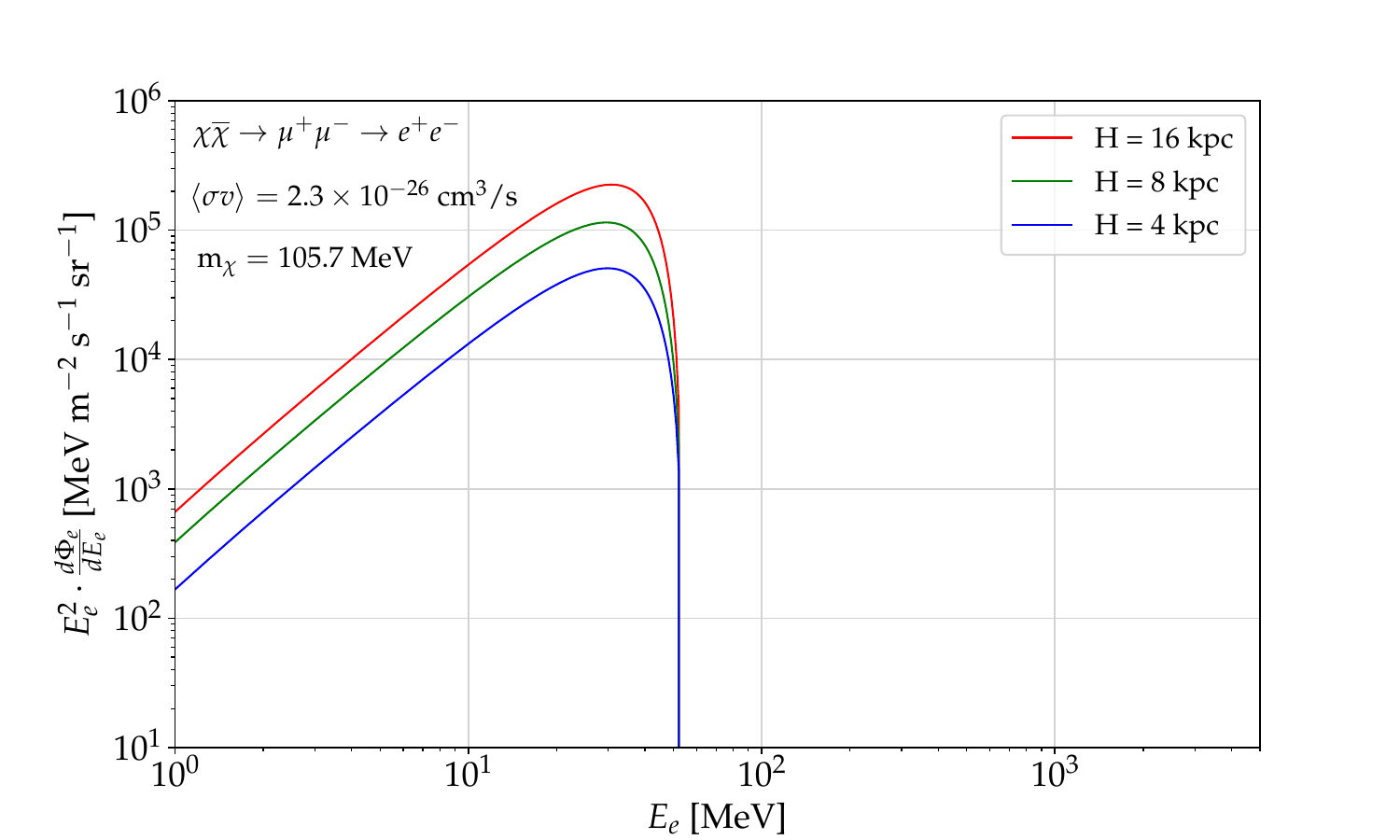}
\includegraphics[width=0.5\linewidth]{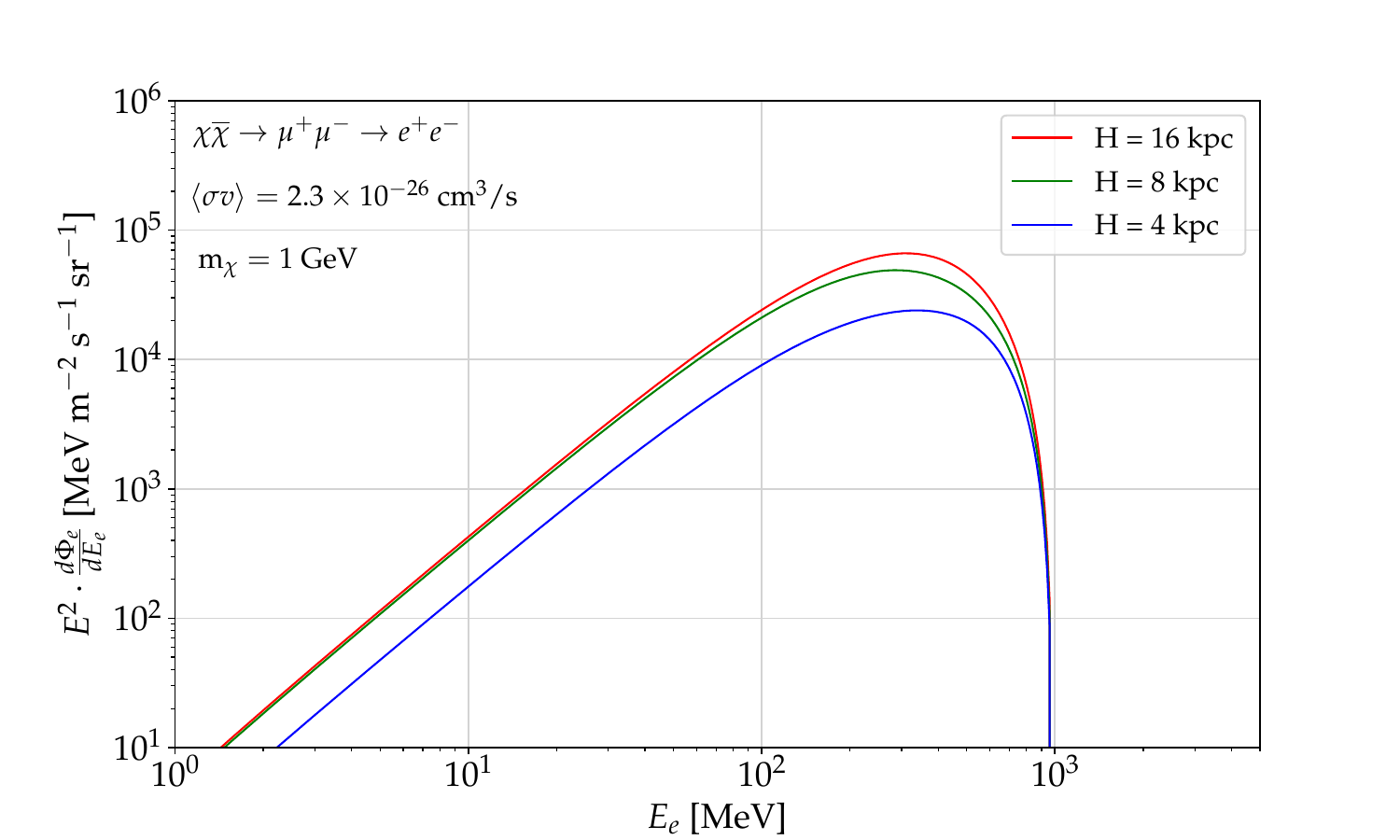}
\caption{Impact of using different values for the halo height $H$ (keeping the $D/H$ ratio constant and $v_A = 0$~km/s) on the $e^\pm$ fluxes. We show the predicted signals from annihilation in the $\mu^\pm$ channel for a mass of $m_\mu=105.7$ MeV (left panel) and $1$~GeV (right panel) for $H=16$, $8$ and $4$~kpc as a red, green and blue line, respectively.}
\label{fig:halosize}
\end{figure*}

The propagation parameters are highly uncertain at sub-GeV energies. Apart from the effect of reacceleration in these signals, which is discussed at length in the main text, one could expect that the halo height value also considerably impacts our evaluations. This is because the amount of DM produced in the Galaxy depends on its volume (and therefore the halo height) and the normalisation of the diffusion coefficient is constrained from the ratio $D/H$, that is found from fits to secondary-to-primary CR species (e.g. B/C). Although the value of $H$ in this work is fixed by the study of $^{10}$Be ratios, this value is highly uncertain due to the uncertainties in cross sections of $^{10}$Be production. Therefore, we illustrate in Fig.~\ref{fig:halosize} different evaluations of the DM $e^{\pm}$ at Earth without including reacceleration, for values of $H=16$~kpc, $8$~kpc (best-fit value) and $4$~kpc, in the $\mu^+ \mu^-$ channel for $m_{\chi}\simeq 100$~MeV (left panel) and $\simeq1$~GeV (right panel).
As we see, modifying the halo height by a factor of $2$ mainly changes the normalisation of the signal (by a factor of a few) and slightly depends on the DM mass (since higher DM mass means there is smaller amount of DM particles for a fixed DM density).
\begin{figure*}[h!]
\includegraphics[width=0.5\linewidth]{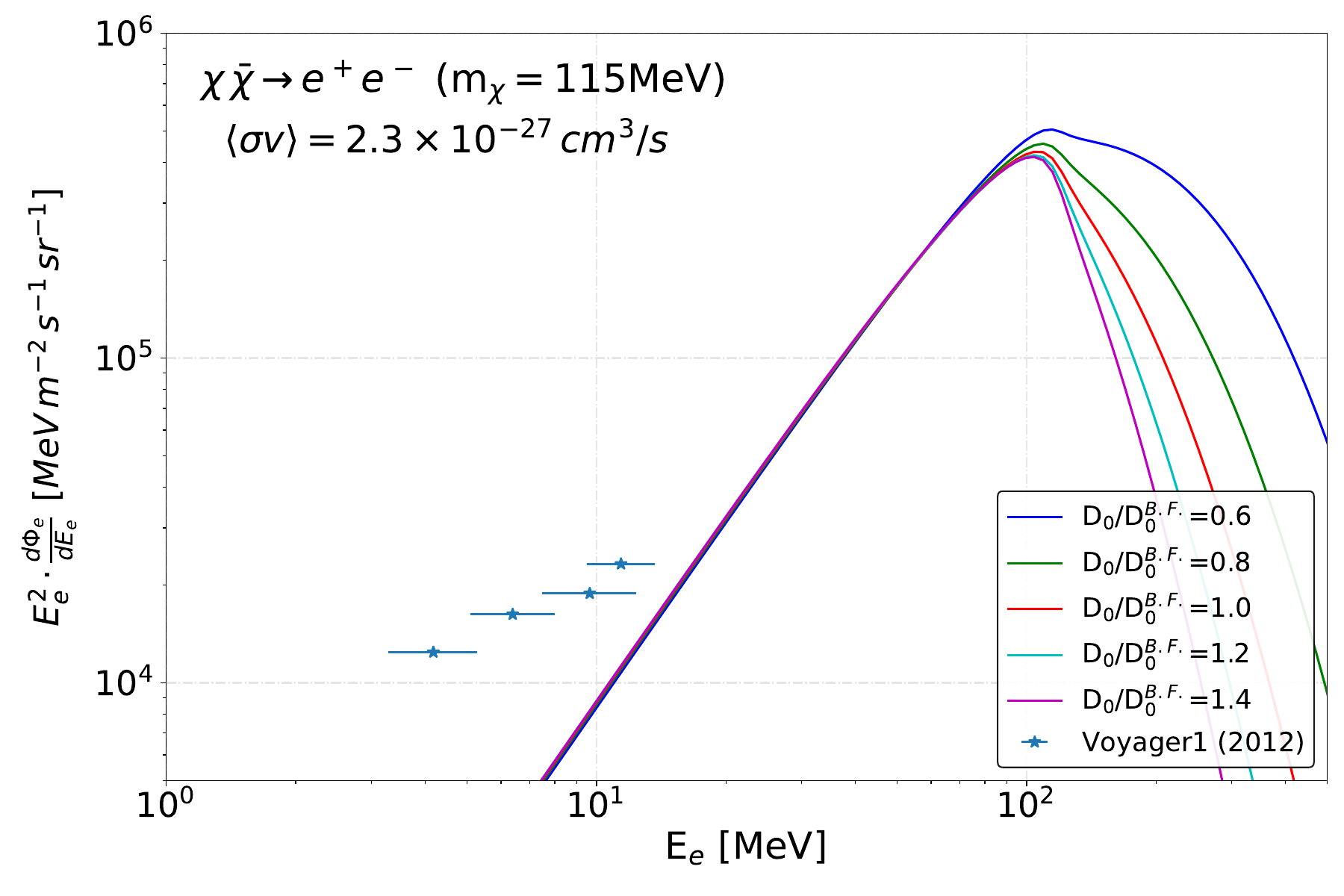}
\includegraphics[width=0.5\linewidth]{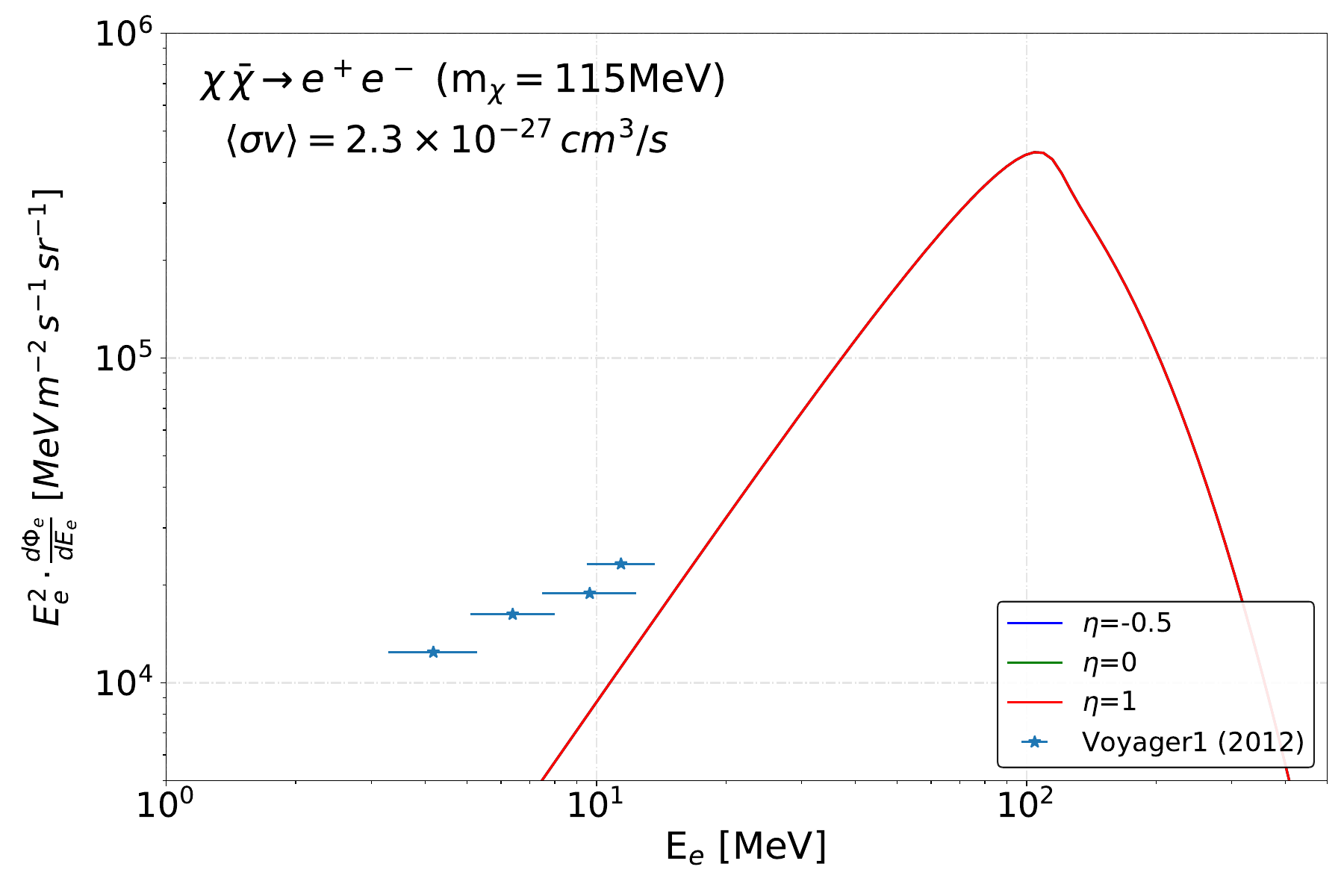}
\begin{center}
\includegraphics[width=0.5\linewidth]{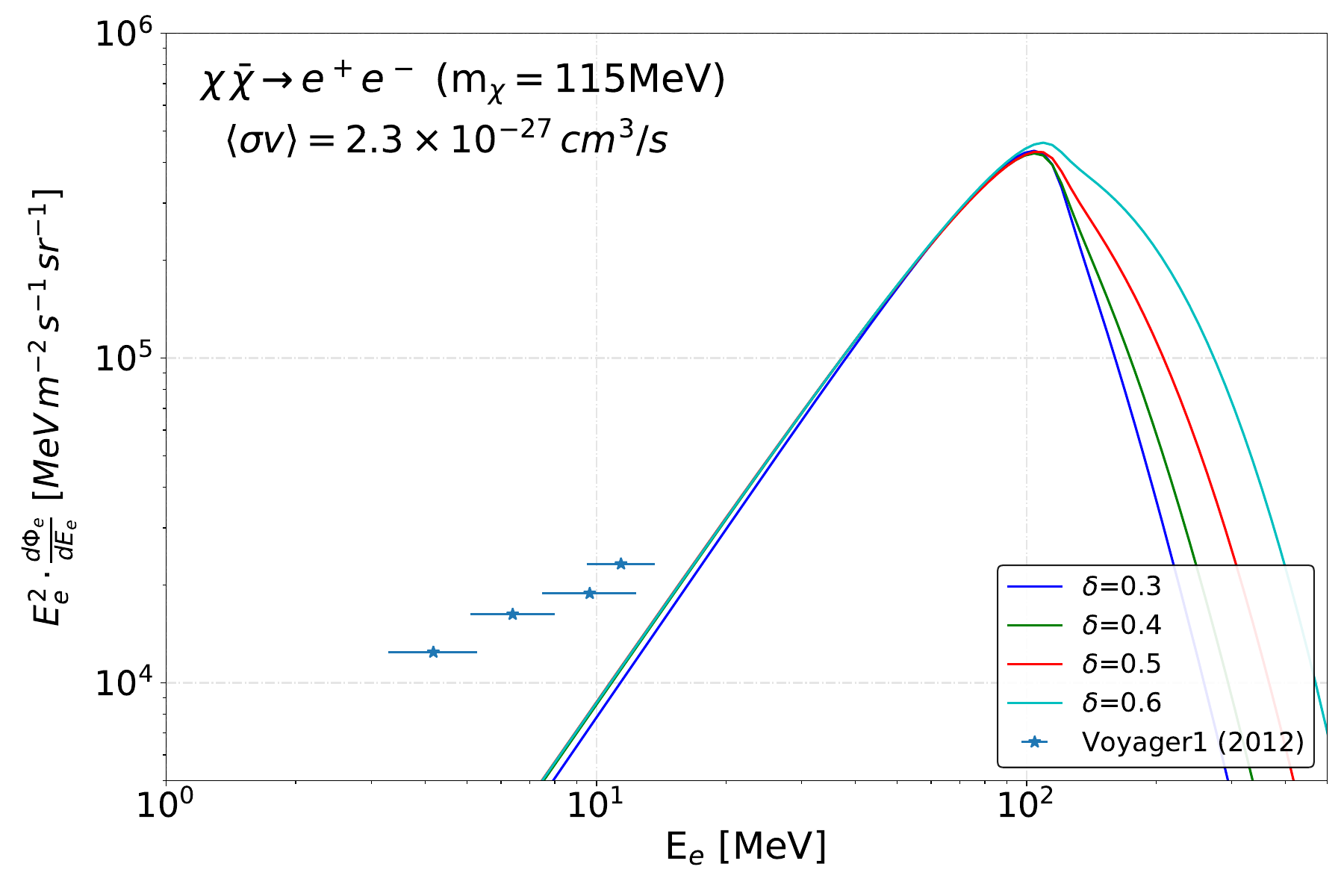}
\end{center}
\caption{Effect of variations on $\eta$, $D_0$ and $\delta$ parameters in the local $e^{\pm}$ spectrum for the direct annihilation of a $115$~MeV DM particle into $e^{\pm}$ for an annihilation cross section of $\langle \sigma v\rangle = 10^{-27}$ cm$^3$/s.}
\label{fig:ScanParams}
\end{figure*}

Uncertainties related to the other propagation parameters are much smaller, always well below a factor of $2$. We have tested this by modifying the diffusion parameters in our main setup in a broad range. One can see that variations of these parameters lead to a very small change in the predicted DM signals in the energy region covered by {\sc Voyager 1} data in Fig.~\ref{fig:ScanParams}. In particular, we show the signals predicted at Earth for different reasonable variations of the $\eta$, $D_0$ and $\delta$ parameters for the direct annihilation of a $\sim100$~MeV DM particle into $e^{\pm}$ for an annihilation cross section of $\langle \sigma v\rangle = 10^{-27}$ cm$^3$/s. 
We show variations of the normalization of the diffusion coefficient, $D_0$, in the upper left panel, where the legend provides the ratio with respect to the best-fit value ($D_0=1.02\times10^{29}$~cm$^2$/s). As we see, variations on this parameter leads to important changes of the spectrum above $\sim 100$~MeV. Below this energy, the differences between the predicted spectra are always below $10\%$ for the variations considered here. In turn, variations of the $\eta$ parameter (top right panel) do not lead to changes above a few hundred of MeV (since $\beta=1$). At Voyager energies, this parameter does not lead to sizeable changes in the predicted spectrum ($\lesssim 2$-$3\%$). However, modifications of this parameter changes our predictions more significantly at lower energies, and can lead to variations of even tens of percent at keV energies. 
Finally, we see that variations on the spectral index of the diffusion coefficient, $\delta$, significantly changes the predicted $e^{\pm}$ flux above around a hundred MeV. In addition, variations in this parameter leads to changes of up to $\sim15\%$ in the predicted spectra at $10$~MeV. As in the case of variations of the other parameters, the changes in the predicted spectra grow at lower energies, but they are always much smaller than the variations expected from the dominant uncertainties coming from the determination of the halo height (leading to uncertainties of up to a factor of a few) and the Alfvén velocity (that leads to changes of orders of magnitude at low energies).
Other common parametrizations of the diffusion coefficient, such as those explored in~\citet{Weinrich_2020} and \cite{silver2024testing}, are expected to produce similar predictions to ours in the {\sc Voyager 1} energy range, but could deviate from our predictions by a few tens of percent at lower energies. We notice that the energy dependence of the diffusion coefficient is not a very relevant issue, but the parametrization of its spatial dependence and the possible strong winds near the GC could significantly affect the constraints from {\sc Xmm-Newton} data (and the IC signals at keV energies). These effects could be interesting to explore in a future work.

\begin{figure*}[t!]
\includegraphics[width=0.54\linewidth]{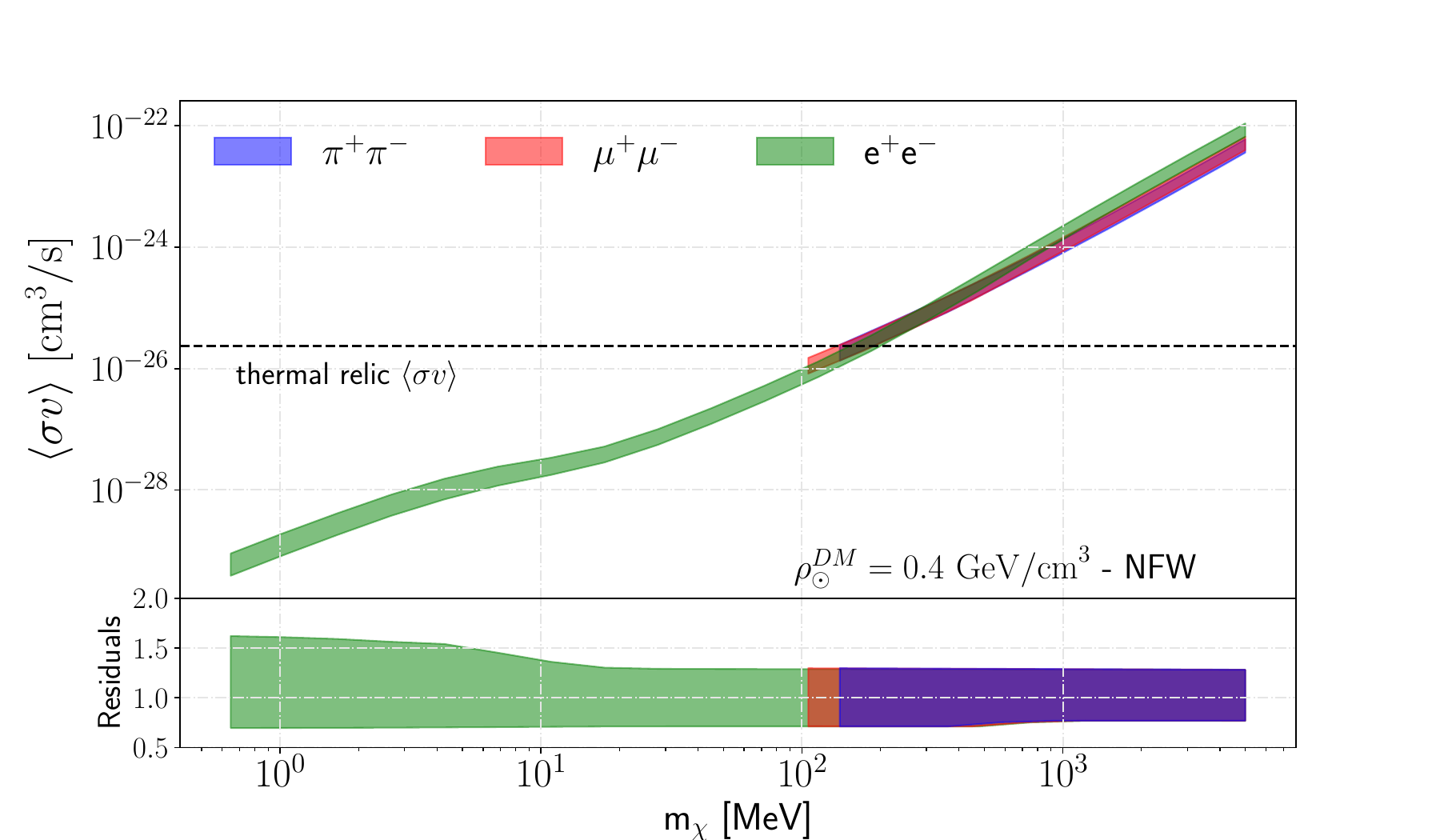} \hspace{-0.7cm}
\includegraphics[width=0.54\linewidth]{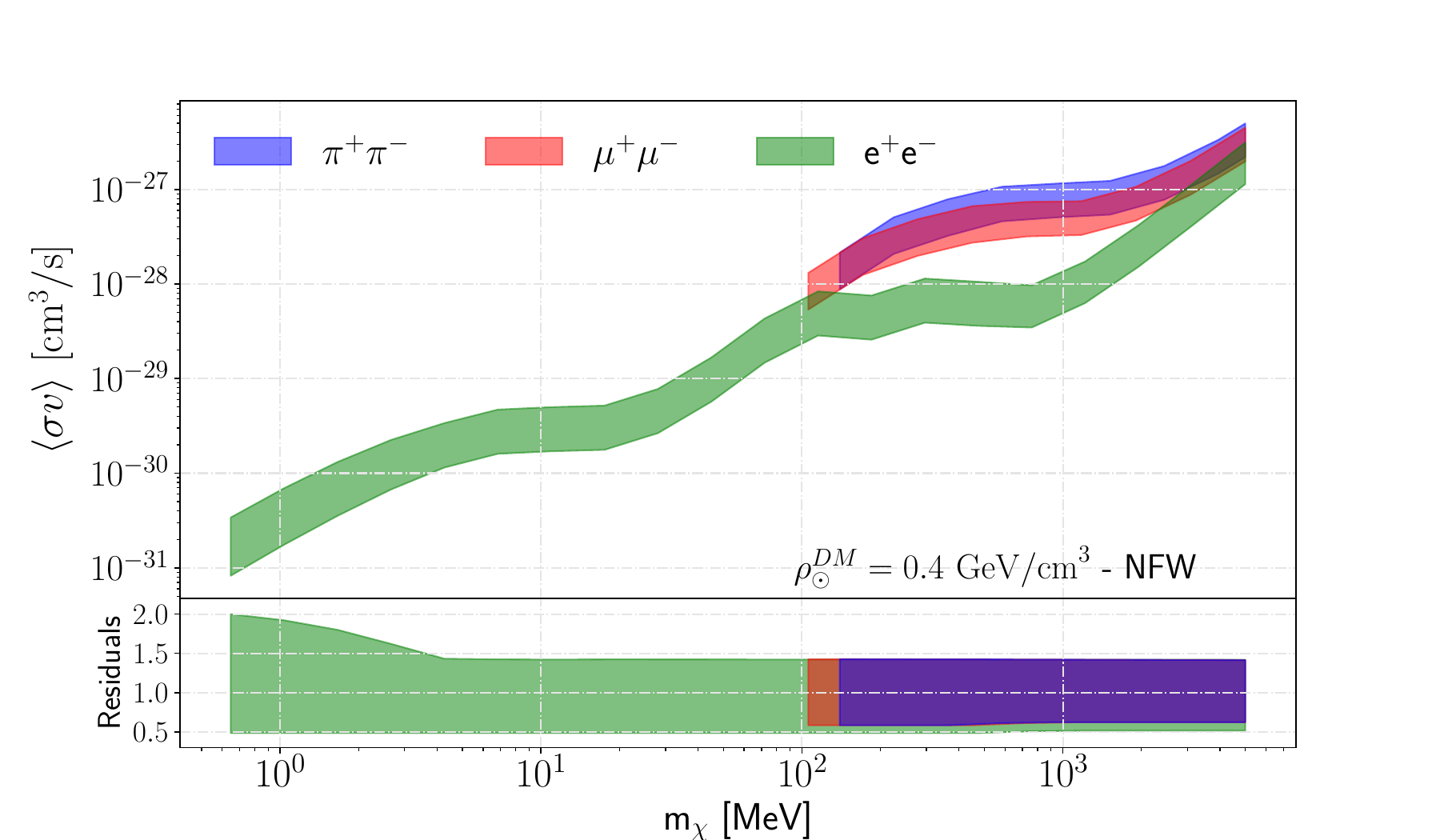}
\caption{{\sc Xmm-Newton} bounds for $v_A = 13.4$ km/s (solid lines) and 0 km/s (dot-dashed line)} with the ones in~\citet{Cirelli:2023tnx} (dashed lines) for the $\pi^+\pi^-$ (blue line), $\mu^+ \mu^-$ (red line) and $e^+ e^-$ (green line) channels respectively. Bottom panels show the estimated uncertainties in our predictions, as discussed in the main text.
\label{fig:Unc_Bounds}
\end{figure*}

In Fig.~\ref{fig:Unc_Bounds} we show the uncertainties in the estimation of the bounds for a NFW profile. These include the same sources of uncertainties as in Fig.~\ref{fig:new+oldbounds}, which include uncertainties in the propagation parameters, normalization of DM density at Earth, gas distribution and energy losses. In addition, in the case of XMM-Newton data, we also include the uncertainty related to the ISRFs (see the discussion at the end of Sec.~\ref{sec:Results}.)

\section{Other bounds from {\sc Xmm-Newton} and final state radiation signals}
\label{sec:appendixC}

\begin{figure*}[h!]
\includegraphics[height=0.23\textheight, width=0.52\linewidth]{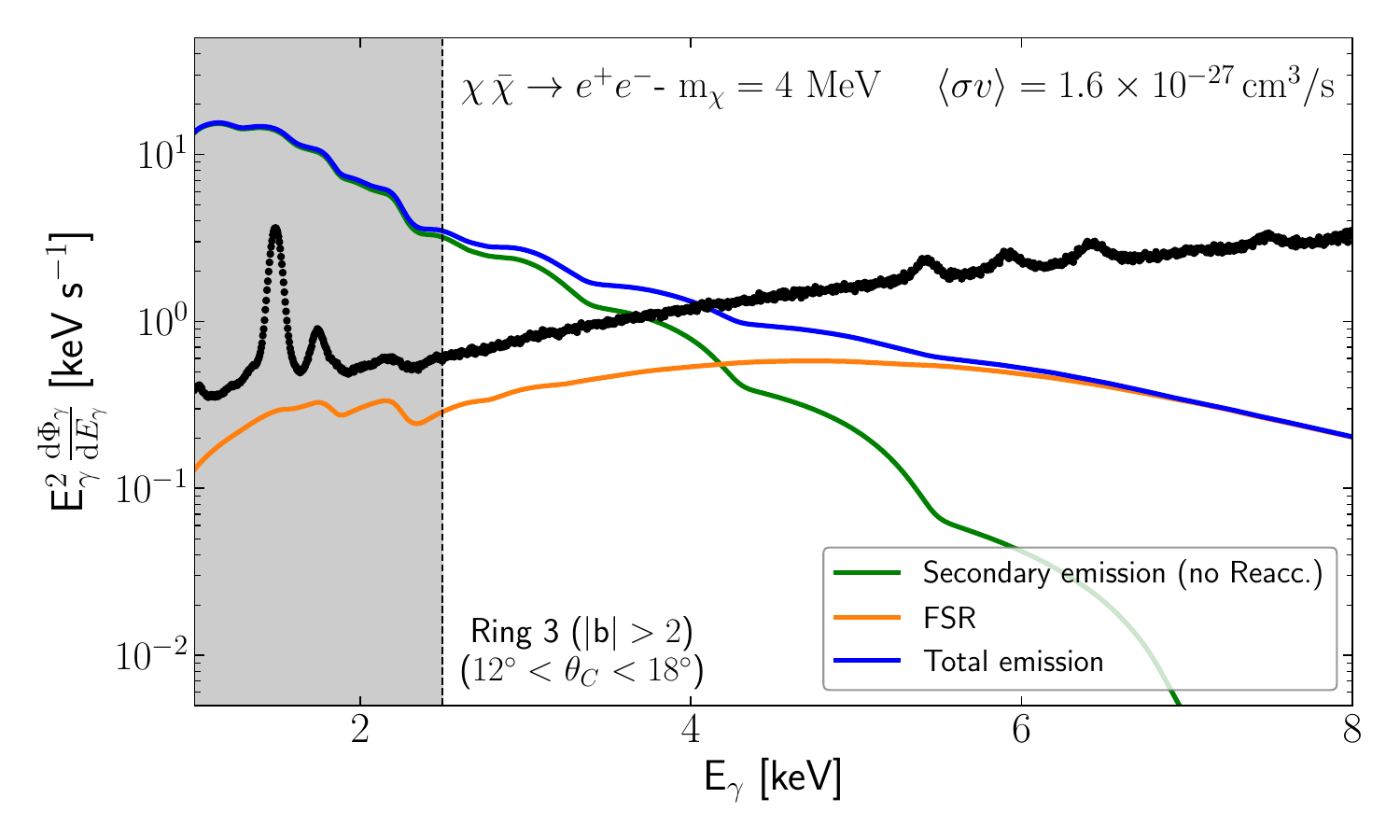}
\includegraphics[width=0.49\linewidth]{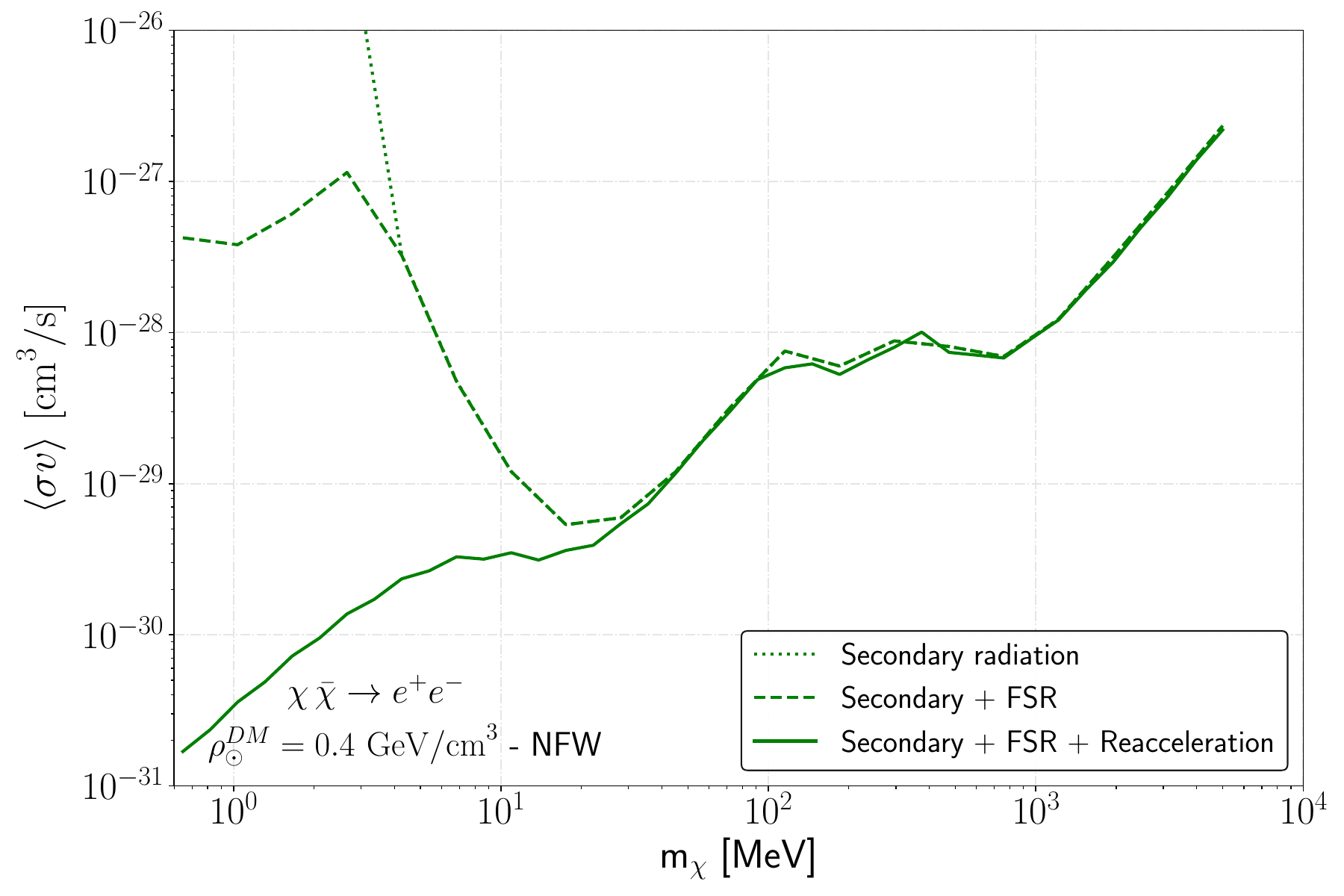}
\caption{Left panel: Prompt (yellow line) and secondary (green line) $X$-ray signals from $\simeq4$~MeV DM annihilating into $e^+e^-$, compared to MOS data at Ring 3~\citep{Foster:2021ngm}. The secondary DM contribution is obtained in the case of no reacceleration (the only case when FSR radiation dominates over the secondary emission). Right panel: Comparison of the bounds (annihilation into $e^+e^-$) derived from the combination of all the ring datasets in the case of only secondary emission with no $e^\pm$ reacceleration (dotted line), the case of FSR emission and secondary emission with no reacceleration (dashed line) and the case including FSR and the secondary emission with $e^\pm$ reacceleration (solid line).}
\label{fig:FSR}
\end{figure*}
We show in Fig.~\ref{fig:FSR}, a comparison between the prompt photon emission (FSR) and secondary photon emission associated with the annihilation of a $\simeq4$~GeV DM particle in the direct $e^+e^-$ channel (left panel). The secondary emission here is obtained in the case of no reacceleration, which is the only case where FSR becomes relevant in our evaluations. In the right panel of this figure, we show the DM annihilation bounds (from the combination of all the ring datasets) in the case of only secondary emission with no $e^\pm$ reacceleration (dotted line), the case of FSR emission and secondary emission with no reacceleration (dashed line) and the case including FSR and the secondary emission with $e^\pm$ reacceleration (solid line).

\begin{figure*}[h!]
\begin{center}
\includegraphics[width=0.6\linewidth]{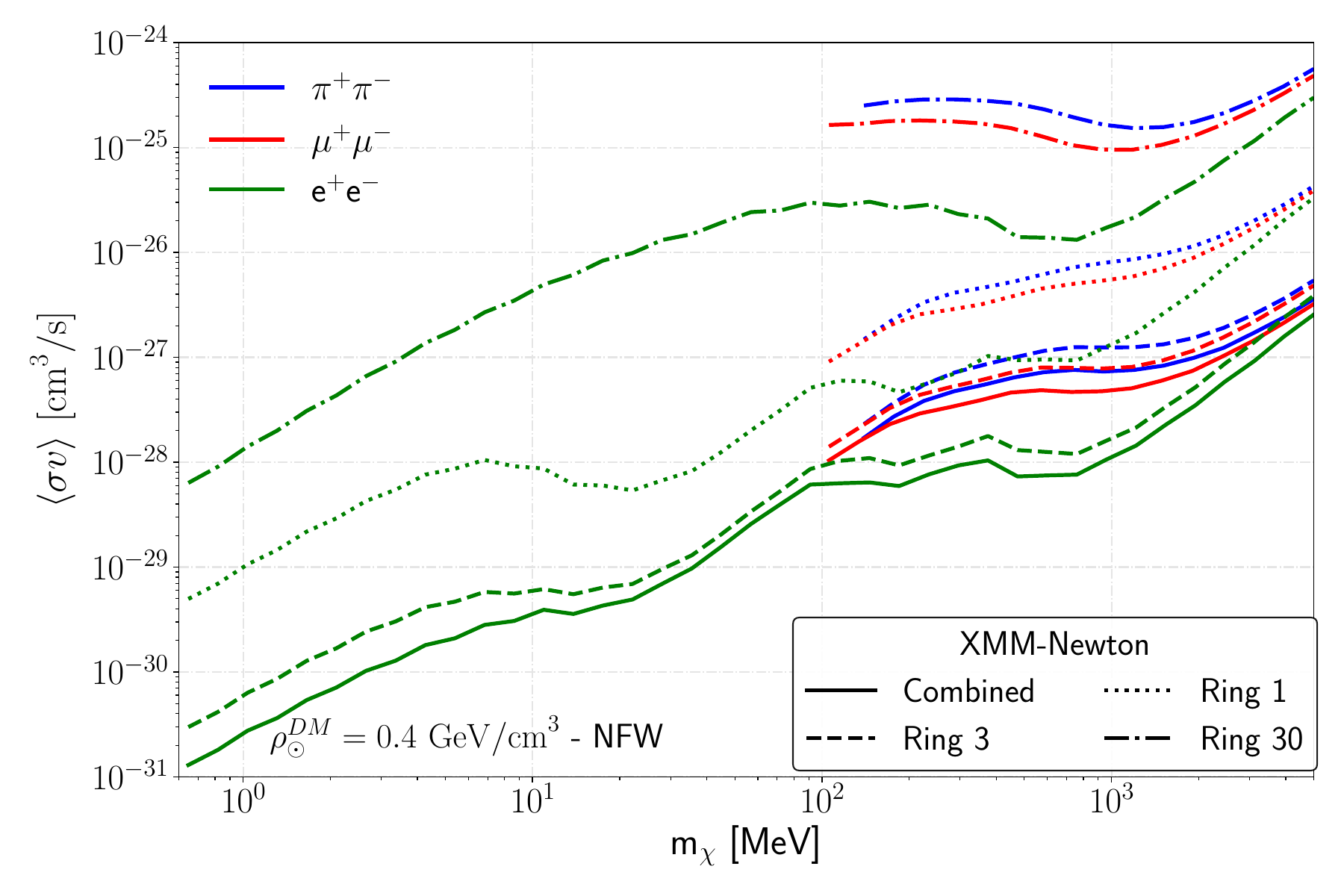}
\caption{Comparison of DM annihilation bounds obtained from different MOS ring datasets~\citep{Foster:2021ngm} for the $\pi^+ \pi^-$ (blue), $\mu^+\mu^-$ (red) and $e^+ e^-$ (green) channels. We show the combined data set (solid lines), Ring 3 (dashed lines), Ring 1 (dotted lines) and Ring 30 (dot-dashed lines).}
\label{fig:lims_XMM_Rings}
\end{center}
\end{figure*}

Finally, we compare in Fig.~\ref{fig:lims_XMM_Rings} the bounds for DM annihilation obtained for different ring datasets. We show the results for the most internal (Ring 1), most external (Ring 30), the most constraining (Ring 3) and the bounds for the combination of all the rings for the channels considered in this work and verify what was found in~\citet{Cirelli:2023tnx}, that the Ring 3 constitutes the most the most constraining ring dataset and that the combination of all the rings improve the limits by a factor $\lesssim2$. 

\clearpage

\bibliography{references.bib}
\bibliographystyle{aasjournal}



\end{document}